\documentclass[10pt,pdftex]{article}
\usepackage[a4paper,top=2cm,bottom=2.5cm,left=4cm,right=4cm]{geometry}
\usepackage[pdftex]{graphicx}
\usepackage{amsmath}
\usepackage{amsfonts}
\usepackage{overpic}
\usepackage{amsthm}
\usepackage{amssymb}
\usepackage{mathrsfs}
\usepackage[T1]{fontenc}
\usepackage{authblk}
\usepackage[english]{babel}
\usepackage{epsfig}
\usepackage{subcaption}
\usepackage{authblk}
\usepackage{lipsum}
\usepackage[dvipsnames]{xcolor}
\usepackage{epstopdf}
\usepackage{algorithmic}
\newtheorem{remark}{Remark}

\makeatletter
\newcommand{\leqnomode}{\tagsleft@true\let\veqno\@@leqno}
\newcommand{\reqnomode}{\tagsleft@false\let\veqno\@@eqno}
\makeatother

\newcommand{\Dumux}{DuMu$^\text{x}$}

\begin{document}

\title{Evaporation-driven density instabilities in saturated porous media}

\author[1]{Carina Bringedal \thanks{carina.bringedal@iws.uni-stuttgart.de}}
	
\author[1]{Theresa Schollenberger}
\author[2]{G. J. M. Pieters}
\author[3]{C. J. van Duijn} 
\author[1]{Rainer Helmig}

\affil[1]{\normalsize Department of Hydromechanics and Modelling of Hydrosystems, University of Stuttgart, Germany}
\affil[2]{Department of Earth Sciences, Utrecht University, Netherlands}
\affil[3]{Department of Mechanical Engineering, Eindhoven University of Technology, Netherlands}

\date{}

\maketitle

\begin{abstract}
   Soil salinization is a major cause of soil degradation and hampers plant growth. For soils saturated with saline water, the evaporation of water induces accumulation of salt near the top of the soil. The remaining liquid gets an increasingly larger density due to the accumulation of salt, giving a gravitationally unstable situation, where instabilities in the form of fingers can form. These fingers can hence lead to a net downwards transport of salt. We here investigate the appearance of these fingers through a linear stability analysis and through numerical simulations. The linear stability analysis gives criteria for onset of instabilities for a large range of parameters. Simulations using a set of parameters give information also about the development of the fingers after onset. With this knowledge we can predict whether and when the instabilities occur, and their effect on the salt concentration development near the top boundary.
\end{abstract}

\section{Introduction}
Evaporation of saline water from soils can cause accumulation of salts in the upper part of the soil, which has a large environmental impact as it hampers plant growth and affects biological activities \cite{salinityreview2016}. As water evaporates, the salts accumulate near the top of the porous medium, which has a negative impact on root water uptake \cite{chaves2008plant}. If the solubility limit of the salt is exceeded, the salts precipitate. In this case, a salt crust at the top of the soil is formed, disconnecting the soil from the atmosphere \cite{chen1992evaporation,jambhekar,emna2017}. The appearance of the salt crust strongly affects the growth conditions for many agricultural plants \cite{pitman2020salinity,singh2015}. In arid regions, for example in Tunisia, soil salinization and soil crust formation is already interfering with and disabling agricultural activities \cite{emna2020}. This is not a new problem, but is gradually causing a greater impact as larger areas are affected \cite{verecken2009agro}.

As salts accumulate at the top of the soil due to water evaporating, the density of the remaining liquid increases with increased salt concentrations. This may lead to a gravitationally unstable setting since the liquid near the top of the soil is the heaviest, due to the accumulated salts \cite{nieldbejan2017}. When the soil is permeable, density instabilities in the form of fingers can be triggered \cite{Elenius,Riaz2006}. The formation of density instabilities in the form of fingers induces a downwards transport of the accumulated salts from the upper part of the soil towards the lower part, where the salt concentration is lower. Hence, when the density instabilities develop, they can hinder the salt concentrations near the top of the soil to exceed its solubility limit. However, for low-permeable soils, these instabilities will typically not develop, or they develop slowly at later times. In this case, salts will continue to accumulate until salt precipitates and a salt crust at the top of the soil is formed. This means that whether salt precipitation occurs, is tightly connected to whether convective instabilities develop. Hence, understanding the process of soil salinization and how the interplay with density instabilities is, are key questions to prevent degradation of soil quality and to ensure food production \cite{shokri2010evaporation,shokri2020evaporation}.

It is well-known that the question of whether the density instabilities occur, can be addressed by a linear stability analysis. Such an analysis has been applied to a wide range of porous-media problems where the density difference creates a gravitationally unstable setting \cite{nieldbejan2017}. The onset of instabilities where an increased salt concentration at the top of the porous domain triggers the instabilities, is analyzed in \cite{Elenius,Riaz2006,vanduijn2019stability}. Fingers are found to appear when the strength of the density difference overcomes the resistance of the porous medium. This is usually expressed and quantified through a Rayleigh number, where a Rayleigh number larger than some critical value means that instabilities can occur. A larger density difference, and hence a larger change in salt concentration, is needed to induce instabilities. Hence, a strong diffusion would hinder the density difference to be strong enough, as the concentration profile is smoothed. A larger resistance of the porous medium, which corresponds to a smaller permeability, makes it more difficult for the density difference to trigger instabilities.

The above-mentioned works on salt-induced instabilities consider a prescribed salt concentration or a prescribed density on the top boundary \cite{Elenius,Riaz2006,vanduijn2019stability}. When considering evaporation from a porous medium, the salt concentration at the top boundary develops with time as the water gradually evaporates and the dissolved salts remain. As we will see later, this corresponds to a Robin-type boundary condition for the salt, which means that the value of the salt concentration is connected to its gradient at the top boundary. For other linear stability problems, such boundary conditions have been considered for example in \cite{barletta2009robin,hattori2015robin}.

Evaporation of water also induces a vertical, upwards throughflow through the domain. The effect of upwards throughflow with an assigned density difference between top and bottom, has been found to have a stabilizing effect on the onset of instabilities \cite{homsy1976throughflow,vanduijn2002stability}. That means, a stronger upwards throughflow would increase the critical Rayleigh number, making it more difficult for the density difference to trigger the formation of downwards-flowing fingers. It is hence not obvious whether an increased evaporation flux of water would have a stabilizing or destabilizing effect: An increased evaporation leads to an increased upwards throughflow, which stabilizes the system, but at the same time the accumulation of salts near the top boundary is increased, which destabilizes.

The method of linear stability gives estimates for the onset of gravitational instabilities and for the moment in time for their appearance. After the instabilities have formed, one has to rely on numerical simulations of the governing system of equations to address the further development of the salt concentration. Numerical simulations can give information on the strength and shape of the appearing convection pattern, as well as their effect on the salt transport.

Throughout this paper we consider the porous medium to be fully saturated with water. This is a simplification, as evaporation usually leads to an unsaturated zone near the top of the porous medium, which again has an impact of the evolution of the salt concentration \cite{shokri2010evaporation,shokri2020evaporation}. However, the proposed analysis gives valuable insight in the idealized case of the porous medium remaining fully water saturated and creates a starting point for further analysis when incorporating an unsaturated zone in the future.

This paper is organized as follows. In Section \ref{sec:mathmodel} we formulate the general model equations together with initial and boundary condition to address evaporation from a porous medium saturated with saline water. In Section \ref{sec:linearstability} we consider a simplified model, for which we perform a linear stability analysis, giving criteria for when instabilities can occur. Section \ref{sec:dumux} explains the numerical framework used to simulate the general model. The results from the linear stability analysis and the numerical simulations are compared and discussed in Section \ref{sec:results}, before final remarks are given in Section \ref{sec:finalremarks}.

\begin{table}
	\small 
\caption{Nomenclature. Dimensions are given for variables and parameters.}
\label{tab:nomenclature}       
\begin{tabular}{lll}
\hline\noalign{\smallskip}
Variables & Explanation & Dimension  \\
\noalign{\smallskip}\hline\noalign{\smallskip}
        $P$ & pressure & kg m$^{-1}$s$^{-2}$ \\
        $\textbf{Q}=(U,V,W)$ & Darcy-velocity & m s$^{-1}$\\
        $X$ & salt mass fraction & $-$ \\
        $\mathsf{x}$ & mole fraction & $-$ \\
        $t$ & time & s \\
        $x,y,z$ & spatial coordinate & m\\
        $p$ & non-dimensional perturbed pressure & $-$ \\
        $\mathbf q=(u,v,w)$ & non-dimensional perturbed Darcy-velocity &$-$ \\
        $\chi$ & perturbed salt mass fraction & $-$\\
        $r$ & reaction rate & mol m$^{-3}$ s$^{-1}$\\
        $\rho(X)$ & liquid density & kg m$^{-3}$ \\
        $\rho_\mathrm{mol}(X)$ & molar liquid density & mol m$^{-3}$ \\
        $\zeta,\eta,\mathcal L$ & basis functions & $-$ \\
        \noalign{\smallskip}\hline
Parameters & Explanation & Dimension  \\
\noalign{\smallskip}\hline\noalign{\smallskip}
        $a$ & wavenumber & m$^{-1}$\\
        $A$ & amplitude of perturbation & $-$\\
        $d$ & depth of the numerical domain & m \\
        $D$ & diffusion coefficient & m$^2$ s$^{-1}$ \\
        $\mathbf e$ & vertical unit vector & $-$\\
		$E$ & evaporation rate & m s$^{-1}$\\
		$E_\mathrm{mol}$ & molar evaporation rate & mol m$^{-2}$ s$^{-1}$\\
		$\gamma$ & salt expansion coefficient & $-$  \\
		$\phi$ & porosity & $-$ \\
		$ g$  & gravity & m s$^{-2}$ \\
		$K$ & permeability & m$^2$\\
		$\lambda$ & wave length & m \\
		$\mu$ & dynamic viscosity & kg m$^{-1}$ s$^{-1}$ \\
		$\mu_\mathrm{\mathsf{x}^\mathrm{NaCl}}$ & mean value of NaCl mole fraction & $-$\\ 
		$m,n$ & integer & $-$ \\
		$M$ & molar mass & kg mol$^{-1}$\\
		$\mathbf n$ & horizontal unit vector & $-$\\
		$\Omega$ & domain & $-$ \\
		$\sigma$ & exponential growth rate & $-$ \\
		$\sigma_\mathrm{\mathsf{x}^\mathrm{NaCl}}$ & standard deviation of NaCl mole fraction & $-$\\
		$\mathcal W$ & half width of domain or wavelength & m \\
\noalign{\smallskip}\hline
Non-dimensional quantities & Explanation & Definition  \\
\noalign{\smallskip}\hline\noalign{\smallskip}
$\hat \beta$ & relative half width or wavelength & $\mathcal WE/D$ \\ 
$k$ & proportionality constant & $Q_\text{ref}\mathcal W/D\pi$ \\
		$R$ & evaporative Rayleigh number& $\gamma\rho_0gKX_0/E\mu$ \\
\noalign{\smallskip}\hline
Subscripts & Explanation &  \\
\noalign{\smallskip}\hline\noalign{\smallskip}
        0 & initial condition & \\
        max & solubility limit & \\
        mol & molar based quantity &\\
        $m,n$ & expansion index & \\
        p & perturbation & \\
        ref & reference value & \\
        solid & solid salt phase &\\
        $x,y,z$ & corresponding direction & \\ 
\noalign{\smallskip}\hline
Superscripts & Explanation &   \\
\noalign{\smallskip}\hline\noalign{\smallskip}
$0$ & ground state solution & \\
$\kappa$ & component  &  \\
        NaCl & salt component  &  \\
        top & top row of grid cells & \\
        w & water component  &  \\
\noalign{\smallskip}\hline
Accents & Explanation & \\
\noalign{\smallskip}\hline\noalign{\smallskip}
$\wedge$ & non-dimensional variable & \\
$\sim$ & Fourier amplitude & \\
\noalign{\smallskip}\hline
\end{tabular}
\end{table}

\section{Mathematical model}\label{sec:mathmodel}
\begin{figure}
\begin{overpic}[width=0.7\textwidth]{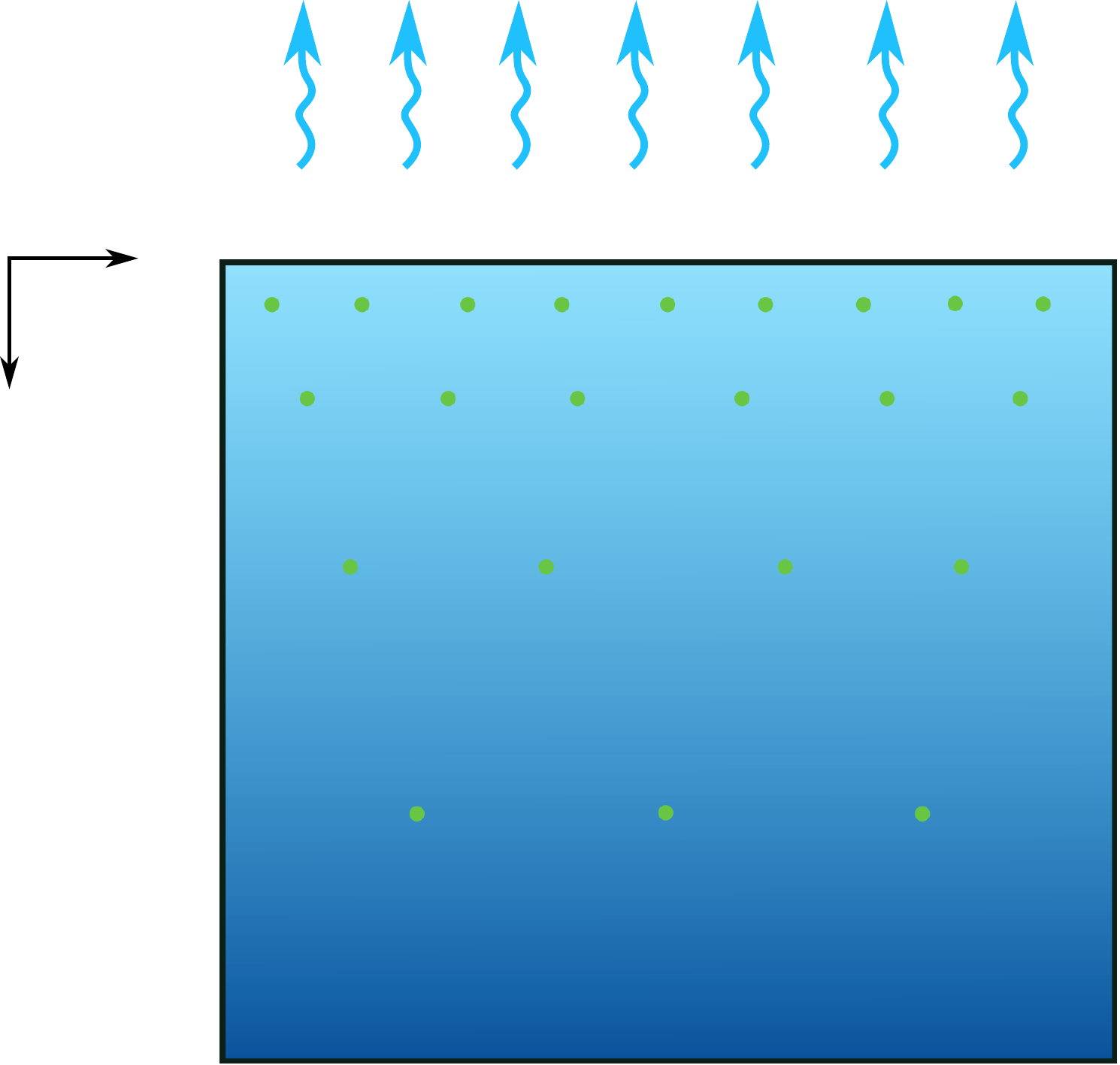}
  \put(13,71.5){$x,y$}
  \put(0,58){$z$}
  \put(45,60){Governing equations:}
  \put(37,55){Conservation of water and salt}
  \put(50,50){Darcy's law}
  \put(45,30){Initial conditions:}
  \put(38,25){Constant salt concentration}
  \put(25,20){Homogeneous and isotropic porous medium}
  \put(40,77.5){Top boundary conditions:}
  \put(27,73){Evaporation flux for water, no flux for salt}
\end{overpic}
\caption{Sketch of evaporation from porous medium and effect on salt concentration}
\label{fig:evaporation}      
\end{figure}
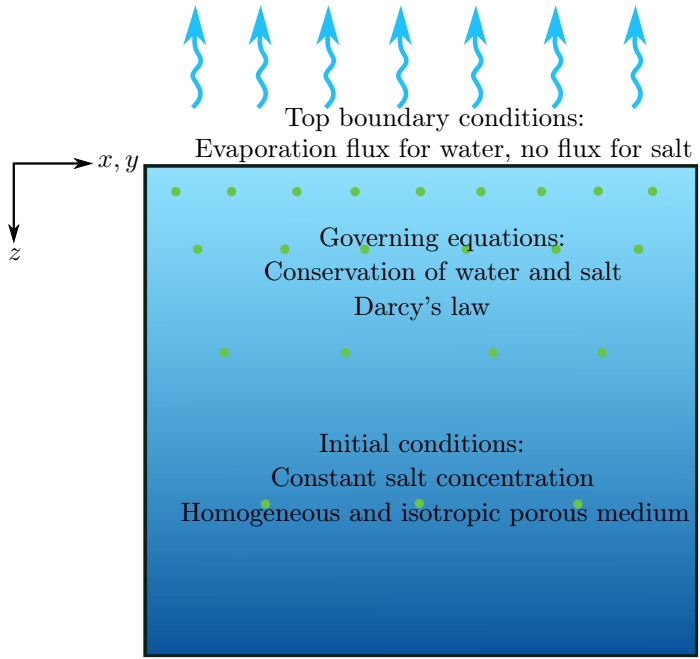

\reqnomode
To describe evaporation from the top of a porous medium and the subsequent changes within the porous medium, we formulate a mathematical model comprising of a domain with model equations, together with initial and boundary conditions. Figure \ref{fig:evaporation} sketches the domain together with the most important model choices. All variables and parameters are summarized in Table~\ref{tab:nomenclature}.

\subsection{Domain and model equations}
The considered domain is unbounded in the vertical direction and either bounded or unbounded in the horizontal directions. Specifically, we consider either
\begin{equation}
    \Omega = \{ (x,y,z)\in\mathbb R : z>0\}.
\end{equation}
or
\begin{equation}
    \Omega = \{ (x,y,z)\in\mathbb R : |x,y|<\mathcal W, z>0\},
\end{equation}
where $\mathcal W$ denotes the horizontal half width. Note that the positive vertical direction is pointing downwards: hence $z=0$ indicates the top of the domain. 
Within our domain we consider mass conservation of water and salt, along with Darcy's law representing the momentum. The porous medium is assumed to be fully saturated with liquid, and the liquid consists of water and dissolved salt. Exemplary the salt sodium chloride NaCl is used. We formulate the conservation of each of these two components independently. Both water and salt are advected with the liquid's velocity and both are subject to diffusion:
\begin{equation}
    \partial_t (\phi \rho_\mathrm{mol} \mathsf{x}^\kappa) =  \nabla \cdot \left(- \rho_\mathrm{mol} \mathsf{x}^\kappa \mathbf Q 
    + D \rho_\mathrm{mol} \nabla \mathsf{x}^\kappa \right) + r^\kappa. \label{eq:molefractions}
\end{equation}
Here, $\kappa \in \{\mathrm{w, NaCl} \}$ represents the two components of the liquid phase and $\mathsf{x}^\kappa$ denotes the mole fraction of the component $\kappa$. Hence, the sum of these two mole fractions is by definition 1, which is also reflected in \eqref{eq:molefractions}. The porosity $\phi$ is constant as long as no salt precipitates. Further, $\rho_\mathrm{mol}$ is the molar density of the liquid phase and $D$ is the effective diffusivity of the components in the mixture. Note that $D$ represents the diffusion of water mixed with salt, which is why we use the same diffusion coefficient for both components. A large diffusivity $D$ would lead to disturbances and gradients in the concentration fields being quickly smoothed away. The reaction term $r^\kappa$ accounts for chemical reactions inside the domain and is only non-zero when salt precipitation takes place within the porous medium. This means, $r^\mathrm{w} =0$ while
\begin{equation}\label{eq:rnacl}
    r^\mathrm{NaCl} = \begin{cases} 0 & \text{ when } \mathsf{x}^\mathrm{NaCl}\leq \mathsf{x}^\mathrm{NaCl}_{\text{max}}, \\ <0 & \text{ when } \mathsf{x}^\mathrm{NaCl}>\mathsf{x}^\mathrm{NaCl}_{\text{max}},\end{cases}
\end{equation}
where $\mathsf{x}^\mathrm{NaCl}_{\text{max}}$ is the solubility limit of NaCl. Hence, we have a sink term for NaCl in \eqref{eq:molefractions} when $\mathsf{x}^\mathrm{NaCl}$ exceeds its solubility limit. 
The Darcy flux $\mathbf Q$ is given by
\begin{equation}
    \mathbf Q = -\frac{K}{\mu} (\nabla P - \rho g \mathbf e_z ),\label{eq:darcy}
\end{equation}
where the permeability $K$ is assumed to be isotropic and constant, and the liquid viscosity $\mu$ is assumed to be constant. Here we use the mass density $\rho(\mathsf{x}^{\mathrm{NaCl}})$ of the liquid phase. Finally, $P$ is the pressure and $g$ is gravity. Note that the unit vector $\mathbf e_z$ points downwards.
Darcy's law quantifies the strength of the liquid volumetric flux through the domain, which depends on the pressure gradient, but also on gravitational influence through the last term of Darcy's law. 
A larger density of the liquid would hence support a stronger downwards flow. The permeability $K$ is a property of the porous medium alone and describes the ability of the medium to transmit fluids. A porous medium with a large permeability hence has a low resistance to flow.

We assume that the liquid density varies with the salt concentration through the linear dependence 
\begin{equation}
    \rho(\mathsf{x}^{\mathrm{NaCl}}) = \rho_{0}(1+\gamma_{\textrm{mol}}(\mathsf{x}^{\mathrm{NaCl}}-\mathsf{x}^{\mathrm{NaCl}}_0)),
    \label{eq:densitymol}
\end{equation}
where $\rho_{0}$ and $\mathsf{x}^{\mathrm{NaCl}}_0$ are the initial liquid density and salt mole fraction, and $\gamma_{\textrm{mol}}$ is a volumetric constant depending on the type of salt. The conversion between molar density and mass density is through
\begin{equation}
    \rho=\rho_{\textrm{mol}}\cdot M,
\end{equation}
where $M$ is the molar mass of the liquid phase. In either case, the liquid density increases with the salt concentrations. 

If the salt exceeds its solubility limit, salt precipitates and becomes part of the solid. In this case the porosity $\phi$ will change with time and we apply a mole balance for the solid salt to describe this process:
\begin{align}
\rho_\mathrm{mol,solid}~ \partial_t \phi = - r_\mathrm{solid}.
\label{eq:molebalancesolid}
\end{align}
Here, $\rho_\mathrm{mol,solid}$ is the molar density of the solid salt phase and $r_\mathrm{solid}$ the reaction term. We have that $r_\mathrm{solid}=-r^{\mathrm{NaCl}}$. 

\subsection{Discussion of physical processes}\label{sec:DiscussionProcesses}
\begin{figure}
    \centering
    \includegraphics[width=0.99\textwidth]{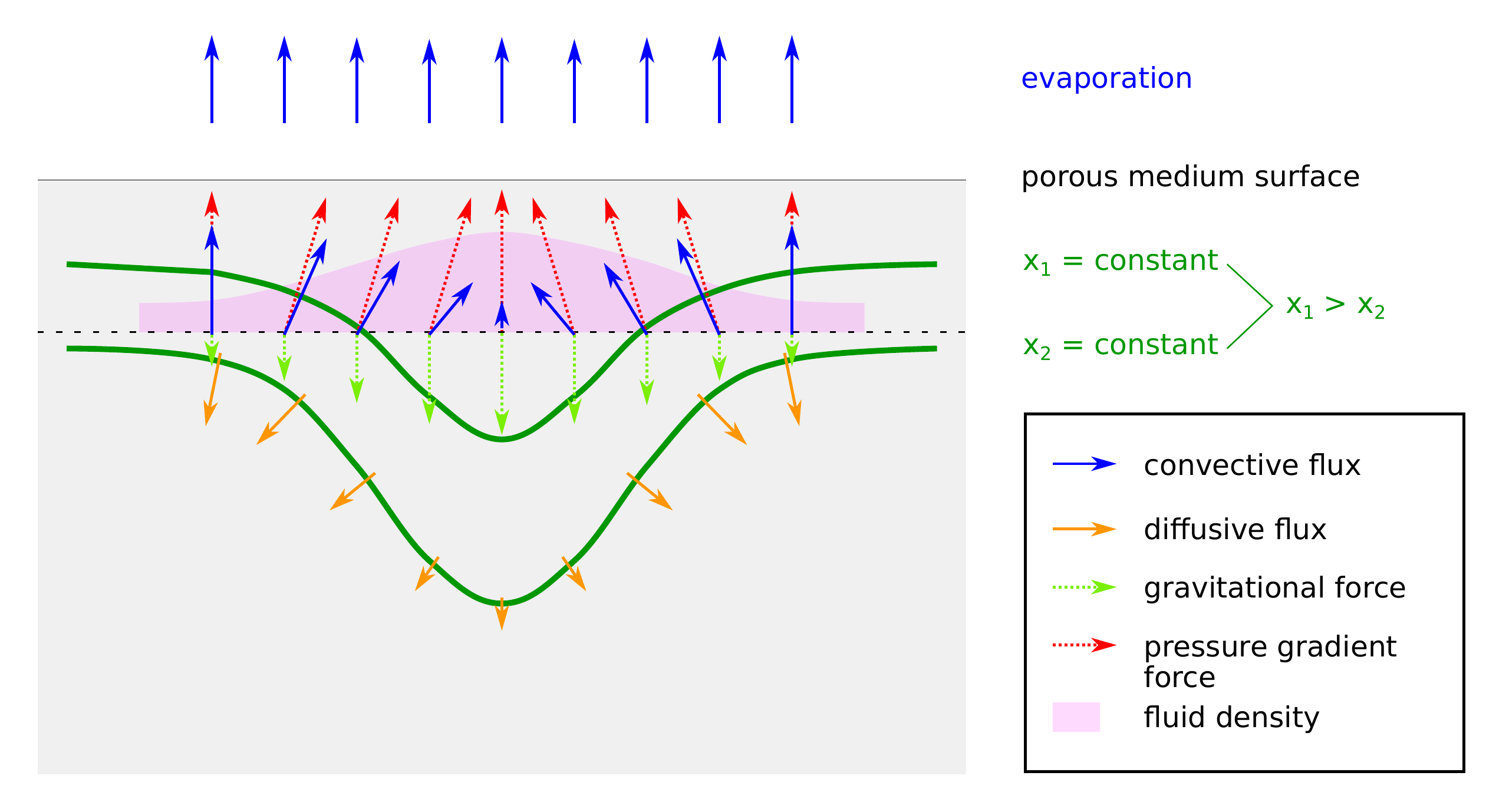}
    \caption{Relevant forces and fluxes for the development of instabilities.}
    \label{fig:ForcesAndFluxes}
\end{figure}
In Figure \ref{fig:ForcesAndFluxes} the relevant fluxes described by Equation \eqref{eq:molefractions} and \eqref{eq:darcy} are illustrated in the context of the development of instabilities. Darcy's law describes the convective flux depending on a pressure gradient force and a gravitational force. In the considered setup the pressure gradient generates an upward force due to evaporation at the top, while the gravitational force however points downwards. Hence, the resulting convective flux depends on the balance of the two counter-effective forces.

Considering a perturbation with increased salt concentration and liquid density, the increased density induces stronger gravitational forces. In case of a still dominating pressure gradient force, the convective flux is in upward direction but slowed down at the location of the perturbation. This leads to a compensation of the flux by the fluxes from the surrounding, which accumulates salt and thus enhances the perturbation. This means the salt accumulation increases the density and gravitational force even more, which can lead to a dominating gravitational force. In this case a resulting convective downward flux is generated, which leads to a development of so-called fingers which transport the accumulated salt downwards. 
In addition, diffusive fluxes are considered in Equation \eqref{eq:molefractions}. The development of instabilities is counteracted by the diffusive transport, which tries to balance out the concentration differences. Hence the balance of the counteracting convective and diffusive fluxes determines if and how fast instabilities develop.

Parameters like the permeability for the porous medium, the diffusion coefficient for the fluid mixture or the evaporation rate as boundary condition have also an important influence on the development of the instabilities. In case of higher permeabilities for example the convective flux is enhanced compared to the diffusive flux. This leads to a faster development of the instabilities.

\subsection{Initial and boundary conditions}\label{sec:mathIcBc}
As initial conditions we take, see \eqref{eq:densitymol} and \eqref{eq:molebalancesolid},
\begin{align}
\mathsf{x}^\mathrm{NaCl}|_{t=0} = \mathsf{x}_0^\mathrm{NaCl},\label{eq:initialmolesalt} \\
\phi|_{t=0} = 0.4.
\end{align}

At the top of the domain we allow water to evaporate while salt remains behind. This corresponds to specifying a given molar flux $E_\mathrm{mol}$ for the water component, while a zero flux is considered for NaCl:
\begin{align}
\left(\rho_\mathrm{mol} \mathsf{x}^\mathrm{w} \mathbf Q - D \rho_\mathrm{mol} \nabla \mathsf{x}^\mathrm{w} \right) \cdot \mathbf e_z|_{z=0}&= -E_\mathrm{mol},\label{eq:evapwatermole}\\
\left(\rho_\mathrm{mol} \mathsf{x}^\mathrm{NaCl} \mathbf Q - D \rho_\mathrm{mol} \nabla \mathsf{x}^\mathrm{NaCl} \right) \cdot \mathbf e_z|_{z=0} &= 0.\label{eq:evapsaltmole}
\end{align} 
Since the vertical unit vector $\mathbf e_z$ points downwards, the evaporation flux $E_\mathrm{mol}>0$ corresponds to an upwards flux of water. In general the Darcy flux $\mathbf Q$ will be non-zero on the top boundary due to the presence of the evaporative flux. This also means that the no-flux boundary for salt \eqref{eq:evapsaltmole} is a Robin boundary condition. Since water escapes through the top boundary while salt remains behind, we expect an accumulation of salt concentration near the top boundary. 
Increasing $E_\mathrm{mol}>0$ yields the concentration of NaCl increasing faster near the top boundary.

For the horizontal extent of the domain, we separate between the bounded and unbounded case. In the unbounded case we do not need any boundary conditions in the horizontal direction.
For the bounded case we apply no-flux boundary conditions at the vertical walls. This corresponds to
\begin{align}
    \mathbf Q\cdot \mathbf n|_{x,y=\pm\mathcal W} = 0,\quad  \nabla \mathsf{x}^\kappa\cdot\mathbf n|_{x,y=\pm\mathcal W} = 0,\label{eq:bcnoflux}
\end{align}
where $\mathbf n$ is the horizontal unit vector pointing out of the sidewalls of the bounded domain $\Omega=\{(x,y,z):|x,y|<\mathcal W,z>0\}$.

The general equations formulated here form the starting point for our further investigation of evaporation from the porous medium and subsequent onset of density instabilities. For the linear stability analysis in Section \ref{sec:linearstability} we use a slightly simplified system of equations, while the numerical simulations described in Section \ref{sec:dumux} uses the model as described above, but using bounded domains.

\section{Linear stability analysis}\label{sec:linearstability}
To address when instabilities can occur and give a criterion for the onset of instabilities depending on the model parameters, we perform a linear stability analysis. We consider here slightly simplified model equations than the general presented in Section \ref{sec:mathmodel}. These simplified equations are non-dimensionalized, before we find a time-dependent stable solution ("the ground state"), which is then perturbed. The linearized problem for the perturbed quantities are then finally formulated as an eigenvalue problem. This will give information on the stability of the ground state as a function of time.

\subsection{Simplified model equations}
We apply the method of linear stability for a simplified version of the model presented in Section \ref{sec:mathmodel}. We assume that the salt is completely dissolved and that its mass fraction is small compared to the mass fraction of water. This corresponds to setting $\mathsf{x}^\mathrm{w}\equiv1$, while keeping a non-zero $\mathsf{x}^{\mathrm{NaCl}}$. Then we invoke the Boussinesq approximation, meaning that the liquid density $\rho$ can be considered constant except in the gravity term of Darcy's law. Further, we disregard salt precipitation, which means that the simplified model is only valid up to the salt reaches its solubility limit. With these simplifications, Equations \eqref{eq:molefractions} reduce to
\begin{align}
    \nabla\cdot\mathbf Q &= 0,\label{eq:massQ}\\
    \phi\partial_tX &= \nabla\cdot(-\mathbf QX+D\nabla X), \label{eq:solutemass}
\end{align}
where $X$ is the mass fraction of salt. Darcy's law is kept as in \eqref{eq:darcy}, and the density of the liquid is still depending linearly on the salt concentration:
\begin{equation}
    \rho(X) = \rho_0(1+\gamma(X-X_0)),
    \label{eq:density}
\end{equation}
where $\gamma$ is a volumetric constant.

As initial condition for the salt we use the corresponding version of \eqref{eq:initialmolesalt}, namely
\begin{align}
    X|_{t=0} &= X_0,
\end{align}
where $X_0$ is the initial mass fraction,  
and the boundary conditions \eqref{eq:bcnoflux} at the sidewalls are hence correspondingly formulated for $X$.

Due to the difference in addressing the mass of water and salt, the boundary conditions \eqref{eq:evapwatermole} and \eqref{eq:evapsaltmole} are replaced with
\begin{align}
    \mathbf Q|_{z=0} &= -E\mathbf e_z\\
    (X\mathbf Q-D\nabla X)|_{z=0}\cdot\mathbf e_z &= 0,
\end{align}
where $E$ is the evaporation rate in terms of a volume flux of water. The two evaporation fluxes $E_{\mathrm{mol}}$ and $E$ are related through
\begin{equation}
    E_{\mathrm{mol}} = E\cdot M\cdot \rho.
\end{equation}

For convenience and to easier identify parameter dependencies, we will recast the equations in a dimensionless form.

\subsection{Non-dimensional model}
The corresponding non-dimensional variables are denoted by a hat:
\begin{align}
(\hat x,\hat y,\hat z) = \frac{(x,y,z)}{\ell_{\text{ref}}},\quad \hat t = \frac{t}{t_{\text{ref}}}\quad \hat P = \frac{P}{P_{\text{ref}}},\\  \hat{\mathbf Q} = \frac{\mathbf Q}{Q_{\text{ref}}},\quad \hat\rho = \frac{\rho}{\rho_{\text{ref}}}, \quad \hat X = \frac{X}{X_{\text{ref}}}.
\end{align}
We choose reference quantities that are meaningful to address the effect of evaporation.
As length reference we choose the ratio between diffusion and evaporation $\ell_{\text{ref}} = D/E$, which quantifies the length scale of which diffusion can smooth out concentration differences caused by the evaporation. As time reference we use $t_{\text{ref}}=\phi\ell_{\text{ref}}/E=\phi D/E^2$. 
The reference velocity is set as the gravitational velocity $Q_{\text{ref}}= \gamma\rho_{\text{ref}}gX_{\text{ref}}K/\mu$ and the reference pressure is chosen to balance the velocity in Darcy's law $P_{\text{ref}}=\mu\ell_{\text{ref}} Q_{\text{ref}}/K$. Finally, the reference concentration and density are chosen as the initial concentration and density; $X_{\text{ref}}=X_0, \rho_{\text{ref}}=\rho_0$. Note that the salt mass fraction is itself non-dimensional, hence the new variable $\hat X$ is only a scaled version of $X$.
We finally introduce the evaporative Rayleigh number $R = Q_{\text{ref}}/E$. Hence, the Rayleigh number describes the ratio between the shear flow and the flow induced by the evaporative flux at the top boundary. Note that a larger evaporation rate corresponds to a smaller Rayleigh number.
\begin{remark}
Note that we here define the Rayleigh number as
\begin{equation*}
    R=\frac{Q_{\text{ref}}}{E} \text{ where } Q_{\text{ref}}=\frac{ \gamma\rho_{0}gX_{0}K}{\mu}.
\end{equation*}
Normally when considering problems with density-driven instabilities, a typical density difference is used in the reference velocity $Q_{\text{ref}}$. However, in our model setup no typical density difference appears as we do not prescribe a density difference through the top boundary condition. Hence our reference velocity is not a shear velocity, but the velocity expected due to gravitational influence since the initial density $\rho_0$ is used. The choice of reference velocity does not affect the linear stability analysis results.
\end{remark}

The non-dimensional model equations are then
\begin{align}
    \hat \nabla\cdot\hat{\mathbf Q} &= 0, \label{eq:Massnondim}\\
    \hat{\mathbf Q} &= -\hat \nabla \hat P+\hat \rho(\hat X)\mathbf e_z, \label{eq:Darcynondim}\\
    \partial_{\hat t}\hat X &= \hat \nabla\cdot(-R\hat{\mathbf Q}\hat X+\hat \nabla \hat X), \label{eq:Saltnondim}
\end{align}
where the non-dimensional density is given as 
\begin{equation}
    \hat\rho(\hat X) = \frac{1}{\gamma X_0}+\hat X-1.
\end{equation}
For the non-dimensional variables we have the initial condition
\begin{align}
    \hat X|_{\hat t=0} &= 1, \label{eq:initialnondim}
\end{align}
and boundary conditions at the top boundary
\begin{align}
    \hat{\mathbf Q}|_{\hat z=0} &= -\frac{1}{R}\mathbf e_z, \label{eq:topQnondim}\\
    (\hat X+\hat\nabla \hat X)|_{\hat z=0}\cdot\mathbf e_z &= 0. \label{eq:topXnondim}
\end{align}
In the bounded case, the no-flux boundary conditions are imposed on $\hat x,\hat y=\pm \hat\beta$, where $\hat\beta = \mathcal W E/D$. Hence
\begin{align}
    \hat{\mathbf Q}\cdot \mathbf n|_{\hat x,\hat y=\pm\hat\beta} = 0,\quad  \hat\nabla \hat X\cdot\mathbf n|_{\hat x,\hat y=\pm\hat\beta} = 0. \label{eq:bcboundednd}
\end{align}

\subsection{Ground state solution}\label{sec:groundstate}
We will investigate the stability of a particular solution of the system of equations \eqref{eq:Massnondim}-\eqref{eq:Saltnondim} under the conditions \eqref{eq:initialnondim}-\eqref{eq:topXnondim}. This solution depends on $\hat z$ and $\hat t$. It is called the \emph{ground state}, and is denoted by $\{\hat{\mathbf Q}^0,\hat X^0,\hat P^0 \}$. 

For the ground state discharge $\hat{\mathbf Q}^0(\hat t,\hat z)$ it is clear from \eqref{eq:Massnondim} together with the boundary condition \eqref{eq:topQnondim} that the only possible solution is
\begin{equation}
    \hat{\mathbf Q}^0(\hat t,\hat z) = -\frac{1}{R}\mathbf e_z,\label{eq:Qstable}
\end{equation}
that is, a constant upwards velocity according to the prescribed evaporation rate.

\begin{figure}
  \includegraphics[width=\textwidth]{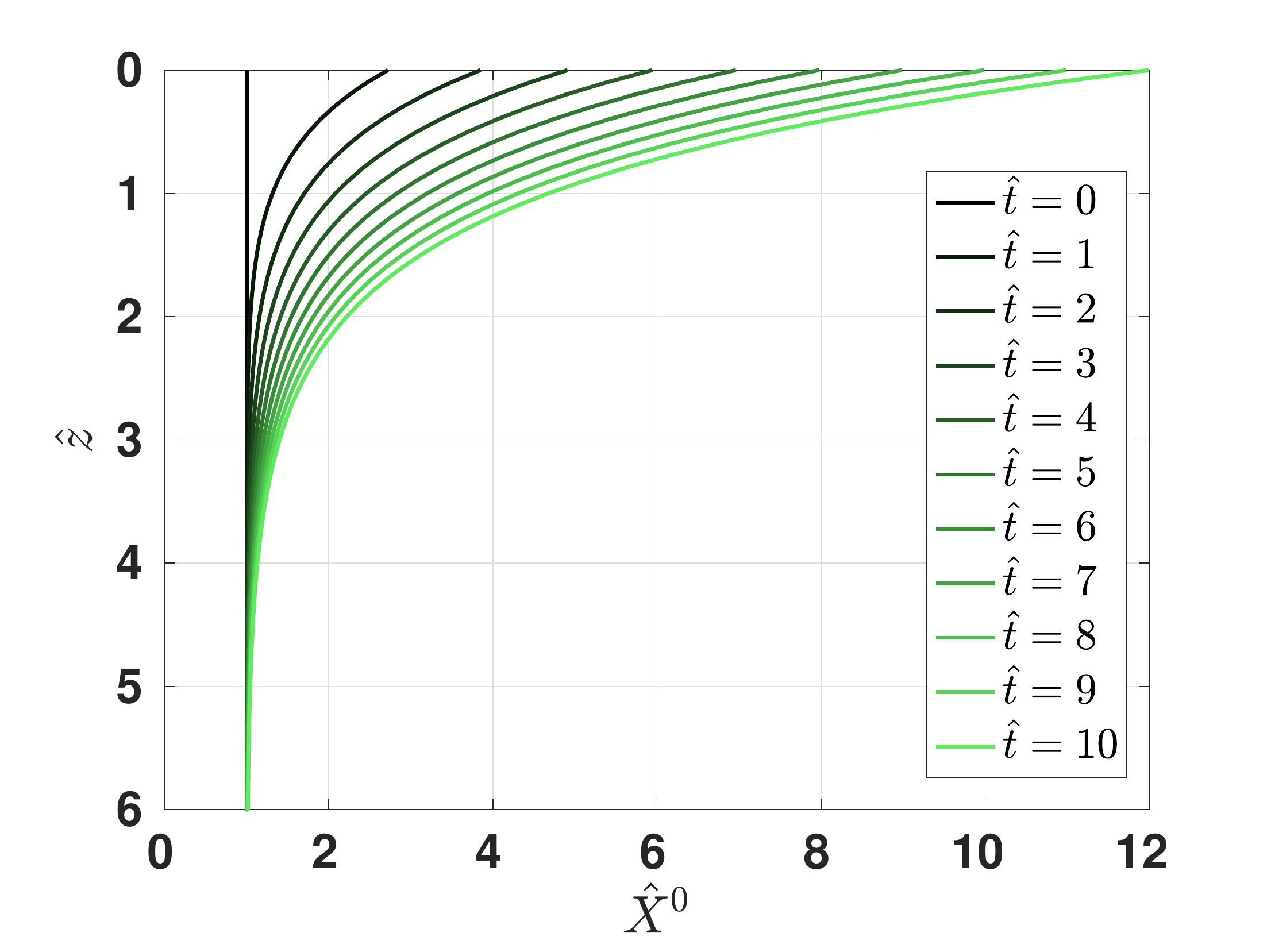}
\caption{Ground state salt mass fractions $\hat X^0$ (horizontal axis) varying with depth $\hat z$ (vertical axis) for various times $\hat t$.}
\label{fig:GroundState_u}      
\end{figure}

The ground state salt mass fraction $\hat X^0(\hat t,\hat z)$ fulfills the following problem:
\begin{align}
    \partial_{\hat t} \hat X^0 = \partial_{\hat z}\hat X^0 + \partial^2_{\hat z}\hat X^0 \quad \hat z>0,\hat t>0, \label{eq:groundstateeq} \\
    \hat X^0 +\partial_{\hat z}\hat X^0 = 0 \quad \hat z=0,\hat t>0,\\
    \hat X^0 = 1 \quad \hat z>0,\hat t=0.
\end{align}
This problem has the explicit solution
\begin{equation}
    \hat X^0(\hat t,\hat z) = 1+\int_0^{\hat t} \partial_{\hat z}f(\theta,\hat z)\ d\theta, \label{eq:ustable}
\end{equation}
where
\begin{equation}
    f(\theta,\hat z) = 1-\frac{1}{2} e^{-\hat z}\text{erfc}\Big(\frac{\hat z-\theta}{2\sqrt\theta} \Big) - \frac{1}{2}\text{erfc}\Big(\frac{\hat z+\theta}{2\sqrt\theta}\Big). \label{eq:ufstable}
\end{equation}
For the derivation of this solution, we refer to Appendix \ref{app:A}. Note in particular that
\begin{equation}
    \partial_{\hat z}f(\theta,\hat z) = \frac{1}{2}e^{-\hat z}\text{erfc}\Big(\frac{\hat z-\theta}{2\sqrt\theta}\Big) + \frac{1}{\sqrt{\pi\theta}}e^{-(\frac{\hat z+\theta}{2\sqrt\theta})^2}, \label{eq:udsfstable}
\end{equation}
enabling a simple evaluation of the integral in \eqref{eq:ustable}. The solution is shown in Figure \ref{fig:GroundState_u}. Clearly, the salt concentration at the top of the domain gradually increases with time and diffuses down through the domain.

Note that since \eqref{eq:Saltnondim} (and hence \eqref{eq:groundstateeq}) does not incorporate the precipitation rate in case the salt mass fraction exceeds the solubility limit, the solution \eqref{eq:ustable} is only valid up to the point where the salt mass fraction at the top reaches the solubility limit. For NaCl, the (non-dimensional) solubility limit is 8.84, which is reached at $\hat t=6.89$. Note however that the ground state \eqref{eq:ustable} does not depend on the type of salt. Our analysis can hence be straightforwardly be applied to any salt type, but only up to times where the corresponding solubility limit is reached.

The ground state pressure $\hat P^0(\hat t,\hat z)$ is such that
\begin{equation}
    \partial_{\hat z}\hat P^0 = \frac{1}{R}  + \hat\rho(\hat X^0).
\end{equation}
As we only have boundary conditions for the value of the velocity, the pressure is only known up to a constant. Hence,
\begin{equation}
    \hat P^0(\hat t,\hat z) = C(\hat t)+ (\frac{1}{R}+\frac{1}{\gamma X_0}-1)\hat z + \int_0^{\hat z} \hat X^0(\hat t,\varsigma)\ d\varsigma,
\end{equation}
where the integration constant $C(\hat t)$ cannot be determined. This is however not a problem as the following analysis does not depend on the value of the pressure.

\subsection{Linear perturbation equation}\label{sec:linearperturb}
To investigate the stability of the ground state $\{\hat{\mathbf Q}^0,\hat X^0,\hat P^0\}$, we write
\begin{align}
    \hat{\mathbf Q}(\hat t,\hat x,\hat y,\hat z) &= \hat{\mathbf Q}^0(\hat t,\hat z)+ \mathbf q(\hat t,\hat x,\hat y,\hat z), \label{eq:Qperturb}\\
    \hat X(\hat t,\hat x,\hat y,\hat z) &=\hat X^0(\hat t,\hat z) + \chi(\hat t,\hat x,\hat y,\hat z), \label{eq:Xperturb} \\
    \hat P(\hat t,\hat x,\hat y,\hat z) &=\hat P^0(\hat t,\hat z) + p(\hat t,\hat x,\hat y,\hat z),\label{eq:Pperturb}
\end{align}
where $\mathbf q = (u,v,w), \chi$ and $p$ are small, perturbed quantities, which we will now study further. Note that although these are all non-dimensional, we write them without the hat to simplify the notation. 
Since $\hat{\mathbf Q},\hat X$ and $\hat P$ still need to solve the original equations and boundary conditions, we achieve equations and boundary conditions for the perturbed quantities. Inserting \eqref{eq:Qperturb}-\eqref{eq:Pperturb} into \eqref{eq:Massnondim}-\eqref{eq:Saltnondim} and into  \eqref{eq:topQnondim}-\eqref{eq:topXnondim} and linearize, we obtain the linear perturbation equations
\begin{align}
    \hat\nabla\cdot\mathbf q &= 0, \label{eq:qperturb}\\
    \mathbf q &= -\hat\nabla p + \chi\mathbf e_z,\label{eq:darcyperturb} \\
    \partial_{\hat t}\chi &= \partial_{\hat z}\chi - Rw\partial_{\hat z}\hat X^0+\hat\nabla^2\chi,\label{eq:xperturb}
\end{align}
with boundary conditions at the top
\begin{align}
    \mathbf q|_{\hat z=0} &= \mathbf 0, \label{eq:topqperturb}\\
    (\chi+\partial_{\hat z}\chi)|_{\hat z=0} &= 0. \label{eq:topxperturb}
\end{align}

\begin{figure}
  \includegraphics[width=\textwidth]{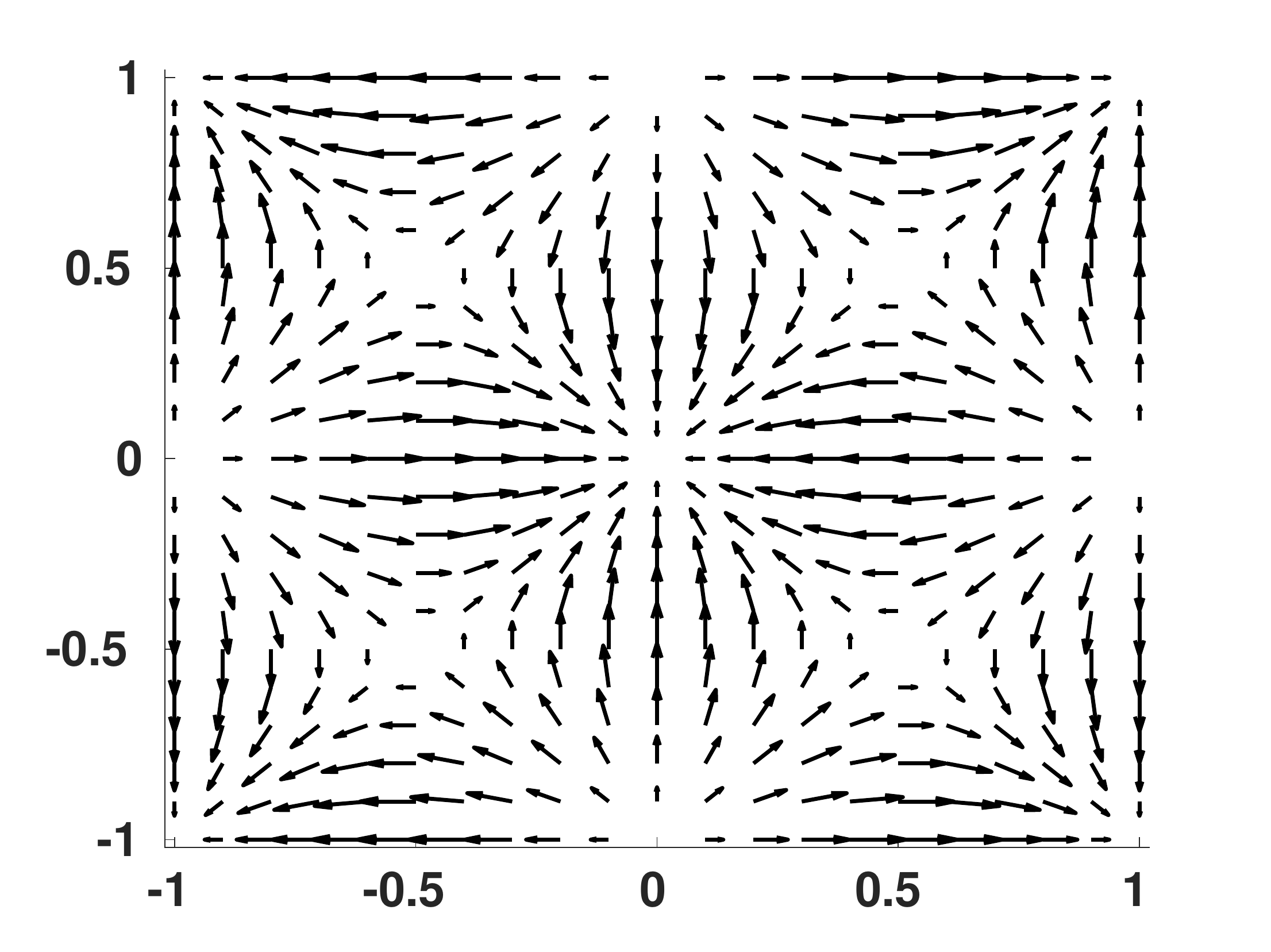}
\caption{Horizontal discharge $(u,v)$ seen from above for a cross section $\{(\hat x,\hat y): |\hat x|,|\hat y|<\hat\beta\}$, here using $\hat\beta=1$.}
\label{fig:horizontal}      
\end{figure}

Equations \eqref{eq:qperturb}-\eqref{eq:xperturb} can be written in terms of $\{w,\chi,p\}$ only, see e.g.~\cite{vanduijn2019stability}. This results in
\begin{align}
    \hat\nabla^2w &= (\partial_{\hat x}^2+\partial_{\hat y}^2)\chi \label{eq:wperturb1}\\
    \partial_{\hat z} w &= (\partial_{\hat x}^2+\partial_{\hat y}^2)p \label{eq:wperturb2}\\
    \partial_{\hat t}\chi &= \partial_{\hat z}\chi - Rw\partial_{\hat z}\hat X^0+\hat\nabla^2\chi,\label{eq:xperturb2}
\end{align}
with boundary conditions
\begin{align}
    w|_{\hat z=0} &= 0, \label{eq:topwperturb}\\
    (\chi+\partial_{\hat z}\chi)|_{\hat z=0} &= 0. \label{eq:topxperturb2}
\end{align}
We seek solutions of this system satisfying
\begin{equation}
    w,\chi \to 0 \text{ as } \hat z\to\infty. \label{eq:wchibottom}
\end{equation}
Since \eqref{eq:wperturb1}-\eqref{eq:wchibottom} is a linear system of equations, we consider solutions of the form
\begin{equation}
    \{w,\chi,p\}(\hat t,\hat x,\hat y,\hat z) = \{\tilde w,\tilde \chi,\tilde p\}(\hat t,\hat z)\cos(\hat a_x\hat x)\cos(\hat a_y\hat y). \label{eq:perturbcos}
\end{equation}
Here $\hat a_x$ and $\hat a_y$ are horizontal wavenumbers. Substituting \eqref{eq:perturbcos} into \eqref{eq:wperturb1}-\eqref{eq:xperturb2} yields for the amplitudes $\{\tilde w,\tilde \chi,\tilde p\}$
\begin{align}
\partial_{\hat z}^2\tilde w -\hat a^2\tilde w &=-\hat a^2\tilde\chi, \label{eq:wperturb1a} \\
\partial_{\hat z}\tilde w &= -\hat a^2\tilde p,\label{eq:wperturb1b}\\
    \partial_{\hat t}\tilde \chi &= \partial_{\hat z}\tilde\chi - R\tilde w\partial_{\hat z}\hat X^0 + \partial_{\hat z}^2\tilde\chi - \hat a^2\tilde\chi,\label{eq:xperturba}
\end{align}
where $\hat a^2=\hat a_x^2+\hat a_y^2$. Expressions for the horizontal discharge components follow from Darcy's law:
\begin{align}
    u(\hat t,\hat x,\hat y,\hat z) &= -\partial_{\hat x}p = -\frac{\hat a_x}{\hat a^2}\partial_{\hat z}\tilde w\sin(\hat a_x\hat x)\cos(\hat a_y\hat y),\label{eq:uperturb}\\
    v(\hat t,\hat x,\hat y,\hat z) &= -\frac{\hat a_y}{\hat a^2}\partial_{\hat z}\tilde w\cos(\hat a_x\hat x)\sin(\hat a_y \hat y).\label{eq:vperturb}
\end{align}
When the domain is unbounded, i.e.~the half space $\{\hat z>0\}$, we consider \eqref{eq:wperturb1a}-\eqref{eq:xperturba}, subject to \eqref{eq:topwperturb}-\eqref{eq:topxperturb2} for any $\hat a>0$. In case of the bounded domain $\{(\hat x,\hat y,\hat z): |\hat x|,|\hat y|<\hat \beta,\hat z>0\}$ we need to choose $\hat a_x$ and $\hat a_y$ so that the boundary conditions \eqref{eq:bcboundednd} are satisfied. This requires 
\begin{equation}
    \hat a_x = n_x\frac{\pi}{\hat\beta},\quad \hat a_y = n_y\frac{\pi}{\hat\beta},\quad n_x,n_y = 1,2,\dots \label{eq:pibeta}
\end{equation}
A sketch of a horizontal discharge field is given in Figure \ref{fig:horizontal}. In this figure we have used $\hat \beta=1$ as well as $\hat a_x=\hat a_y=\pi$, corresponding to one full oscillation in both $\hat x$ and $\hat y$ direction.

It is clear that it suffices to consider only Equations \eqref{eq:wperturb1a} and \eqref{eq:xperturba}, subject to boundary conditions \eqref{eq:topwperturb}-\eqref{eq:wchibottom}. Once they are solved, the pressure results from \eqref{eq:wperturb1b} and the horizontal discharge from \eqref{eq:uperturb}-\eqref{eq:vperturb}.

\subsection{Eigenvalue problem}\label{sec:eigenvalueproblem}
We want to address the stability of the evolving ground state $\hat X^0(\hat t,\hat z)$. This amounts to a stability analysis of the linear amplitude Equations \eqref{eq:wperturb1a} and \eqref{eq:xperturba}. The standard approach is to look for solutions of the form
\begin{equation}
    \{\tilde w,\tilde \chi\}(\hat t,\hat z) = \{\hat w,\hat \chi\}(\hat z)e^{\sigma \hat t},\label{eq:sigma}
\end{equation}
where $\sigma$ is the exponential growth rate in time, while $\hat w$ and $\hat \chi$ describes the variability of the perturbation with $\hat z$. For $\sigma<0$, small perturbations decay in time and the ground state is stable. For $\sigma>0$, perturbations grow in time and the ground state is unstable. 

Since $\hat X^0 = \hat X^0(\hat t,\hat z)$, the separation of variables as proposed in \eqref{eq:sigma} does not directly apply. This is circumvented by assuming that $\hat X^0(\hat t,\hat z)$ changes slower in time than any exponentially growing instability. This is known as the quasi steady state approach (QSSA) or the
"frozen profile" approach \cite{Riaz2006,vanduijn2002stability}. For any fixed $\hat t_*>0$, we consider $\hat \tau = \hat t-\hat t_*$ as the new time variable. Then \eqref{eq:xperturba} becomes
\begin{equation}
    \partial_{\hat \tau}\tilde \chi = \partial_{\hat z}\tilde\chi - R\tilde w\partial_{\hat z}\hat X^0(\hat t_*+\hat\tau,\hat z) + \partial_{\hat z}^2\tilde\chi - \hat a^2\tilde\chi.
\end{equation}
In this equation we take $\hat \tau$ small, $0<\hat\tau\ll1$, and write $\hat X^0(\hat t_*+\hat \tau,\hat z)\approx\hat X^0(\hat t_*,\hat z)$, i.e.~the frozen profile. Setting now 
\begin{equation}
    \{\tilde w,\tilde \chi\}(\hat \tau,\hat z) = \{\hat w,\hat \chi\}(\hat z)e^{\sigma \hat \tau},\label{eq:sigmatau}
\end{equation}
we obtain
\begin{align}
    \partial_{\hat z}^2 w-\hat a^2\hat w &=-\hat a^2\hat\chi \\
    \sigma\hat\chi &= \partial_{\hat z}\hat\chi - R\hat w\partial_{\hat z}\hat X^0(\hat t_*,\hat z) + \partial_{\hat z}^2\hat\chi - \hat a^2\hat\chi,
\end{align}
for $0<\hat z<\infty$ subject to \eqref{eq:topwperturb}-\eqref{eq:wchibottom}.

For given $\hat a>0,\hat t_*>0$ and $\sigma\in\mathbb{R}$, this is an eigenvalue problem in terms of $\{\hat w,\hat\chi\}$ and $R$. The object is to determine the smallest positive eigenvalue $R=R_\ast(\hat a,\hat t_*,\sigma)$. We verified numerically, see Appendix \ref{app:B} and \cite{vanduijn2019stability}, that there is exchange of stability; i.e.,
\begin{equation}
    R_\ast(\hat t,\hat t_*,\sigma) \gtrless R_\ast(\hat a,\hat t_*,0) \text{ if and only if } \sigma \gtrless 0.
\end{equation}
This means that if the parameters of the problem are such that $R>R_\ast(\hat a,\hat t_*,0)$, then at time $\hat t_*$ a perturbation with wavenumber $\hat a$ will emerge, implying that the ground state looses its stability at time $\hat t_*$.

Based on this observation it suffices to analyze the eigenvalue problem for the case of neutral stability; that is, $\sigma=0$. Thus we need to consider the problem (dropping the asterisk in $\hat t_*$ and $R_\ast$):\\
Given $\hat a>0$, $\hat t>0$, and $\hat X^0=\hat X^0(\hat t,\hat z)$ by \eqref{eq:ustable}, find the smallest $R=R(\hat a,\hat t)>0$ such that
\begin{equation}
    \label{eq:eigenvaluescaled}
    \tag{${\text{EP}}$}
    \leqnomode
    \left\{\begin{array}{lr}  \hat w''+\hat a^2\hat \chi-\hat a^2\hat w = 0 & \hat z>0,\\
    \hat\chi' - R\hat w\partial_{\hat z}\hat X^0 + \hat\chi'' - \hat a^2\hat\chi =0 & \hat z>0,\\
    \text{where } \hat w \text{ and } \hat \chi \text{ fulfill } & \\
    \hat w=0, \hat\chi+\hat\chi'=0 & \hat z=0,\\
    \hat w\to 0, \hat\chi\to 0 & \hat z\to\infty,\\
    \end{array}
    \right.
\end{equation}
has a non-trivial solution.

Note that $'$ denotes the derivative with respect to $\hat z$.

\begin{remark}
In the work of Riaz et al.~(2006)\cite{Riaz2006}, the ground state results from a diffusion process and depends on $\hat\xi=\frac{\hat z}{\sqrt{\hat t}}$ only. This allows them to perform the coordinate transformation $(\hat t,\hat z)\to (\hat t,\hat\xi)$ and then apply the QSSA to the transformed linear problem. This yields a sharp stability bound. In our case the ground state results from diffusion and upward convection due to evaporation. Consequently the ground state \eqref{eq:ustable} does not have such a simple dependence. Therefore we apply the QSSA directly to the original coordinates $(\hat t,\hat z)$.
\end{remark}

\subsection{Solution strategy for the eigenvalue problem}\label{sec:linsolstrategy}
\begin{figure}
  \includegraphics[width=\textwidth]{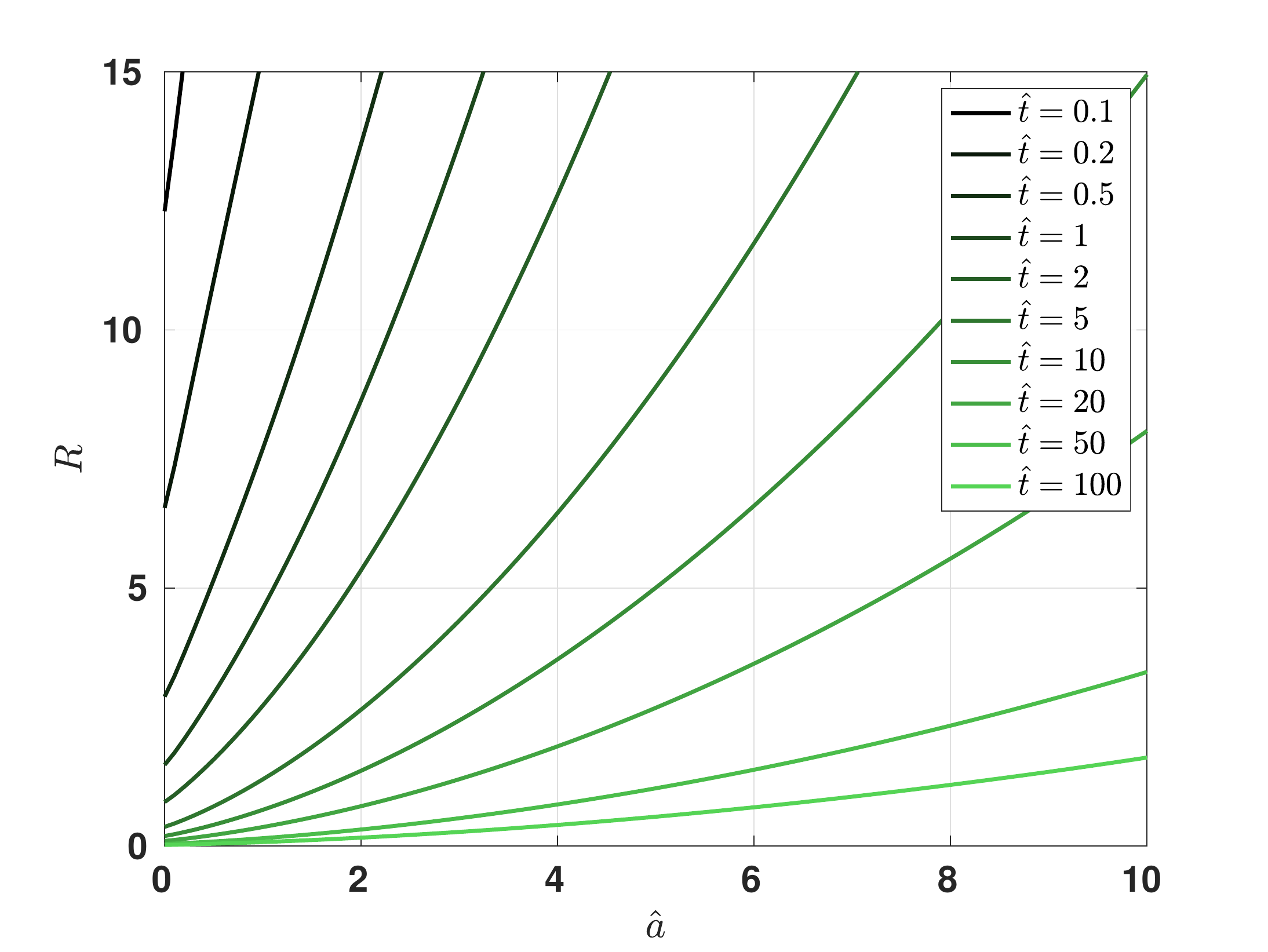}
\caption{Resulting minimal eigenvalue $R$ (vertical axis) as a function of $\hat a$ (horizontal axis) for various $\hat t$.}
\label{fig:RLvsa}       
\end{figure}

The eigenvalue problem \eqref{eq:eigenvaluescaled} is solved via a Laguerre-Galerkin method. We let
\begin{equation}
    \hat \chi(\hat z) = \sum_{n=0}^\infty \hat\chi_n\zeta_n(\hat z), \quad \hat w(\hat z) = \sum_{n=0}^\infty \hat w_n\eta_n(\hat z),
\end{equation}
where the basis functions are given by
\begin{equation}
    \zeta_n(\hat z) = e^{-\frac{\hat z}{2}}\Big(\mathcal L_n(\hat z)+\frac{1-2n}{1+2n}\mathcal L_{n+1}(\hat z)\Big) \text{ and } \eta_n(\hat z) = e^{-\frac{\hat z}{2}}\Big(\mathcal L_n(\hat z)-\mathcal L_{n+1}(\hat z)\Big),\label{eq:Laguerre}
\end{equation}
and where $\mathcal L_n$ is the Laguerre polynomial of degree $n$ \cite{temme1996special}; i.e.~
\begin{equation}
    \mathcal L_n(\hat z) = \sum_{\ell=0}^n(-1)^\ell\begin{pmatrix} n\\ \ell\end{pmatrix} \frac{\hat z^\ell}{\ell!} \text{ with } \mathcal L_n(0)=1 \text{ and } \mathcal L'_n(0)=-n.
\end{equation}
The special combinations in \eqref{eq:Laguerre} are chosen so that that $\zeta_n(\hat z)$ and $\eta_n(\hat z)$ satisfy the boundary conditions for $\hat\chi$ and $\hat w$ in \eqref{eq:eigenvaluescaled}$_{4,5}$, respectively. Inserting \eqref{eq:Laguerre} into \eqref{eq:eigenvaluescaled}$_{1,2}$, multiplying with $\zeta_m$ and $\eta_m$ and integrating with respect to $\hat z$ yields
\begin{align*}
    \sum_{n=0}^{\infty}\hat w_n\int_0^\infty \eta_n''\eta_m d\hat z + \hat a^2\sum_{n=0}^\infty\hat\chi_n\int_0^\infty\zeta_n\eta_m d\hat z -\hat a^2\sum_{n=0}^\infty\hat w_n\int_0^\infty \eta_n\eta_m d\hat z &= 0,\\
    \sum_{n=0}^\infty\hat\chi_n\int_{0}^\infty\zeta_n'\zeta_m d\hat z - R\sum_{n=0}^\infty\hat w_n\int_{0}^\infty\partial_{\hat z} \hat X^0\eta_n\zeta_m d\hat z &\ \\ +\sum_{n=0}^\infty\hat\chi_n\int_{0}^\infty\zeta_n''\zeta_m d\hat z - \hat a^2\sum_{n=0}^\infty\hat\chi_n\int_{0}^\infty\zeta_n\zeta_m d\hat z &=0.
\end{align*}
We truncate this expression at a large number $n=N$. Inspecting the convergence of the terms, we found $N=32$ to be sufficient.
The integrals can be determined analytically using the properties of Laguerre polynomials. The only exception is the integral also involving $\partial_{\hat z}\hat X^0$, which is approximated using a Gauss-Laguerre quadrature rule. This results in a system of matrix-vector equations. Through the eigenvalues of the resulting matrix, we can find the corresponding minimal $R$ as a function of $\hat a$ for given $\hat t$, i.e.~$R=R(\hat a,\hat t)$. Results are shown in Figure \ref{fig:RLvsa}. From the figure we observe that the system becomes gradually more unstable for increasing $\hat t$. With the Rayleigh number specified by model parameters $R_s=\frac{\gamma\rho_0 gK X_0}{E\mu}$, the system remains stable when $R_s<R$ in Figure \ref{fig:RLvsa}. As the curves for $R$ move downwards with increasing time, we can for given parameters find a corresponding onset time, which is when $R_s=R$. 
For a fixed time $\hat t$, the minimum $R$ always appears for $\hat a$ approaching 0. This corresponds to longer wavelengths being more unstable. For an explanation of the behavior for small wavenumbers we refer to Appendix \ref{app:c}. In this appendix we also derive an approximation (in fact a lower bound) for the value of $R$ at $\hat a=0$. We find
\begin{equation*}
    R(0,\hat t) \geq \frac{2}{\hat X^0(\hat t,0)-1} \text{ for all } \hat t>0.
\end{equation*}
Hence, the evaporation problem is stable as long as
\begin{equation*}
    R_s<\frac{2}{\hat X^0(\hat t,0)-1}.
\end{equation*}

\begin{figure}
  \includegraphics[width=\textwidth]{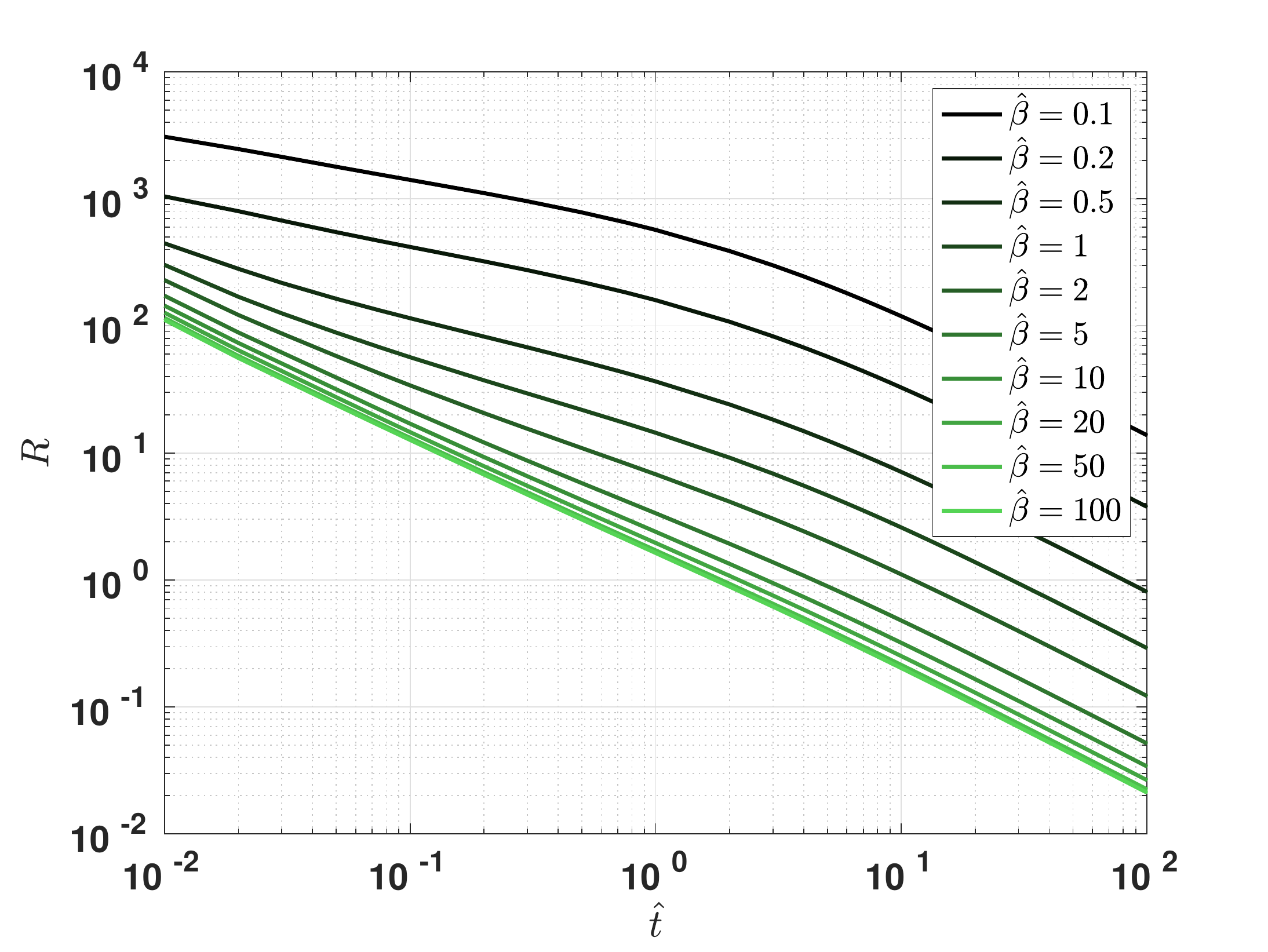}
\caption{Critical Rayleigh number $R$ (vertical axis) as a function of $\hat t$ (horizontal axis) for various $\hat \beta$.}
\label{fig:Rvshatt}       
\end{figure}

In Figure \ref{fig:RLvsa} we treat $\hat a$ as a continuous variable. This is allowed when the domain is unbounded. For bounded domains, with zero-flux boundary conditions on the sidewalls, only values of $\hat a$ that are multiples of $\pi/\hat\beta$ are allowed (see \eqref{eq:pibeta}). We observe in Figure \ref{fig:RLvsa} that for fixed $\hat t$, the minimum $R$ always occurs for the lowest possible $\hat a$. Hence in the bounded case, we would only need to solve the eigenvalue problem for $\hat a=\pi/\hat \beta$. For a fixed $\hat \beta$ (and hence a fixed $\hat a$), the system becomes generally more unstable for later times $\hat t$. In Figure \ref{fig:Rvshatt} we show the critical Rayleigh number for various choices of $\hat \beta$ as a function of $\hat t$, which hence are the $R$-values from Figure \ref{fig:RLvsa} corresponding to $\hat a=\pi/\hat \beta$. Note that for larger values of $\hat\beta$, the critical Rayleigh numbers tend to those for $\hat a\to0$, which corresponds to an infinitely wide domain or an infinitely long wavelength, where no effect of the imposed boundary conditions at the sidewalls are found. These results imply that we can find unique times for onset of instabilities for a given set of parameters, both in the bounded and unbounded case. For a given $\hat\beta$ and model parameters $R_s$, the corresponding unique onset time can be found in Figure \ref{fig:Rvshatt} by finding the corresponding time where $R=R_s$ for our choice of $\hat \beta$. Similarly, for a given $\hat a$ and model parameters $R_s$, we can find the corresponding unique onset time in Figure \ref{fig:RLvsa}.

Note that although we show results for a large range of non-dimensional times in Figures \ref{fig:RLvsa} and \ref{fig:Rvshatt}, the results are only physically meaningful when the stable salt concentration does not reach its solubility limit. For NaCl the maximum solubility is 8.84, which the stable solution \eqref{eq:ustable} reaches after $\hat t=6.89$. Hence, for NaCl, the eigenvalue results are only meaningful before this non-dimensional time. However, since the eigenvalue problem \eqref{eq:eigenvaluescaled} can be used for any salt, results are shown for a larger times as well.

\subsection{Effect of varying the evaporation rate}
Although the solutions of the eigenvalue problem \eqref{eq:eigenvaluescaled} can be used to discuss the effect of all model parameters, we discuss here in particular the effect of the evaporation rate $E$. The expected influence of $E$ on the stability is a-priori not obvious as it affects the system in two different ways: Firstly, a larger evaporation rate corresponds to a stronger vertical, upwards throughflow, and secondly, a larger evaporation rate means that the accumulation of salts at the top of the domain is faster. As shown by \cite{homsy1976throughflow,vanduijn2002stability}, an increased throughflow corresponds to the system being more stable. However, an increased accumulation of salts at the top of the domain is expected to destabilize the system as the density difference is then larger.

\begin{figure}
    \begin{subfigure}{0.49\textwidth}
        \includegraphics[width=\textwidth]{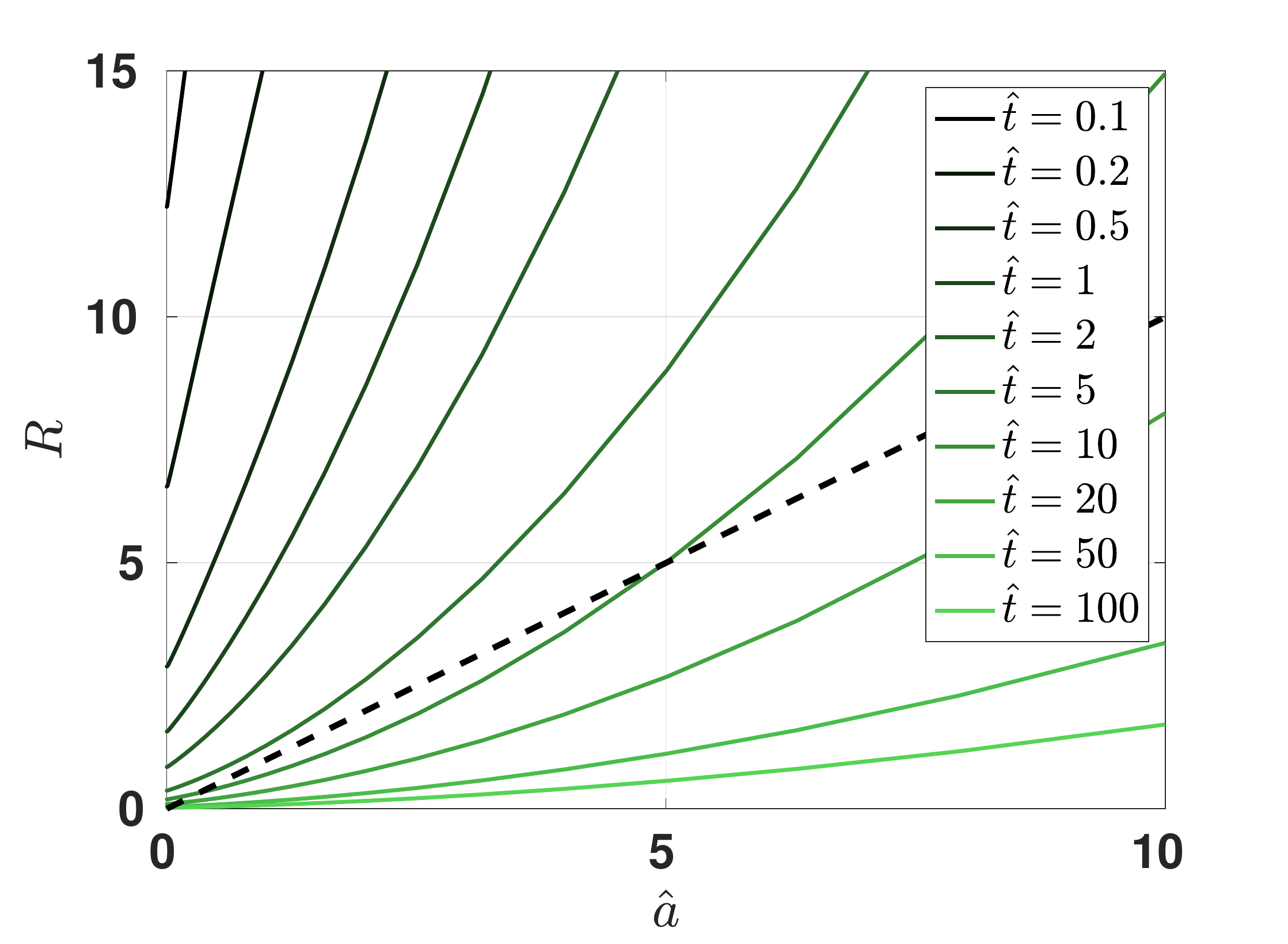}
        \subcaption{Repetition of Figure \ref{fig:RLvsa} including the line $R=k\hat a$ for $k=1$ as dashed line.}
        \label{fig:RLvs_k}
    \end{subfigure}
        \begin{subfigure}{0.49\textwidth}
        \includegraphics[width=\textwidth]{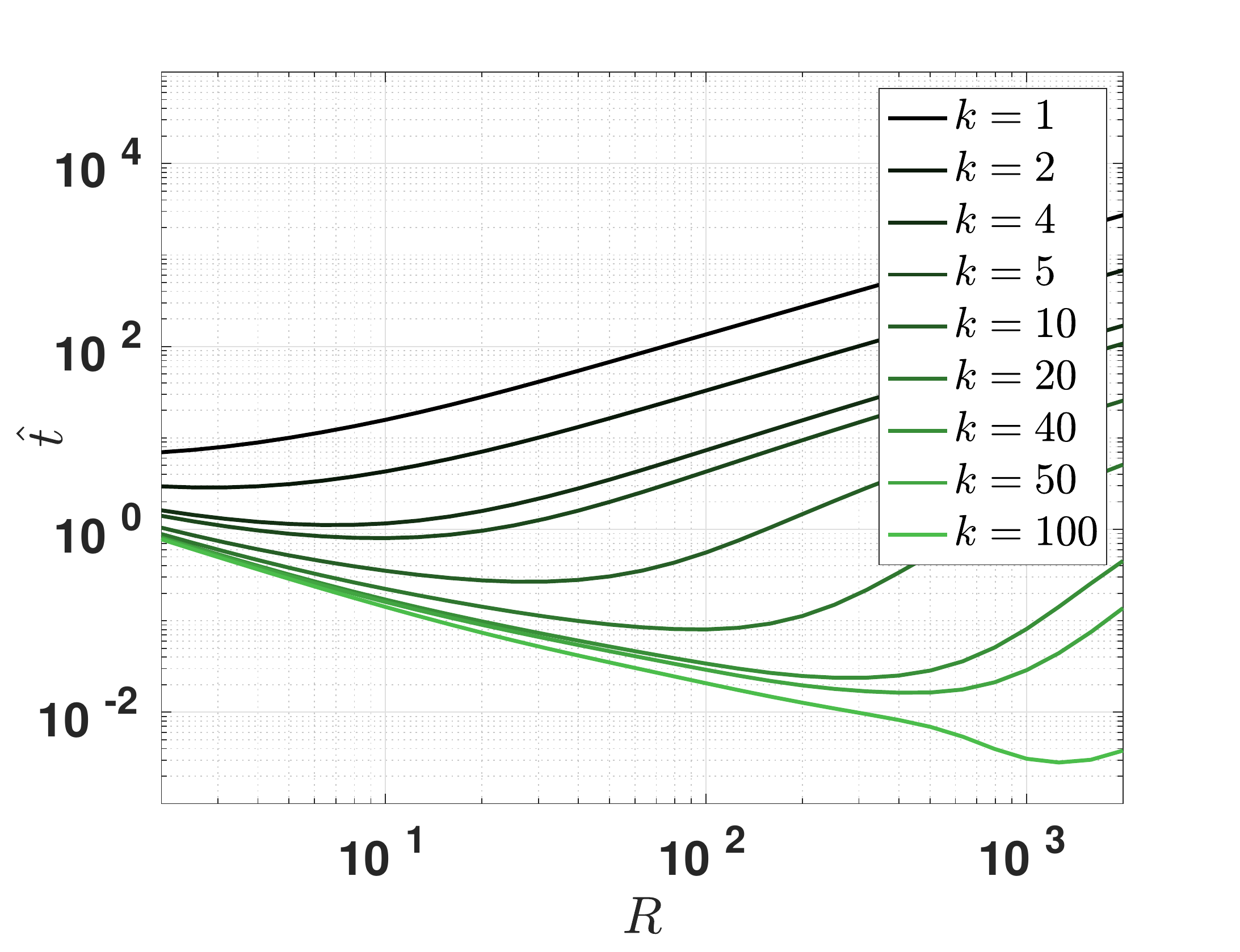}
        \subcaption{Non-dimensional onset time $\hat t$ (vertical axis) as a function of Rayleigh number $R$ (horizontal axis) for various $k$.}
        \label{fig:Rvst_varyE}
    \end{subfigure}
\caption{Relation between onset time and critical Rayleigh number for varying evaporation rate.}
\label{fig:hattvsR_varyE}    
\end{figure}

To investigate the overall effect of increasing evaporation rate, we recall the definitions
\begin{equation}
    R=\frac{Q_{\text{ref}}}{E},\quad \hat\beta=\frac{\mathcal W E}{D},\quad \hat a=\frac{\pi}{\hat\beta}=\frac{\pi D}{\mathcal W E},\label{eq:nondimnum2}
\end{equation}
where $\hat a$ is the lowest possible wavenumber in the bounded domain case and hence the most unstable mode. Eliminating $E$ from \eqref{eq:nondimnum2} yields
\begin{equation}
    R = k \hat a \quad \text{with } k=\frac{Q_{\text{ref}}\mathcal W}{D\pi}.\label{eq:Rka}
\end{equation}
Returning to Figure \ref{fig:RLvsa}, this means that for a given $k$ we look for the points where the curves for fixed times cross the curves $R=k\hat a$. As a given $E$ corresponds to a given $\hat a$ (and also $R$), we can find a corresponding onset time $\hat t$ for each $E$. Due to the shape of the curves from Figure \ref{fig:RLvsa}, there will be an evaporation rate where a minimum non-dimensional onset time is found. In Figure \ref{fig:hattvsR_varyE} we re-plot Figure \ref{fig:RLvsa} together with the line corresponding to $k=1$, as well as the corresponding onset times as a function of $R$ for various choices of $k$. The minimum onset times seen in Figure \ref{fig:Rvst_varyE} correspond to the times where the line $R=k\hat a$ is tangential to a line corresponding to a fixed time in Figure \ref{fig:RLvs_k}.

When the evaporation rate increases, $R$ decreases. Hence, as $E$ increases, we follow a specific curve (corresponding to the chosen value of $k$) in Figure \ref{fig:Rvst_varyE}, from right to left. We see that as evaporation increases, the corresponding non-dimensional onset time decreases until it reaches a minimum, but then increases as $E$ increases even further. Hence, it appears that for large evaporation rates, increasing $E$ further has a stabilizing effect since the onset time increases. However, Figure \ref{fig:Rvst_varyE} shows \emph{non-dimensional} onset times. To re-dimensionalize the onset times, we multiply with $t_{\text{ref}} = {\phi D}/{E^2}$. Hence, for increased evaporation rates, the time scaling is smaller. Therefore, when considering the dimensional onset times, these are found to always decrease as the evaporation rate increases. This is independent of the choice of $k$. Hence, an increased evaporation rate always has a destabilizing effect on the system, since instabilities can appear earlier in dimensional time.

\section{Numerical methods}\label{sec:dumux}
Additionally a numerical, REV-scale model is used to simulate the problem described in Section \ref{sec:mathmodel}. Inline with the formulation in Section \ref{sec:mathmodel}, and due to the applied numerical model using dimensional variables, we now consider the dimensional model. The numerical model is able to give insights in the distribution and development of the salt concentration in the entire domain. This means that the onset as well as the development of instabilities can be observed in detail. However it is computationally cost intensive as quite a fine spatial and temporal discretization is necessary to reduce the influence of numerical diffusion and resolve the instabilities. The numerical model is implemented in the open-source simulator $\mathrm{DuMu^x}$ \cite{koch2020dumux} for multi-phase, multi-component flow and transport in porous media. It is a research code written in C++ and based on Dune \cite{Bastian2021}, a scientific numerical software framework.

The model considers the precipitation of salt if the solubility limit is exceeded. In this section we therefor first define the reaction term. Additional boundary and initial conditions are specified as we here consider a vertically bounded domain and introduce initial perturbations for the salt mole fraction. Further the spatial and temporal discretizations are described. The domain is discretizied by a cell-centered finite volume scheme and an implicit Euler method is used for the time discretization. In the end the evaluation method of the simulation results is described, the numerical onset time is defined and the influence of different initial perturbations is shown.

\subsection{Salt precipitation reaction in the numerical model}\label{sec:reactionrate}
In the numerical model the precipitation of solid salt is considered if the solubility limit is exceeded. Therefor the reaction term in the mole balance of the soluted and solid salt (Equation \eqref{eq:molefractions} and \eqref{eq:molebalancesolid}) has to be specified.
For the reaction terms of NaCl $r_\mathrm{solid} = -r^\mathrm{NaCl}$ an equilibrium reaction is assumed. 
This is based on the assumption that the chemical reaction is fast so that every mol of NaCl above the solubility limit $\mathsf{x}^\mathrm{NaCl}_\mathrm{max}$ can precipitate within one numerical time step~$\Delta t$. 
\begin{align}
r_\mathrm{solid} &= -r^\mathrm{NaCl} = \rho_{\mathrm{mol}} \phi ~ \frac{\mathsf{x}^\mathrm{NaCl} - \mathsf{x}^\mathrm{NaCl}_\mathrm{max}}{\Delta t}.\label{eq:reactionTerm}
\end{align}
Due to the precipitation, the porosity of the porous medium and thus the permeability decreases. For the calculation of the permeability a Kozenzy-Carman-type relationship based on the initial values $\phi_0$, $K_0$, is used:
\begin{align}
K = K_0\left( \frac{1-\phi_0}{1-\phi} \right)^2 \left( \frac{\phi}{\phi_0} \right)^3.
\end{align}
For the investigation of the onset of the instabilities only the period of increasing salt concentration with concentrations below the solubility limit is considered. Hence, the solubility limit is mainly of importance as critical value for the start of precipitation and a more detailed representation of the precipitation process is not necessary.

\subsection{Boundary and initial conditions}\label{sec:NumericBcic}
The numerical model uses additional boundary and initial conditions to the ones described in Section \ref{sec:mathIcBc}.
As the domain is finite at the bottom for the numerical simulation, boundary conditions for the bottom are necessary. A Dirichlet boundary condition sets the pressure in the liquid phase corresponding to the hydro-static pressure at the domain depth and the salt concentration is set equal to the initial mole fraction $\mathsf{x}_0^\mathrm{NaCl}$. The depth $d$ of the domain is chosen so that at the time of onset no increase in concentration is observable at the bottom and so the influence of the bottom boundary is assumed to be negligible. 
\begin{align}
P|_{z=d} = P_0 + \rho(\mathsf{x}_0^\mathrm{NaCl}) g d, \\
\mathsf{x}^\mathrm{NaCl}|_{z=d} = \mathsf{x}_0^\mathrm{NaCl}.
\end{align}

As initial condition additionally a hydro-static pressure profile is used which leads to better convergence at the beginning of the simulations:
\begin{align}
P|_{t=0} = P_0 + \rho(\mathsf{x}_0^\mathrm{NaCl}) g z.
\end{align}
The numerical model uses also an initial perturbation for $\mathsf{x}^\mathrm{NaCl}$, denoted $\mathsf{x}^\mathrm{NaCl}_{0,\mathrm{p}}$, to correspond to the perturbations used in the analytical analysis. If no initial perturbations are applied, instabilities are triggered by tiny numerical errors in the order of machine precision.
Instead, two different types of perturbations are used, a periodic and a random one, which can either be applied in the top row of cells or in the whole domain. 
The periodic perturbation is applied by using a cosine function along the $x$-coordinate: 
\begin{align}
\mathsf{x}^\mathrm{NaCl}_{0,\mathrm{p}}(x) =\mathsf{x}_\mathrm{0}^\mathrm{NaCl} + A \cdot \mathrm{cos} \left(\frac{2\pi}{\lambda} \cdot x \right),\label{eq:periodicperturb}
\end{align}
with the wavelength $\lambda$ and the amplitude $A$. The wavelength and amplitude need to be prescribed. Which amplitude to use will be discussed in Section \ref{sec:perturbinfluence}, while the choice of wavelength is discussed in Section \ref{sec:resultswidth}.
Alternatively a random perturbation for $\mathsf{x}^\mathrm{NaCl}_{0,\mathrm{p}}$ is used. In this case, values for $\mathsf{x}^\mathrm{NaCl}_{0,\mathrm{p}}$ are randomly picked for every discrete cell from a normal distribution $\mathcal{N}$ using a mean value of $\mu_\mathrm{\mathsf{x}^\mathrm{NaCl}}=\mathsf{x}_0^\mathrm{NaCl}$ and a standard deviation $\sigma_\mathrm{\mathsf{x}^\mathrm{NaCl}}$:
\begin{align}
\mathsf{x}^\mathrm{NaCl}_{0,\mathrm{p}} \sim \mathcal{N} \left(\mu_\mathrm{\mathsf{x}^\mathrm{NaCl}}, \sigma_\mathrm{\mathsf{x}^\mathrm{NaCl}}^2 \right).
\end{align}
The choice of standard deviation is discussed in Section \ref{sec:perturbinfluence}.

\subsection{Discretization}
For the spatial discretization a cell-centered finite volume scheme applying the two-point flux approach is used, with a first order upwind scheme for the convective flux and a second order scheme for the diffusive fluxes \cite{Helmig1997}. A first order, implicit Euler method is used for time discretization \cite{Helmig1997}.
A convergence study for the spatial and temporal discretization is conducted (see Appendix \ref{app:d}), ensuring that there is a negligible influence of numerical diffusion effects. 
For the spatial discretization it is important to use finer grid cells than the expected wavelength in order to resolve the instabilities: $\Delta x \ll \lambda$. 
Through preliminary testing we found that the expected wavelength depends on the permeability and the vertical density difference, and is smaller for higher permeabilities. 
At least 10 cells are used per wavelength of the highest investigated permeability.

As the instabilities are initiated at the top, a fine resolution in $z$-direction is important at the top ($z=0$). Further, with a REV-scale model the increase in salt mole fraction and density due to evaporation refers to the cell volume. Thus the evaporative water loss applied to cells at the top boundary would lead to smaller increase in $\mathsf{x}^\mathrm{NaCl}$ for bigger $\Delta z$. To lower the influence of this effect a relatively fine $\Delta z$ is used at the top with $\Delta z^\mathrm{top} = 3.3 \cdot 10^{-4}~\mathrm{m}$.
To lower the computational costs in the lower parts of the domain coarser cell sizes can be used. Hence, $\Delta z$ increases continuously towards the bottom. 

A time step of $\Delta t = 50~\mathrm{s}$ is used. This means that the grid velocity of the top cell in $z$-direction ${\Delta z^\mathrm{top}}/{\Delta t}$ is higher than the evaporation rate $E$ and thus is able to capture the evaporation process correctly.

\subsection{Evaluation of numerical simulations}\label{sec:NumericEvaluation}
To estimate the numerical onset time, the mean value $\mu_\mathrm{\mathsf{x}^\mathrm{NaCl}}^\mathrm{top}$ and the standard deviation $\sigma_\mathrm{\mathsf{x}^\mathrm{NaCl}}^\mathrm{top}$ of the salt mole fraction $\mathsf{x}^\mathrm{NaCl}$ of the grid cells in the top row are calculated. The standard deviation is a measure for the variation of the salt mole fraction. As the instabilities are characterized by the development of fingers with higher salt mole fraction separated by regions with lower salt mole fractions, an increase of the standard deviation describes the onset of the instabilities. Thus, the onset time for the numerical simulations is defined as the time when the standard deviation is at its minimum. 
Physically this corresponds to the case where the convective flux starts to dominate the diffusive fluxes in horizontal direction, which leads to the enhancement of the perturbations. The development of $\mu_\mathrm{\mathsf{x}^\mathrm{NaCl}}^\mathrm{top}$ and $\sigma_\mathrm{\mathsf{x}^\mathrm{NaCl}}^\mathrm{top}$ over time can be seen in Figure \ref{fig:NumericPhases} and will be discussed in detail in Section \ref{sec:instabilityDevelopment}.

\subsection{Influence of initial perturbation on onset}\label{sec:perturbinfluence}
\begin{figure}
    \begin{subfigure}{0.49\textwidth}
    \includegraphics[width=\textwidth]{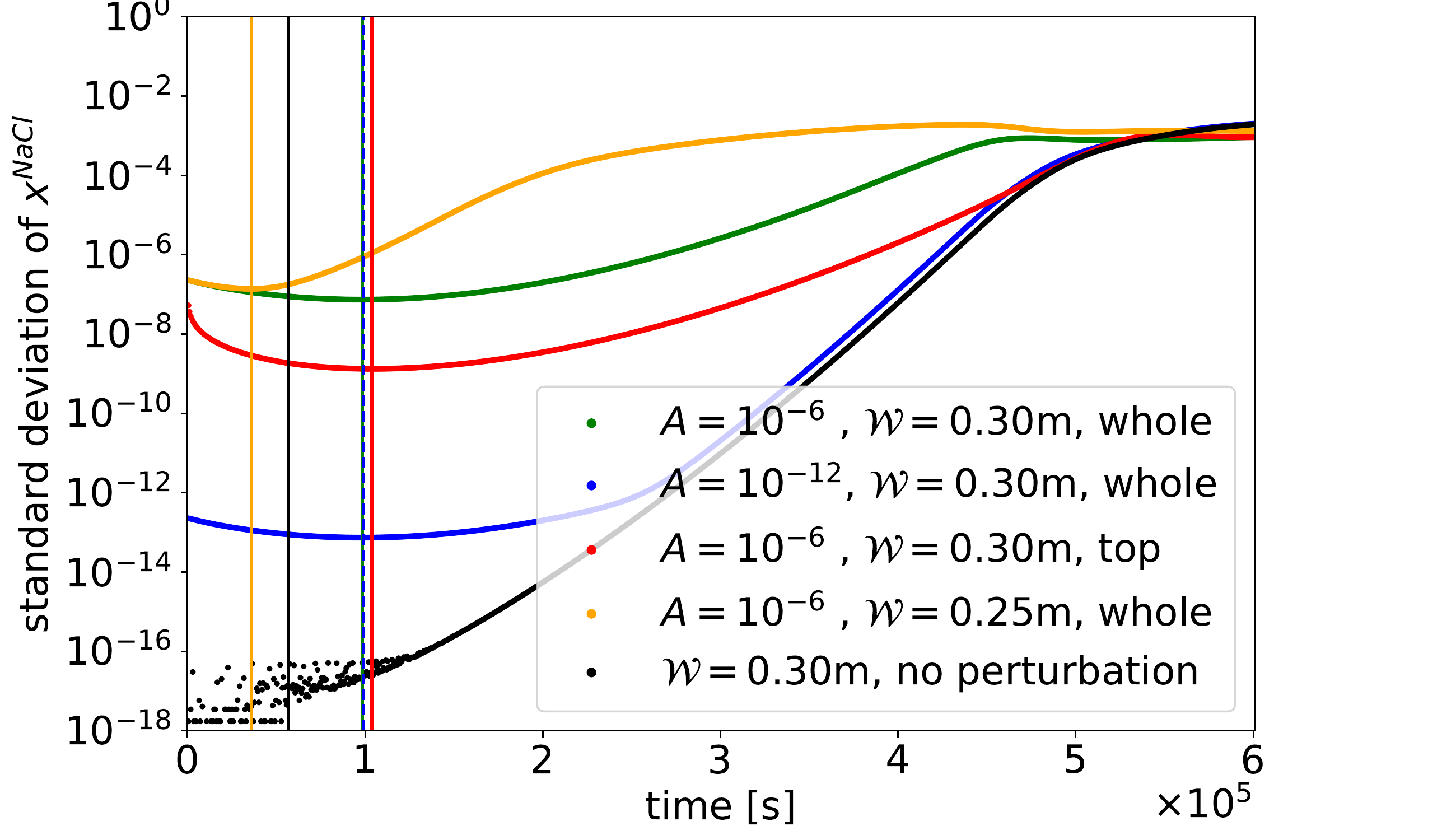}
    \subcaption{Periodic initial pertubations}
    \label{fig:NumericInitialPeriodic}
    \end{subfigure}
    \begin{subfigure}{0.49\textwidth}
    \includegraphics[width=\textwidth]{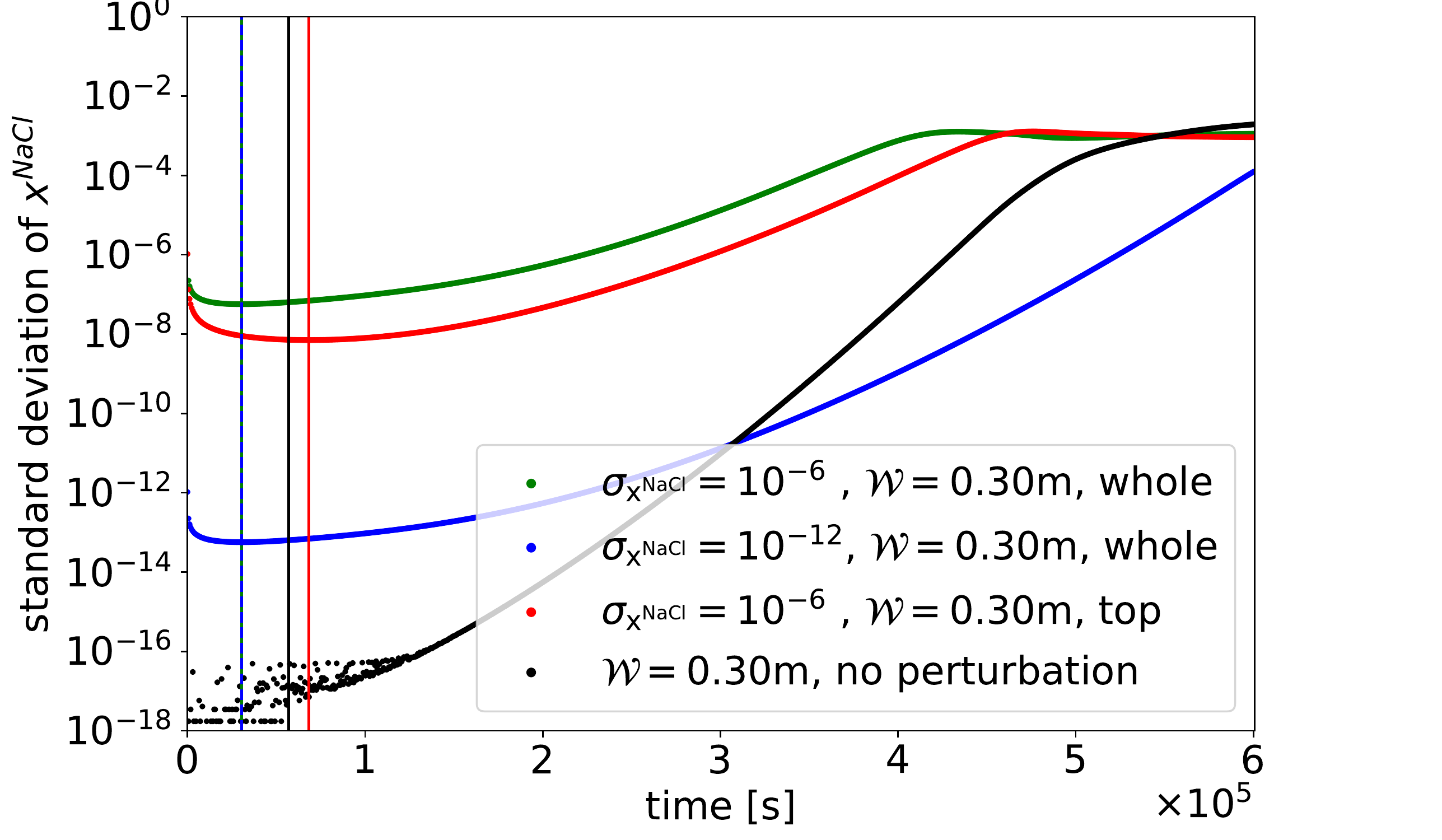}
    \subcaption{Random initial pertubations}
    \label{fig:NumericInitialRandom}
    \end{subfigure}
    \caption{Influence of different initial perturbations for the numeric simulation on the onset time for periodic initial perturbations (a) and random initial perturbations (b). For the simulations the parameters listed in Table \ref{tab:parameters} are used with a permeability of $K=10^{-11}~\mathrm{m^2}$. The parameters amplitude $A$, half domain width $\mathcal{W}$, standard deviation $\sigma_\mathrm{\mathsf{x}^\mathrm{NaCl}}$ as described in the legend as well as the application area (top or whole). The vertical lines indicate the time of onset for the respective cases. Note that the blue and green vertical lines are in both cases on top of each other.}
\label{fig:NumericInitial}     
\end{figure}

Figure \ref{fig:NumericInitial} shows the development of the standard deviation $\sigma_\mathrm{\mathsf{x}^\mathrm{NaCl}}^\mathrm{top}$ and onset time for different parameters for the periodic initial perturbation (Figure \ref{fig:NumericInitialPeriodic}) and the random initial perturbation (Figure \ref{fig:NumericInitialRandom}). Also the case without perturbations is shown, where the perturbations are triggered by tiny numerical errors.
This figure shows that the amplitude and the standard deviation of the initial perturbation do not affect the onset time.
If the initial perturbation is applied only to the top of the domain and not in the whole domain, the onset is later for both perturbation types. However, for the periodic perturbations the difference in onset is relatively small. For the periodic perturbations \eqref{eq:periodicperturb} it is of importance that the width of the domain is a multiple of the initial wavelength. Here, simulations with $\lambda = 0.03$m is used, hence $\mathcal{W}=0.30$m is a multiple of it, while $\mathcal{W}=0.25$m is not. For the latter case the onset time is earlier. 

In the following investigations we use a perturbation in the whole domain. This corresponds better to the manner perturbations are applied in the linear stability analysis and hence benefits the comparison of the onset times between the linear stability analysis and the numerical simulations. Further an amplitude of $A= 10^{-6}$ is used for the periodic perturbations and a standard deviation of $\sigma_\mathrm{\mathsf{x}^\mathrm{NaCl}}= 10^{-6}$ for the random perturbations.

\section{Onset and development of density instabilities}\label{sec:results}
Here we compare results from linear stability analysis and numerical simulations with respect to predicted onset times for instabilities and with respect to the behavior of the salt concentration before onset of instabilities. Most parameters are for simplicity kept fixed, and are as specified in Table \ref{tab:parameters}. These parameters are chosen to realistically represent saline water in a porous domain, with evaporation corresponding to 34.7 cm/year. 
The permeability $K$ of the porous medium is varied. We consider the cases $K=10^{-10}$ m$^2$, $K=10^{-11}$ m$^2$, $K=10^{-12}$ m$^2$ and $K=10^{-13}$ m$^2$. 

\begin{table}
\caption{Fixed parameter choices}
\label{tab:parameters}       
\begin{tabular}{lll}
\hline\noalign{\smallskip}
Parameter & Value & Dimension \\
\noalign{\smallskip}\hline\noalign{\smallskip}
$X_0$ or $\mathsf{x}_0$ & 0.035 or 0.011 & -\\
$\rho_0$  & 1025 & kg m$^{-3}$ \\
$\mu$ &  $1.1\cdot 10^{-3}$ & kg m$^{-1}$ s$^{-1}$ \\
$\phi$ & 0.4 & -\\
$\gamma$ & 0.7 & - \\
$g$ & 9.8 & m s$^{-2}$ \\
$E$ or $E_{\mathrm{mol}}$& $1.08\cdot 10^{-8}$ or $6.165\cdot 10^{-4}$& m s$^{-1}$ or mol m$^{-2}$ s$^{-1}$ \\
$D$ & $4.42\cdot 10^{-10}$ & m$^2$ s$^{-1}$\\
\noalign{\smallskip}\hline\noalign{\smallskip}
\multicolumn{3}{l}{Additional numerical parameters} \\
\noalign{\smallskip}\hline\noalign{\smallskip}
$d$ &  0.2 & m \\
$P_0$ & $1.0 \cdot 10^{5}$ & Pa \\
$\mathsf{x}_\mathrm{max}^\mathrm{NaCl}$ & $0.0977$  & - \\
\noalign{\smallskip}\hline
\end{tabular}
\end{table}

Note that the specified evaporation rate is used for the numerical simulations as an input parameter. For the linear stability analysis, the choice of evaporation rate translates into a Rayleigh number, which determines the onset time, as described in Section \ref{sec:linsolstrategy}. We separate between the bounded case, where the domain has a fixed width, and the unbounded case, where we investigate the onset of a specific wavelength and use periodic boundary conditions. For both cases we can compare the onset times from the linear stability analysis and from the numerical simulations.

\subsection{Onset times for a domain of fixed width}\label{sec:resultswidth}
\begin{figure}
    \begin{subfigure}{0.49\textwidth}
        \includegraphics[width=\textwidth]{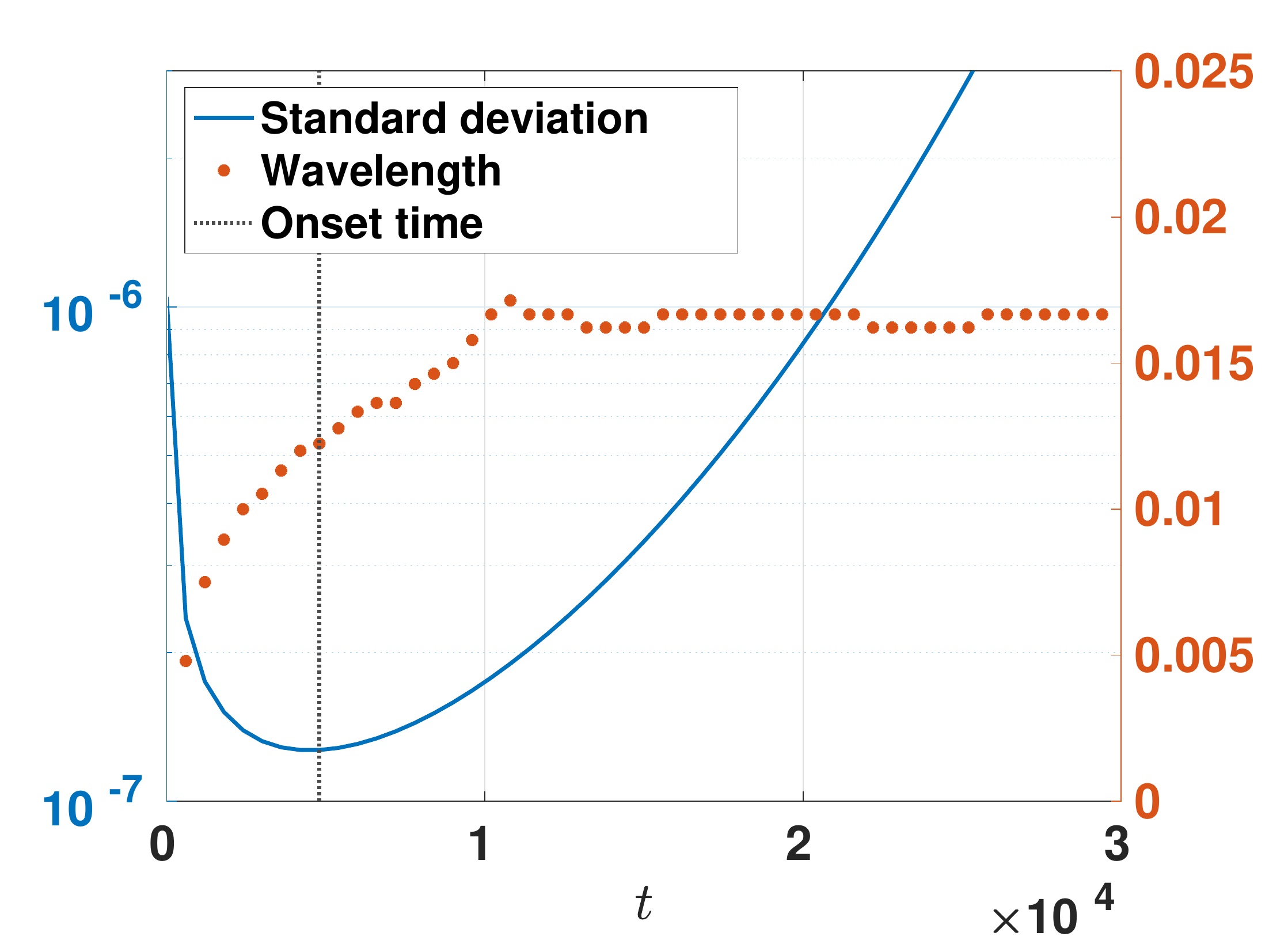}
        \subcaption{$K=10^{-10}$ m$^2$}
    \end{subfigure}
    \begin{subfigure}{0.49\textwidth}
        \includegraphics[width=\textwidth]{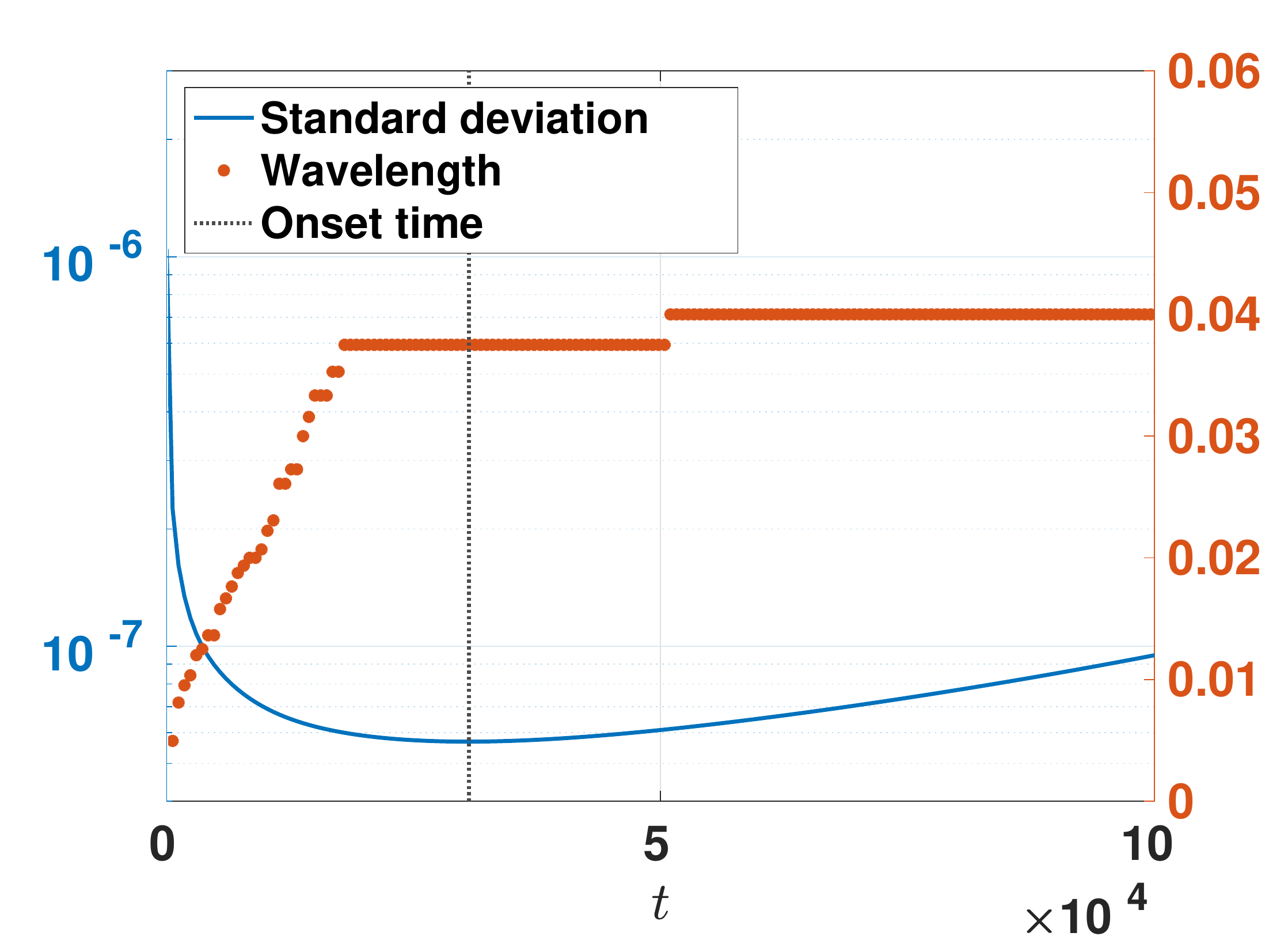}
        \subcaption{$K=10^{-11}$ m$^2$}
    \end{subfigure}
    \begin{subfigure}{0.49\textwidth}
        \includegraphics[width=\textwidth]{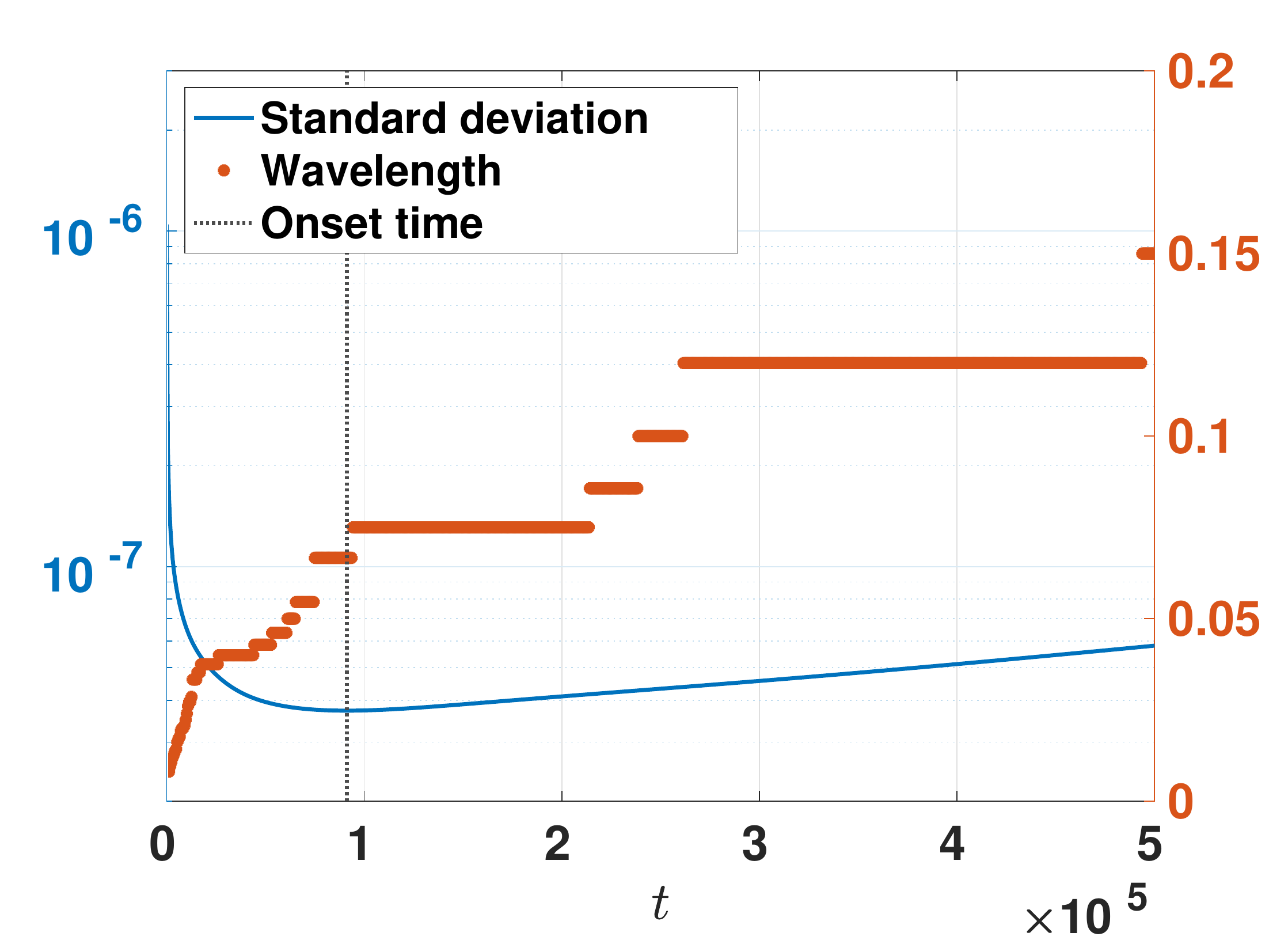}
        \subcaption{$K=10^{-12}$ m$^2$}
    \end{subfigure}
    \begin{subfigure}{0.49\textwidth}
        \includegraphics[width=\textwidth]{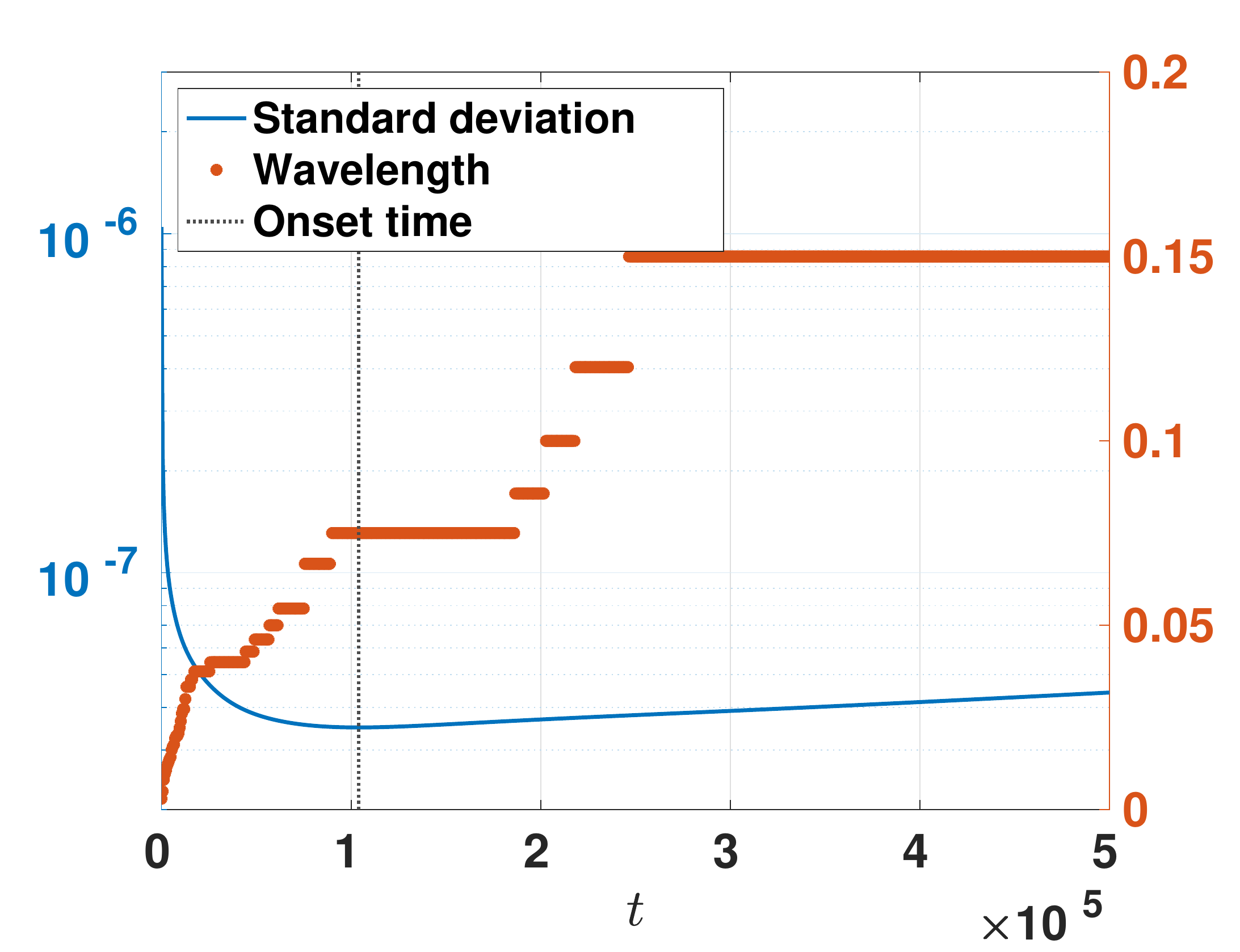}
        \subcaption{$K=10^{-13}$ m$^2$}
    \end{subfigure}
    \caption{Standard deviation and average wavelength (in m) over time for fixed width for $K=10^{\text{-} 10}$~m$^2$ (top left), $K=10^{\text{-} 11}$~m$^2 $ (top right), $K=10^{\text{-} 12}$~m$^2 $ (bottom left) and $K=10^{\text{-} 13}$~m$^2 $ (bottom right). The dotted vertical line indicates time of minimum standard deviation, which is used as the onset time.}
\label{fig:NumericPlotsRandom}  
\end{figure}

We here investigate the onset of instabilities in the bounded case. We use a domain with a fixed width of 60 cm. In the linear stability analysis we assume that the most unstable wavelength will be dominating and we find the corresponding onset time for this wavelength, as described in Section \ref{sec:linsolstrategy}. In the numerical simulations we use an initial random perturbation, which is assumed to trigger the onset of the most unstable wavelength. The development of the standard deviation and the appearing (average) wavelength is shown in Figure \ref{fig:NumericPlotsRandom}. The resulting onset times are in Table \ref{tab:onsetTimesAnalysisWidth}. Both the analytic and numerical approach show that lower permeabilities correspond to later onset times. Although the numbers are of the same order of magnitude, the deviation between onset times estimated by the linear analysis and numerical simulations is quite large. 

\begin{table}
\caption{Onset times from the linear stability analysis for a domain of fixed width. A width of 60 cm has been used for all cases.}
\label{tab:onsetTimesAnalysisWidth}    
\begin{tabular}{lll}
\hline\noalign{\smallskip}
permeability & analytic onset time & numeric onset time \\[1mm]
\noalign{\smallskip}\hline\noalign{\smallskip}
  $K=10^{- 10}$ m$^2$ & $6.57 \cdot 10^{2}$ s & $4.80 \cdot 10^{3}$ s\\
  $K=10^{-11}$ m$^2 $ & $1.16 \cdot 10^{4}$ s & $3.06 \cdot 10^{4}$ s\\
  $K=10^{-12}$ m$^2 $ & $1.34 \cdot 10^{5}$ s & $9.12 \cdot 10^{4}$ s\\
  $K=10^{-13}$ m$^2 $ & $2.11 \cdot 10^{6}$ s & $1.04 \cdot 10^{5}$ s\\
  \noalign{\smallskip}\hline
  \end{tabular}
\end{table}

This analysis reveals some fundamental differences in the underlying assumptions in the two approaches. The linear stability analysis indicates that the most unstable wavelength should be the longest one that will fit into the domain, as lower wavenumbers $\hat a$ are more unstable, as seen in Figure \ref{fig:RLvsa}. For this case that would correspond to a wavelength of 60 cm. 
In the simulations, we can observe that different wavelengths are dominating before and after the estimated onset time. Since we use a random perturbation, several wavelengths are represented and they can also interact with each other. The appearing wavelengths after onset are generally found to be shorter for increasing permeability, as seen in Figure \ref{fig:NumericPlotsRandom}. 
For the lower permeabilities $K=10^{-13}$ m$^2$ and $K=10^{-12}$ m$^2$, the early appearing dominating wavelengths are 15 cm and 12 cm, respectively, which is close to the ones assumed by the linear stability analysis. 
For the larger permeabilities $K=10^{-11}$ m$^2$ and $K=10^{-10}$ m$^2$, the appearing dominating wavelengths are 4 cm and 1.5 cm, respectively. 
Although the linear stability analysis indicates that the wavelength of 60 cm should be more unstable and hence preferred, the initial random perturbation have triggered modes that are much shorter in wavelength. The wavelengths that do appear in the numerical simulations depend on the permeability, following a similar trend as observed in Riaz et al.~(2006) \cite{Riaz2006}. This is remarkable since the setup is different. 
Also, using a random perturbation in the numerical simulations can give nonlinear effects as the different wavelengths interact with each other, possibly affecting the resulting onset mode and time. However, the linear stability analysis assumes that the perturbation is a specific wavelength and hence does not account for any interaction between different wavelengths. This motivates to rather use a specific wavelength for the numerical perturbation and compare with the onset time of this wavelength from the linear stability analysis.

\subsection{Onset times of a fixed wavelength}\label{sec:resultswavelength}
\begin{figure}
    \begin{subfigure}{0.49\textwidth}
        \includegraphics[width=\textwidth]{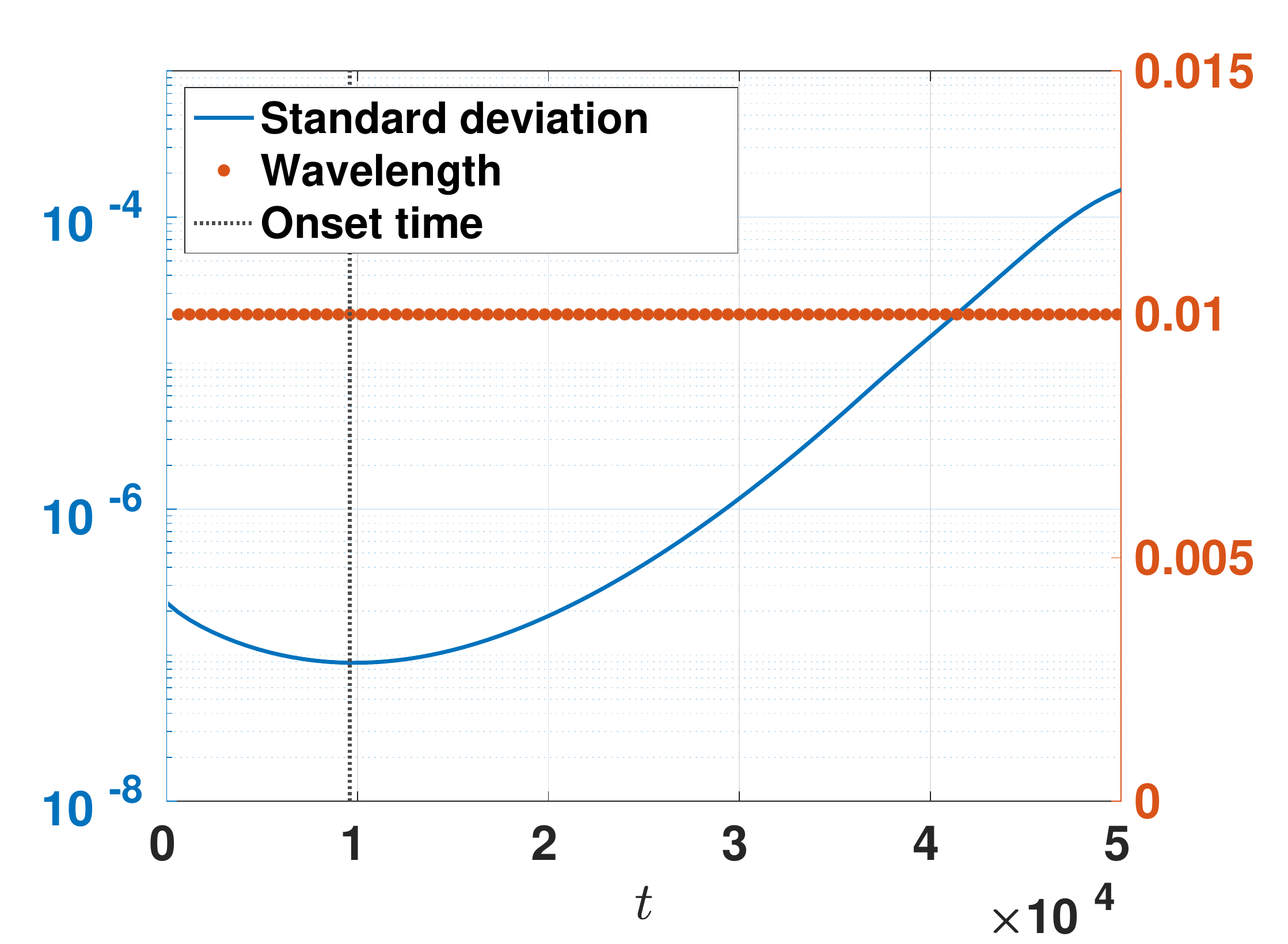}
        \subcaption{$K=10^{-10}$ m$^2$ and wavelength 0.01 m}
    \end{subfigure}
        \begin{subfigure}{0.49\textwidth}
        \includegraphics[width=\textwidth]{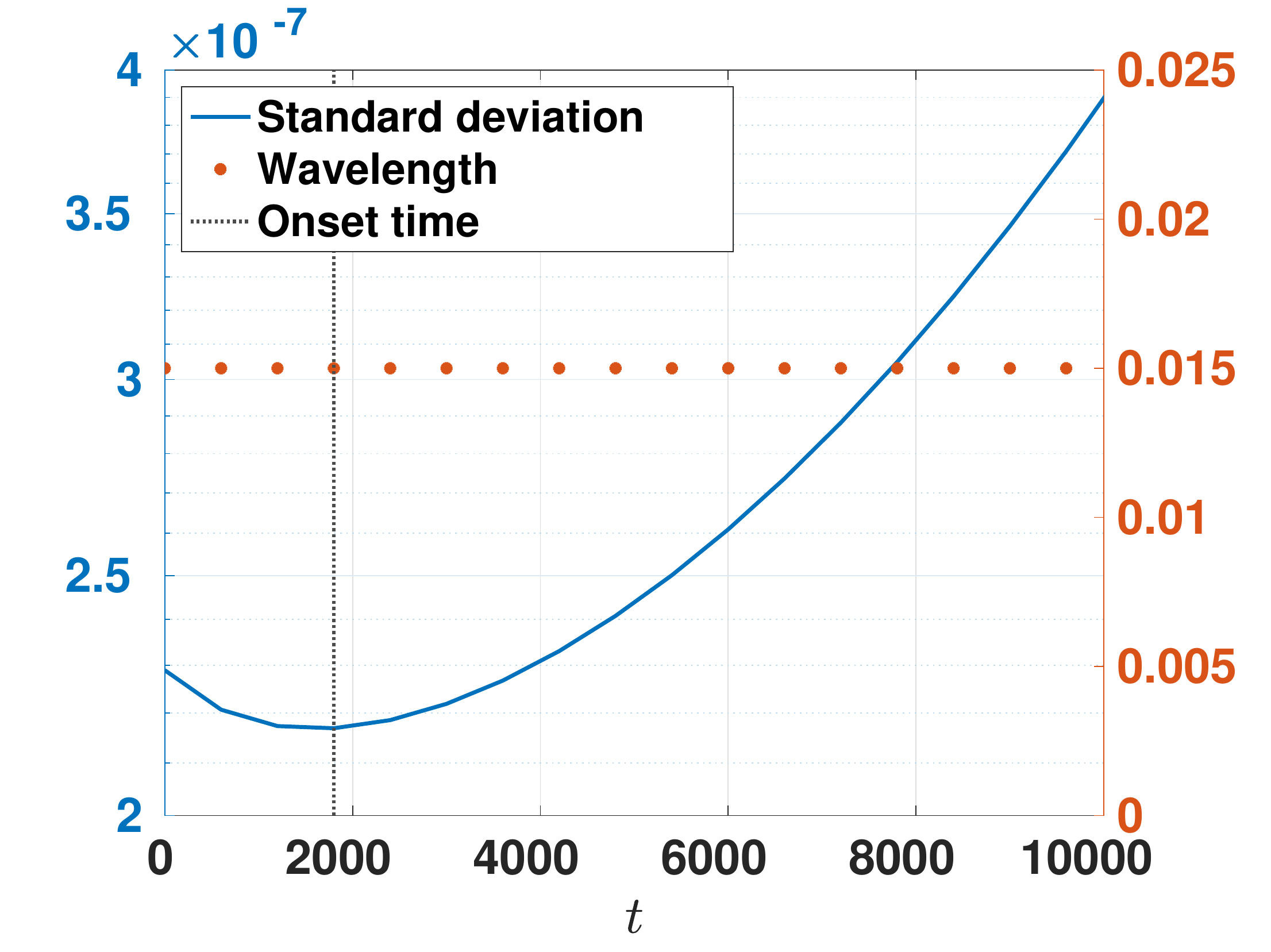}
        \subcaption{$K=10^{-10}$ m$^2$ and wavelength 0.015 m}
    \end{subfigure}
    \begin{subfigure}{0.49\textwidth}
        \includegraphics[width=\textwidth]{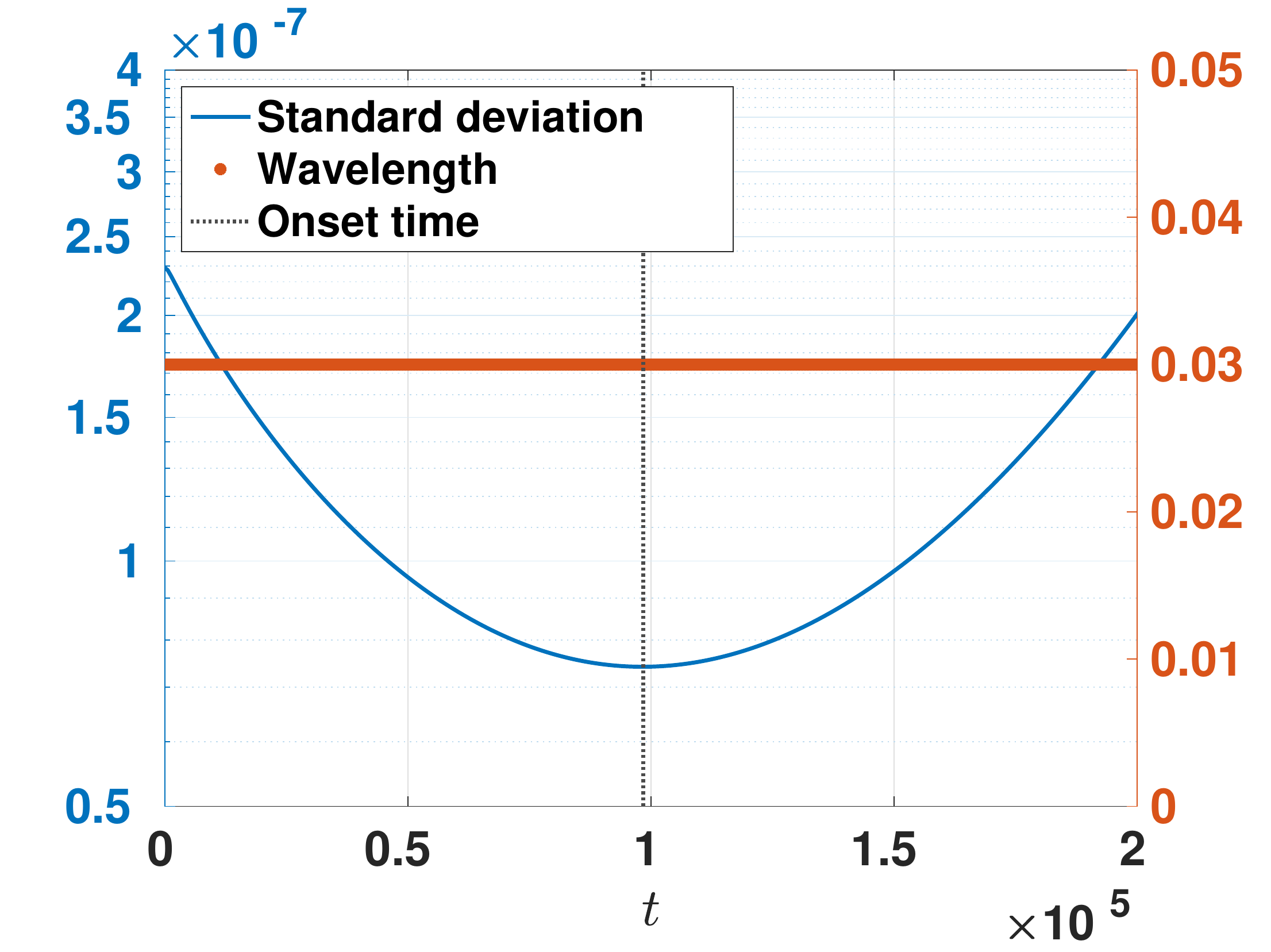}
        \subcaption{$K=10^{-11}$ m$^2$ and wavelength 0.03 m}
    \end{subfigure}
        \begin{subfigure}{0.49\textwidth}
        \includegraphics[width=\textwidth]{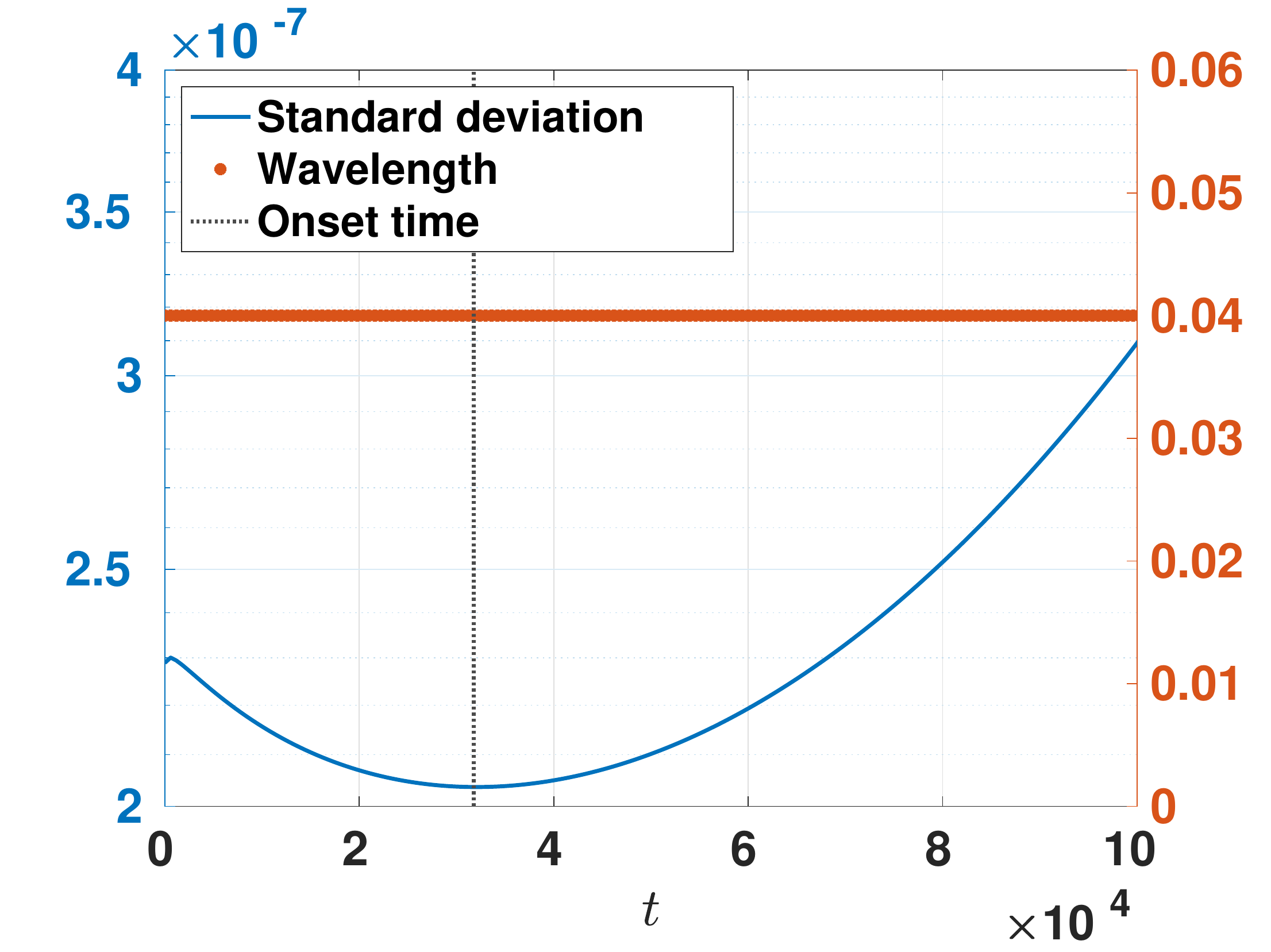}
        \subcaption{$K=10^{-11}$ m$^2$ and wavelength 0.04 m}
    \end{subfigure}
    \begin{subfigure}{0.49\textwidth}
        \includegraphics[width=\textwidth]{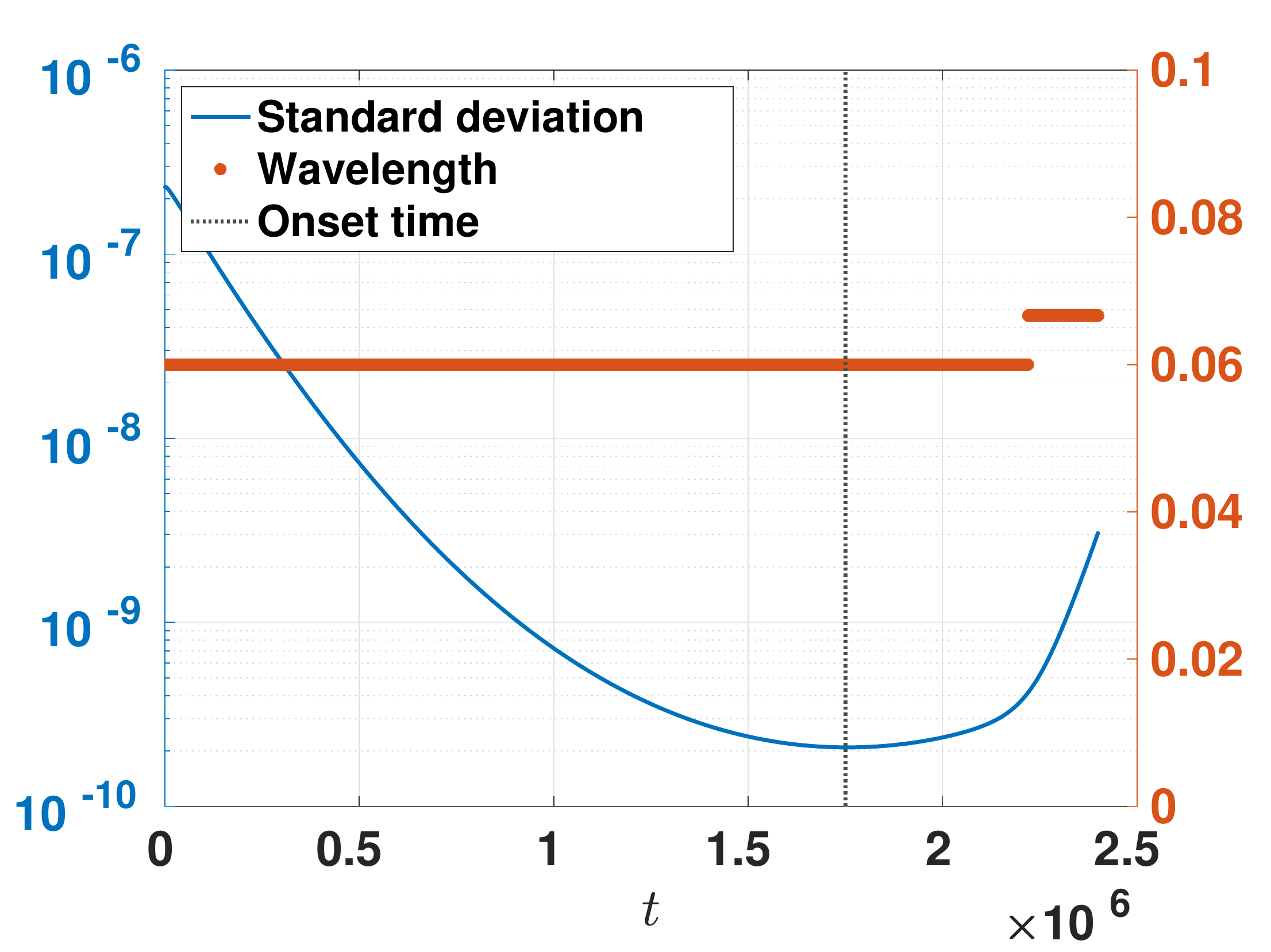}
        \subcaption{$K=10^{-12}$ m$^2$ and wavelength 0.06 m}
    \end{subfigure}
        \begin{subfigure}{0.49\textwidth}
        \includegraphics[width=\textwidth]{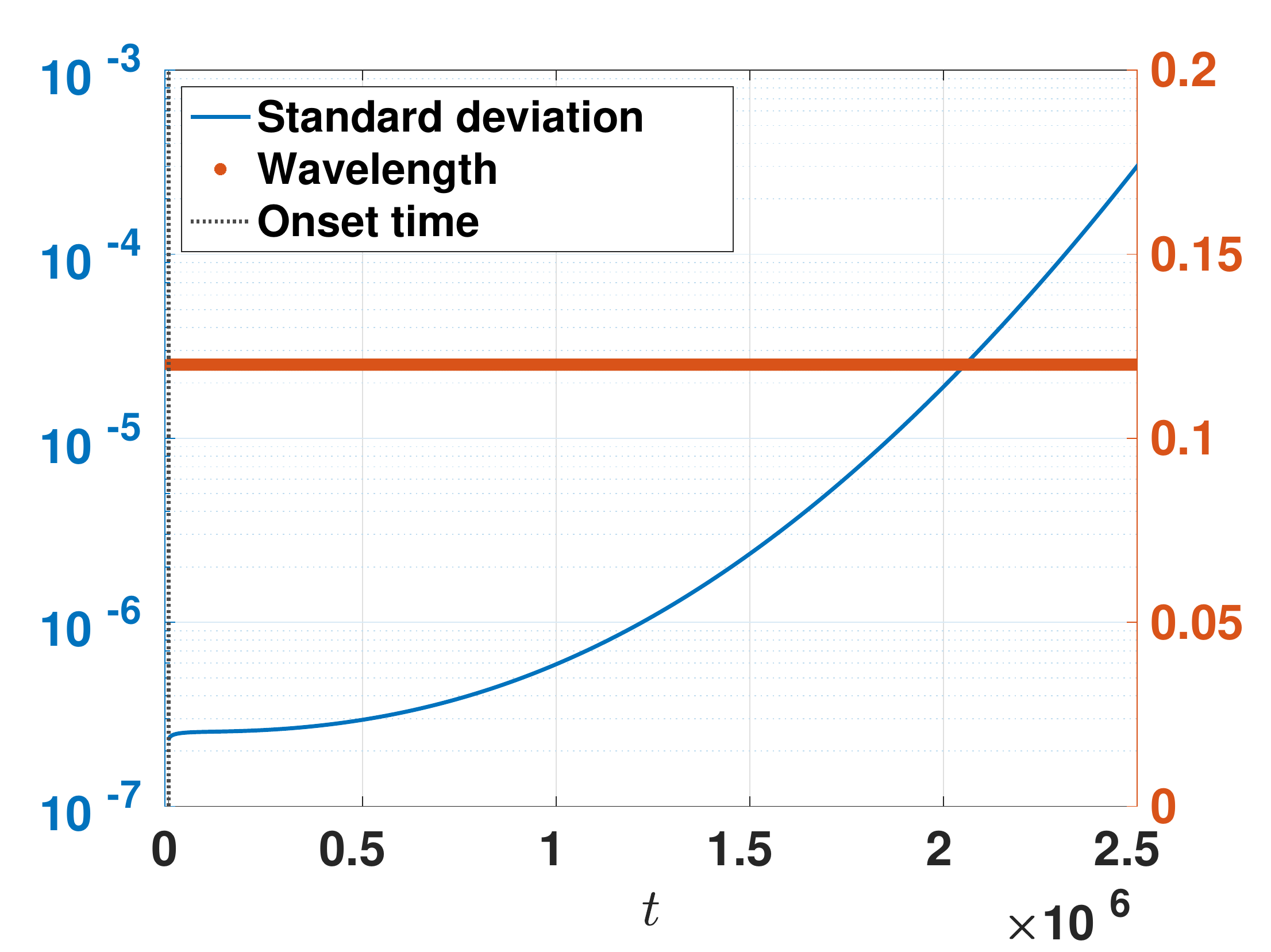}
        \subcaption{$K=10^{-12}$ m$^2$ and wavelength 0.12 m}
    \end{subfigure}
    \begin{subfigure}{0.49\textwidth}
        \includegraphics[width=\textwidth]{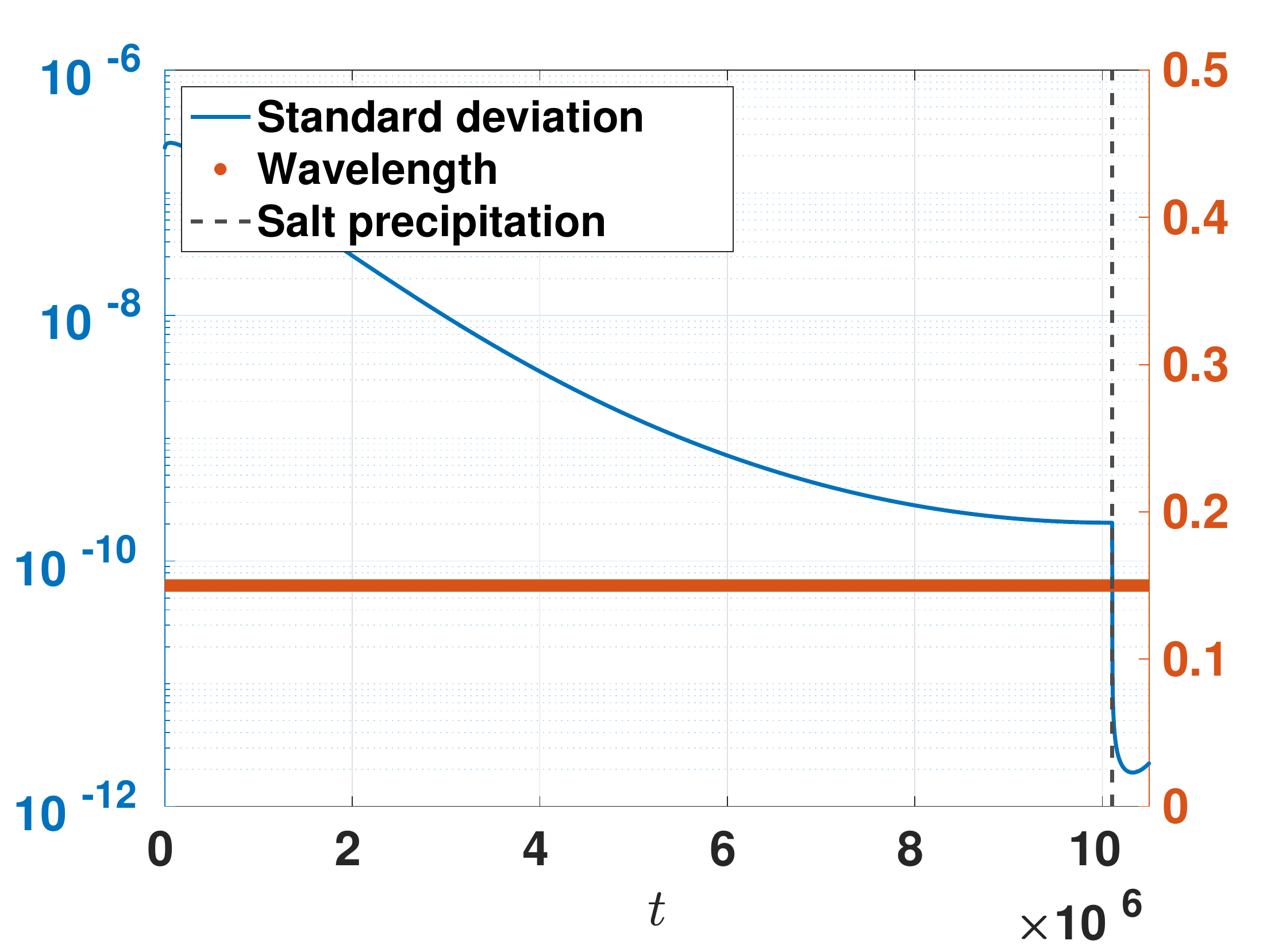}
        \subcaption{$K=10^{-13}$ m$^2$ and wavelength 0.15 m}
        \label{fig:WaveK13_15}
    \end{subfigure}
    \begin{subfigure}{0.49\textwidth}
        \includegraphics[width=\textwidth]{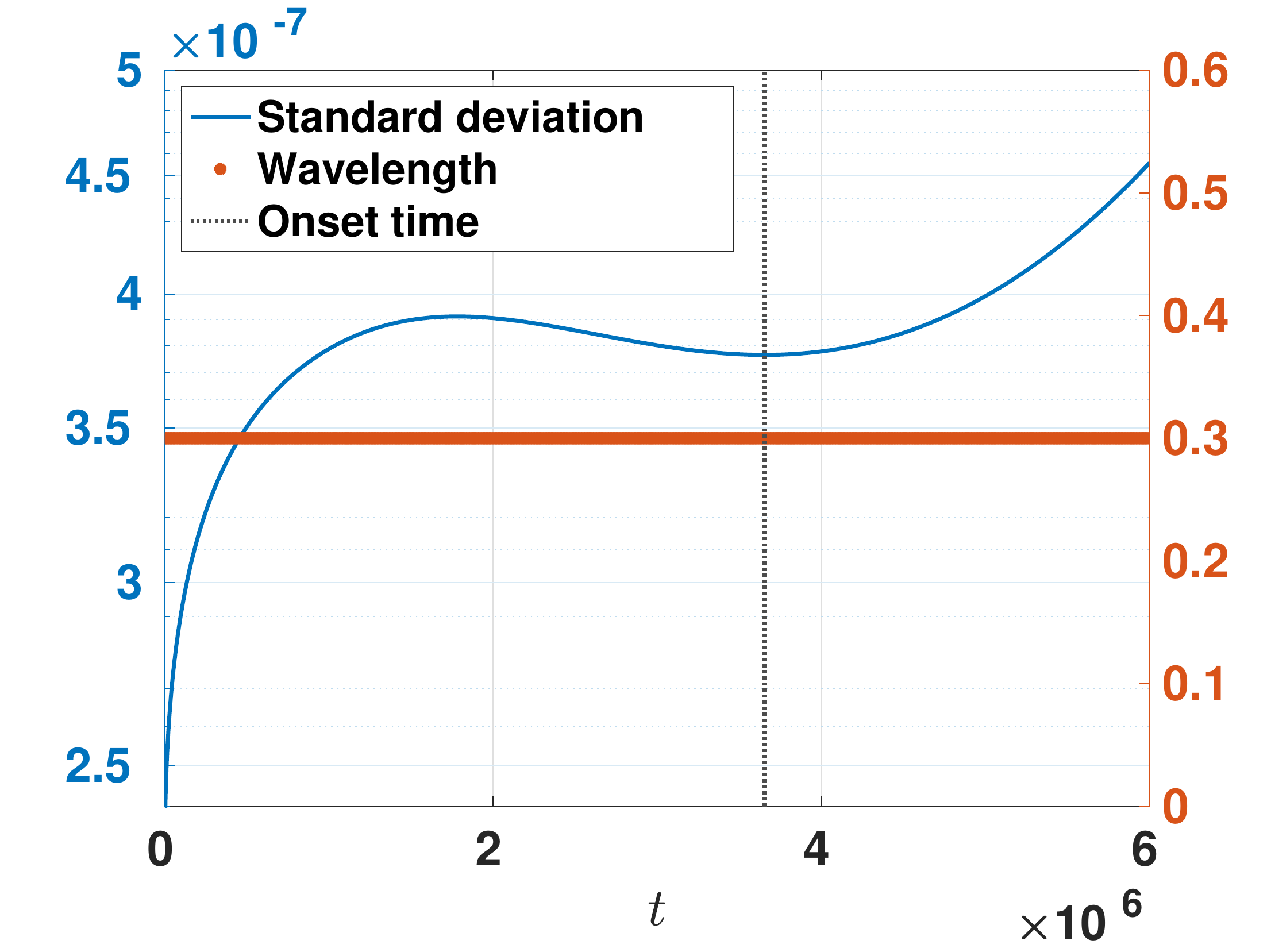}
        \subcaption{$K=10^{-13}$ m$^2$ and wavelength 0.3 m}
    \end{subfigure}
    \caption{Standard deviation and average wavelength (in m) over time for fixed wavelengths for $K=10^{\text{-} 10}$~m$^2 $ (top row), $K=10^{\text{-} 11}$~m$^2 $ (second row), $K=10^{\text{-} 12}$~m$^2 $ (third row) and $K=10^{\text{-} 13}$~m$^2 $ (bottom row) for two chosen wavelengths (left and right). The dotted vertical line indicates time of minimum standard deviation, which is used as the onset time. For the case of $K=10^{-13}$ m$^2$ and wavelength 0.15 m (bottom right) time of initial salt precipitation is marked.}
\label{fig:NumericPlots}      
\end{figure}

We here investigate the onset of instabilities in the unbounded case. Although we use a domain of a width 60 cm for the numerical simulations, we apply periodic boundary conditions on the sidewalls to mimic the domain being unbounded. By using an initial perturbation width a fixed wavelength in the numerical simulations, we investigate the onset of this particular wavelength. This means that the same type of perturbation is used for both numerical simulation and for the linear stability analysis.
We use wavelengths based on the those that appeared after onset in the numerical simulations in Section \ref{sec:resultswidth}, but adjusted such that they fit within the domain. 
For the linear stability analysis we then investigate the onset of this particular wavelength, as explained in Section \ref{sec:linsolstrategy}. For the numerical simulations, a small amplitude of the cos-perturbation is used, and the development of the standard deviation and average appearing wavelength is seen in Figure \ref{fig:NumericPlots}. The used wavelengths and resulting onset times are found in Table \ref{tab:onsetTimesNumeric}. For two of the lower permeabilities, no onset time could be found from the numerical simulations. For one case ($K=10^{-12}$ m$^2$ and wavelength 0.12 m), the standard deviation increases throughout the simulation, which means that no local minimum could be detected. For another case ($K=10^{-13}$ m$^2$ and wavelength 0.15 m), salt precipitation occurred before onset of instabilities.

For the cases where numeric onset times could be determined, the onset times generally agree well with the ones predicted by the linear stability analysis. For the larger permeabilities, the deviations between analytical and numeric onset time are 1-40 $\%$, while for the lower permeabilites, the deviations are 0.5-10 $\%$. For the $K=10^{-10}$ m$^2$ simulations we see that the onset of instabilities occurs shortly after the perturbation is applied, and it could be that the size of the perturbation was too large, since only a small density increase is needed for fingers to develop (cf.~Figure \ref{fig:massfractioncompwavelength}), and hence triggered the development of the instabilities at an earlier time than expected. In general we see that lower permeabilities correspond to later onset times, as also observed by Riaz et al.~(2006) \cite{Riaz2006}.

In the numerical simulations, the appearing wavelength at onset is the one used in the initial perturbation. For the permeability $K=10^{-12}$ m$^2$ and using a wavelength of 0.06 m, a slightly longer average wavelength appears shortly after onset as two waves merge. The merging of waves is a common development after onset of instabilities, although it usually appears some time after the instabilities are developed.

\begin{table}
\caption{Onset times from the linear stability analysis and the numerical simulations for specific wavelengths.}
\label{tab:onsetTimesNumeric}   
\begin{tabular}{llll}
\hline\noalign{\smallskip}
permeability &  fixed wavelength & analytic onset time& numeric onset time \\[1mm]
\noalign{\smallskip}\hline\noalign{\smallskip}
  $K=10^{-10}$ m$^2$ & $0.01$ m & $1.78 \cdot 10^{4}$ s& $9.60 \cdot 10^{3}$ s\\
  $K=10^{-10}$ m$^2$ & $0.015$ m & $3.02 \cdot 10^{3}$ s& $1.80 \cdot 10^{3}$ s\\
  $K=10^{-11}$ m$^2$ & $0.03$ m & $9.72 \cdot 10^{4}$ s& $9.84 \cdot 10^{4}$ s\\
  $K=10^{-11}$ m$^2$ & $0.04$ m & $5.14 \cdot 10^{4}$ s& $3.18 \cdot 10^{4}$ s\\
  $K=10^{-12}$ m$^2$ & $0.06$ m & $1.59 \cdot 10^{6}$ s& $1.75 \cdot 10^{6}$ s\\
  $K=10^{-12}$ m$^2$ & $0.12$ m & $4.33 \cdot 10^{5}$ s& $0.0$ s\\
  $K=10^{-13}$ m$^2$ & $0.15$ m & $8.16 \cdot 10^{6}$ s& - \\
  $K=10^{-13}$ m$^2$ & $0.3$ m & $3.63 \cdot 10^{6}$ s& $3.65 \cdot 10^{6}$ s\\
  \noalign{\smallskip}\hline
  \end{tabular}
\end{table}

\subsection{Behavior of top salt concentration before and near onset of instabilities}
Using the explicit solution \eqref{eq:ustable} for the ground state salt concentration, we can address the expected development of the salt concentration over time before onset of instabilities. For convenience the comparison is shown using salt mole fractions, but the numbers could also be converted to salt mass fractions.
The largest salt concentration is always found at the top of the domain, hence we focus on this one in the following. Salt precipitates if exceeding $\mathsf{x}^\mathrm{NaCl}_\mathrm{max}=0.11$ (corresponding to $X_{\text{max}}=0.26$). That means, if the ground state salt concentration \eqref{eq:ustable} at the top exceeds this $\mathsf{x}^\mathrm{NaCl}_\mathrm{max}$ before onset of instabilities is expected, then salt will instead precipitate. In this case, instabilities will not develop, as the salt mole fraction cannot extend beyond $\mathsf{x}^\mathrm{NaCl}_\mathrm{max}$, hence the corresponding density difference is not large enough to trigger instabilities. This occurred for the numerical simulation of  $K=10^{-13}$ m$^2$ and using a wavelength of 0.015 m, as seen in Figure \ref{fig:WaveK13_15}.
Note however, as seen from the numerical simulations, the salt concentration at the top of the domain can still increase after the onset of instabilities, before the instabilities are too strong. Hence, one could have salt precipitating after instabilities develop. The linear stability analysis can however only determine whether salt precipitation would occur before onset of instabilities.

We compare the salt mole fraction found from the explicit solution \eqref{eq:ustable} with the one from the numerical simulations. 
For the numerical simulations we take the average over all the cells in the top row. Since \Dumux\  uses a cell-centered scheme, this means that we are not looking at the salt mass fraction at the top, but $\frac{1}{2}\Delta z_{\text{top}}$ away from the top. Hence, the explicit solution is therefore also evaluated at the height corresponding to the centre of the top grid cells. 

\begin{figure}
    \begin{subfigure}{0.49\textwidth}
        \includegraphics[width=\textwidth]{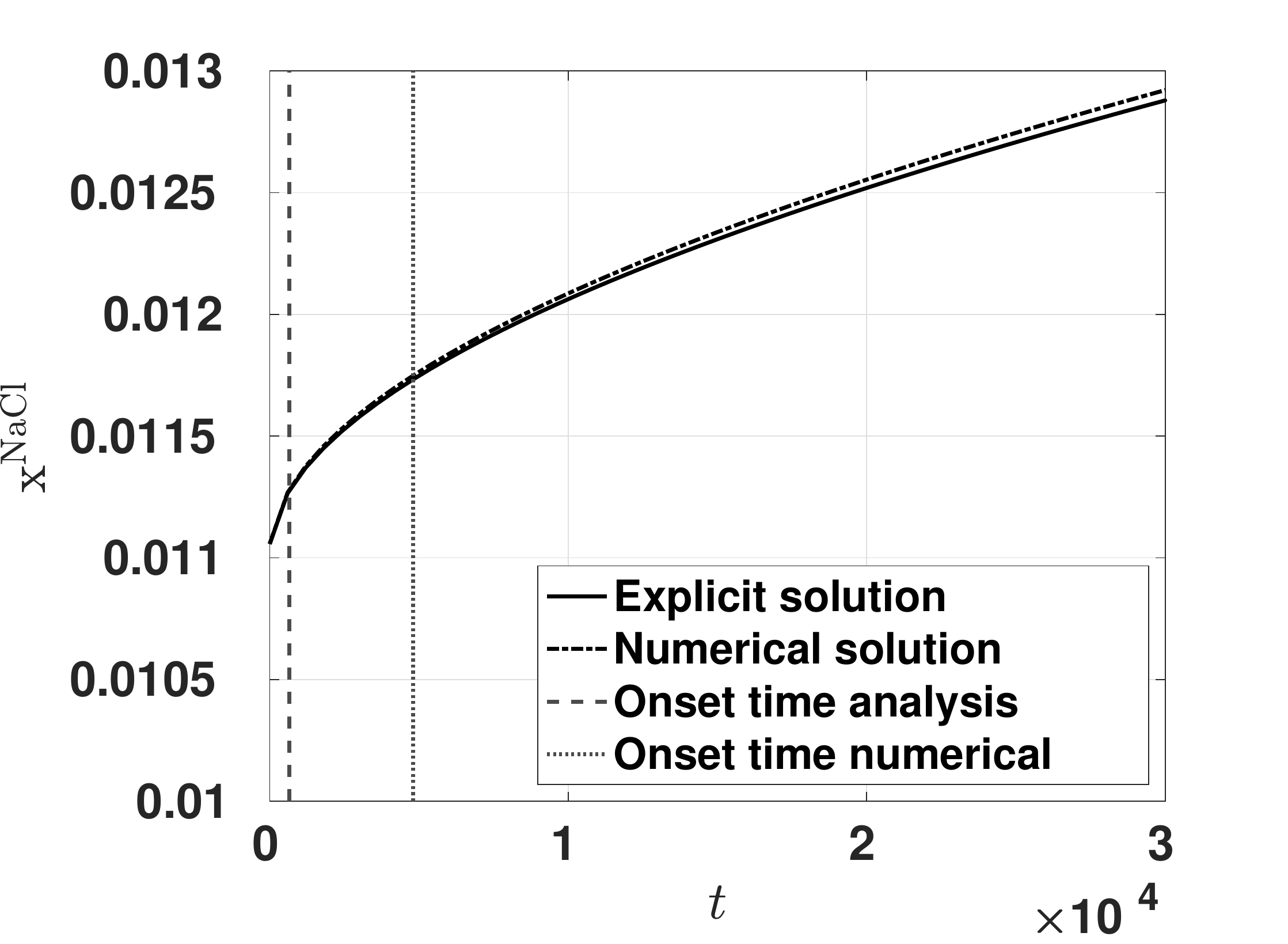}
        \subcaption{$K=10^{-10}$ m$^2$}
    \end{subfigure}
        \begin{subfigure}{0.49\textwidth}
        \includegraphics[width=\textwidth]{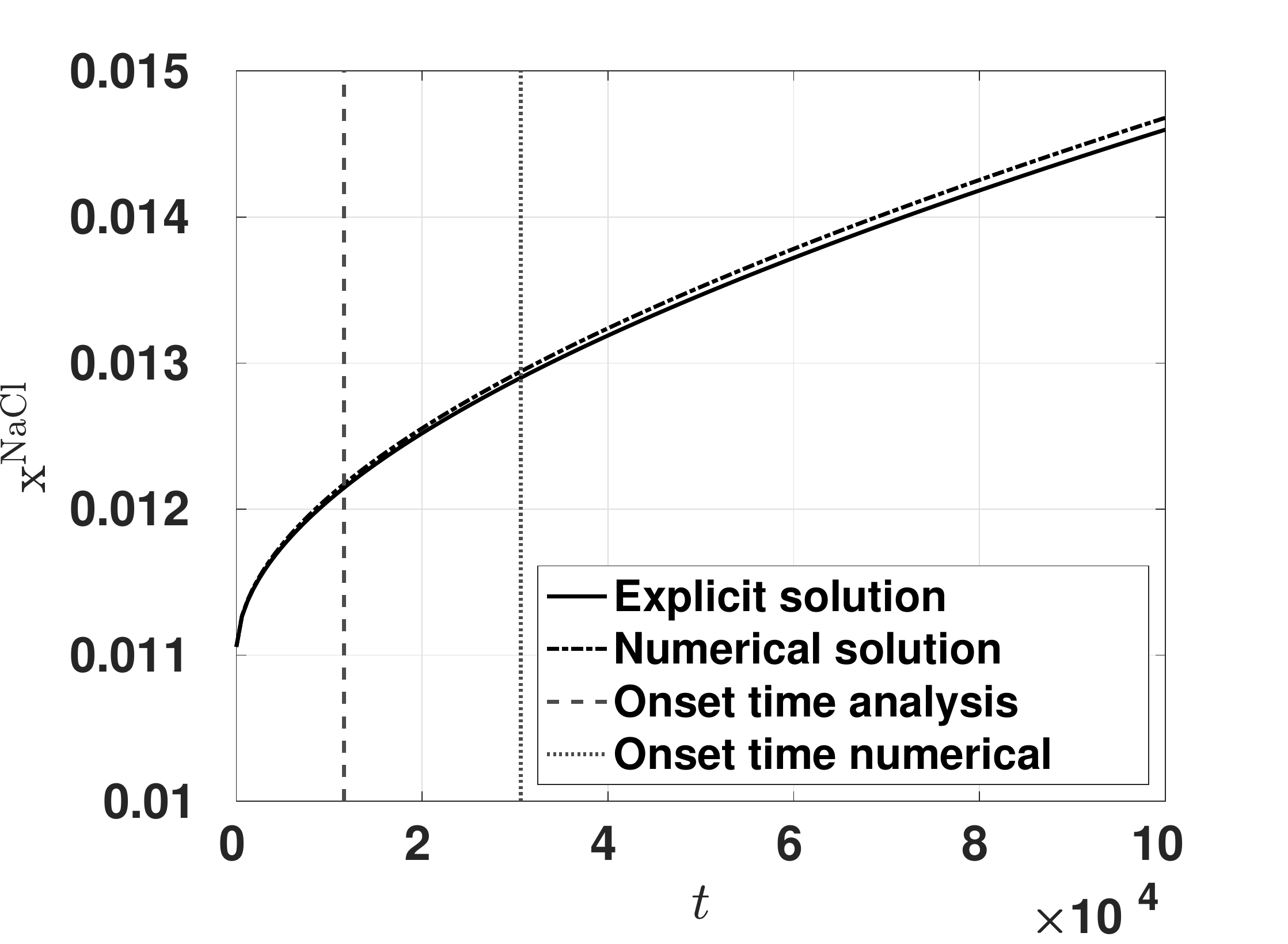}
        \subcaption{$K=10^{-11}$ m$^2$}
    \end{subfigure}
        \begin{subfigure}{0.49\textwidth}
        \includegraphics[width=\textwidth]{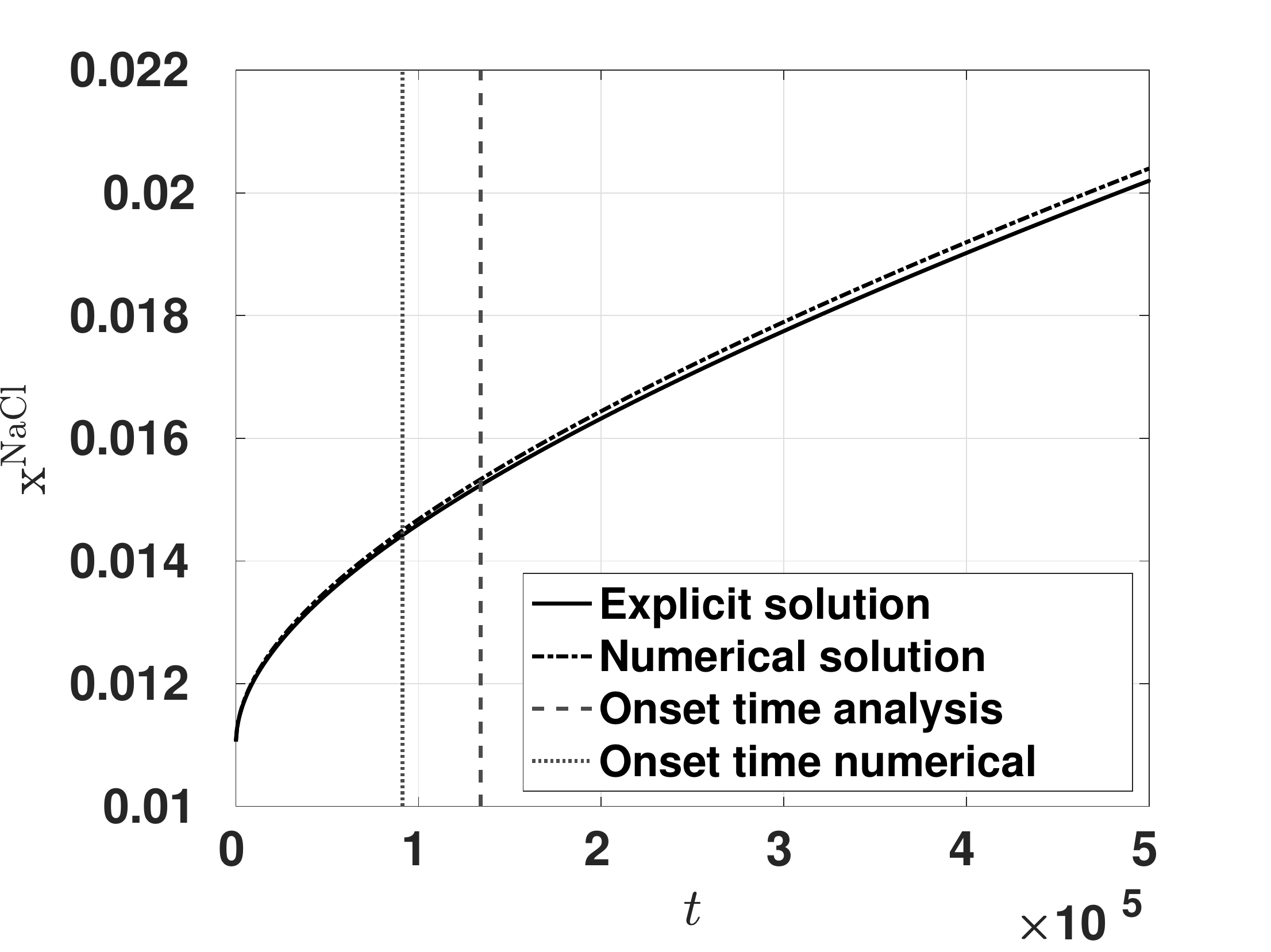}
        \subcaption{$K=10^{-12}$ m$^2$}
    \end{subfigure}
        \begin{subfigure}{0.49\textwidth}
        \includegraphics[width=\textwidth]{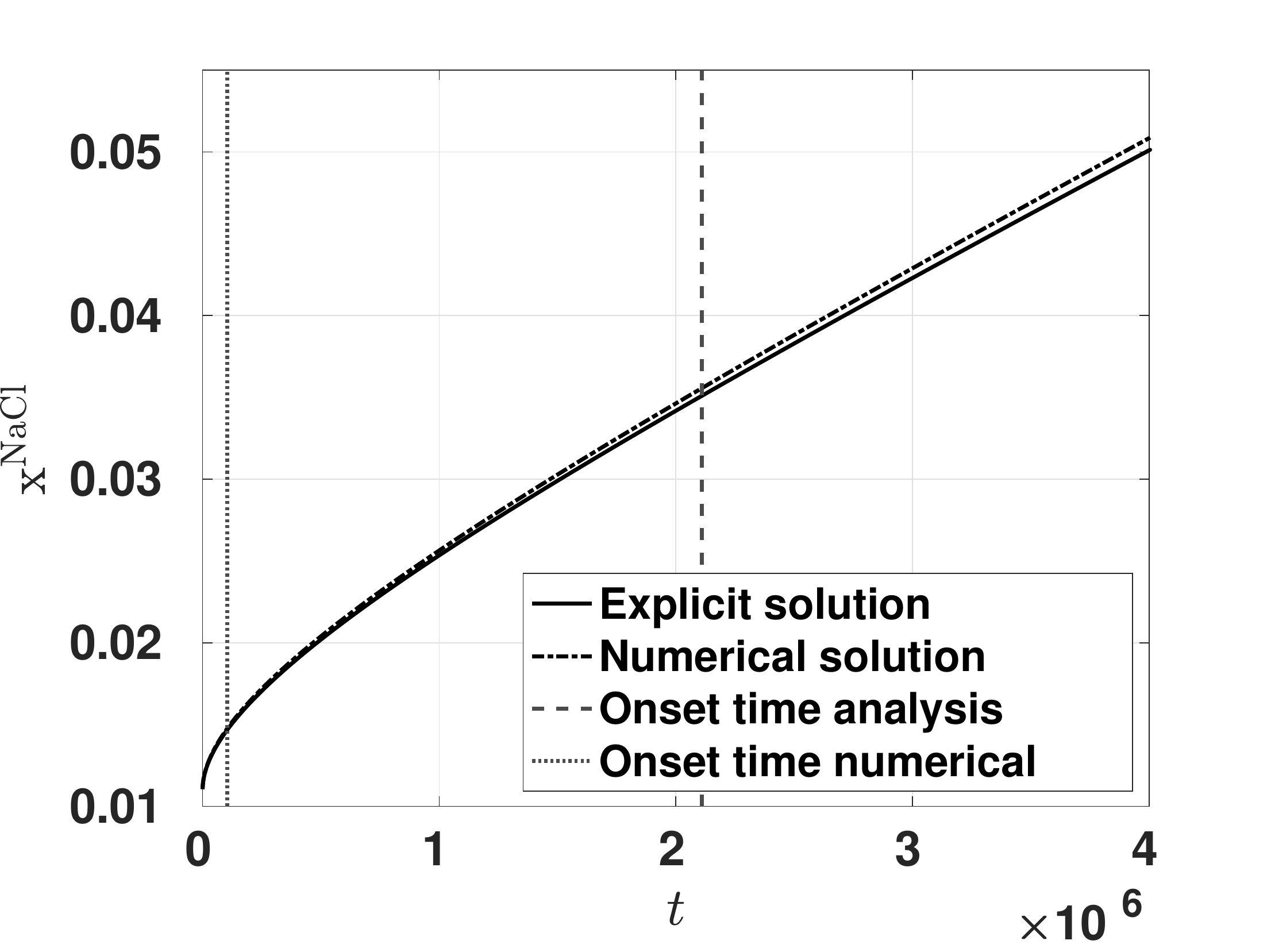}
        \subcaption{$K=10^{-13}$ m$^2$}
    \end{subfigure}
\caption{Salt mole fraction $\mathsf{x}^\mathrm{NaCl}$ at top of the domain as a function of time (in seconds), for $K=10^{-10}$ m$^2$ (top left), $K=10^{-11}$ m$^2$ (top right), $K=10^{-12}$ m$^2$ (bottom left), $K=10^{-13}$ m$^2$ (bottom right), when a specific width is used. The estimated onset times are marked as vertical lines.}
\label{fig:massfractioncompwidth}      
\end{figure}

\begin{figure}
    \begin{subfigure}{0.49\textwidth}
        \includegraphics[width=\textwidth]{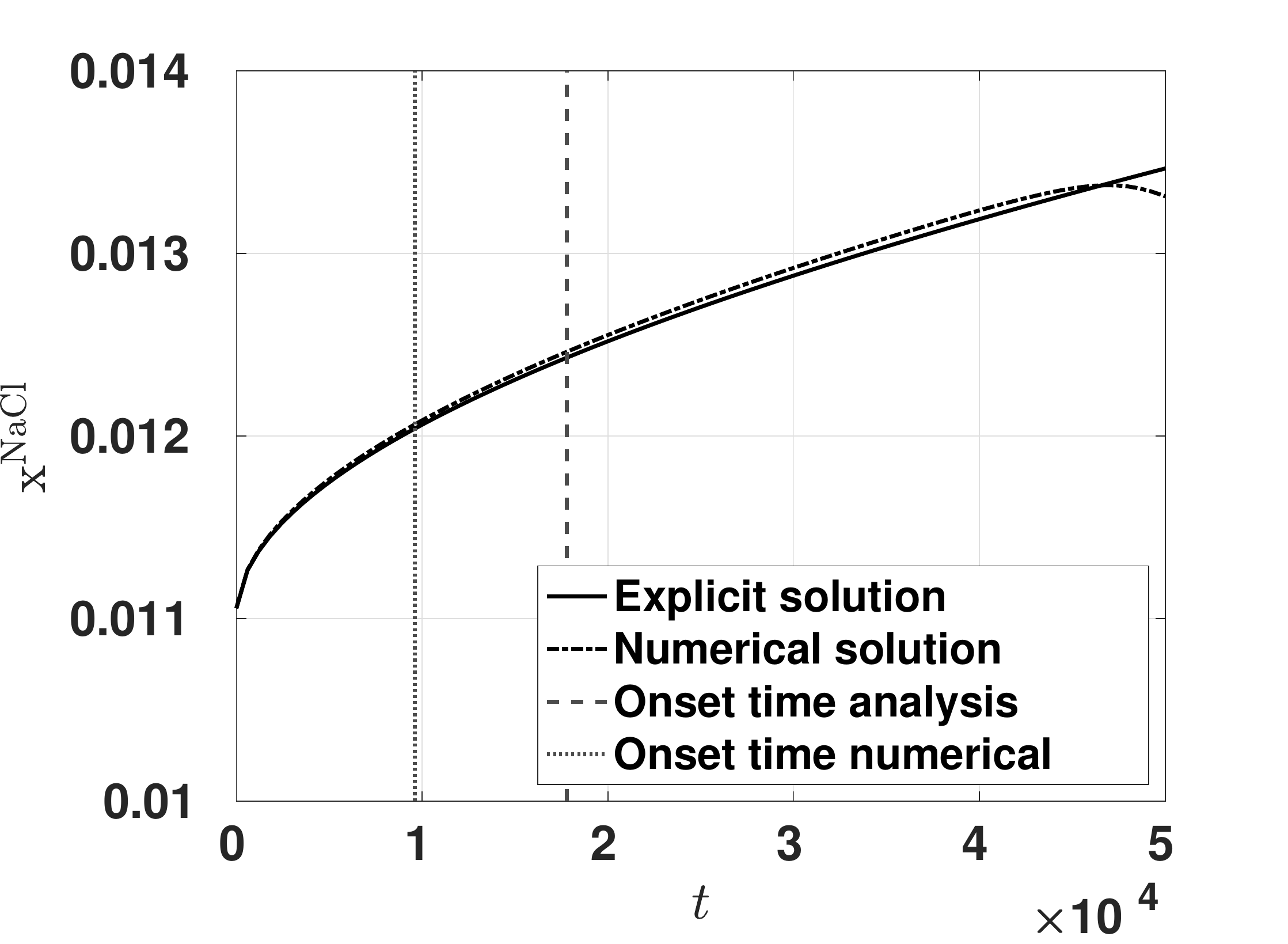}
        \subcaption{$K=10^{-10}$ m$^2$}
    \end{subfigure}
        \begin{subfigure}{0.49\textwidth}
        \includegraphics[width=\textwidth]{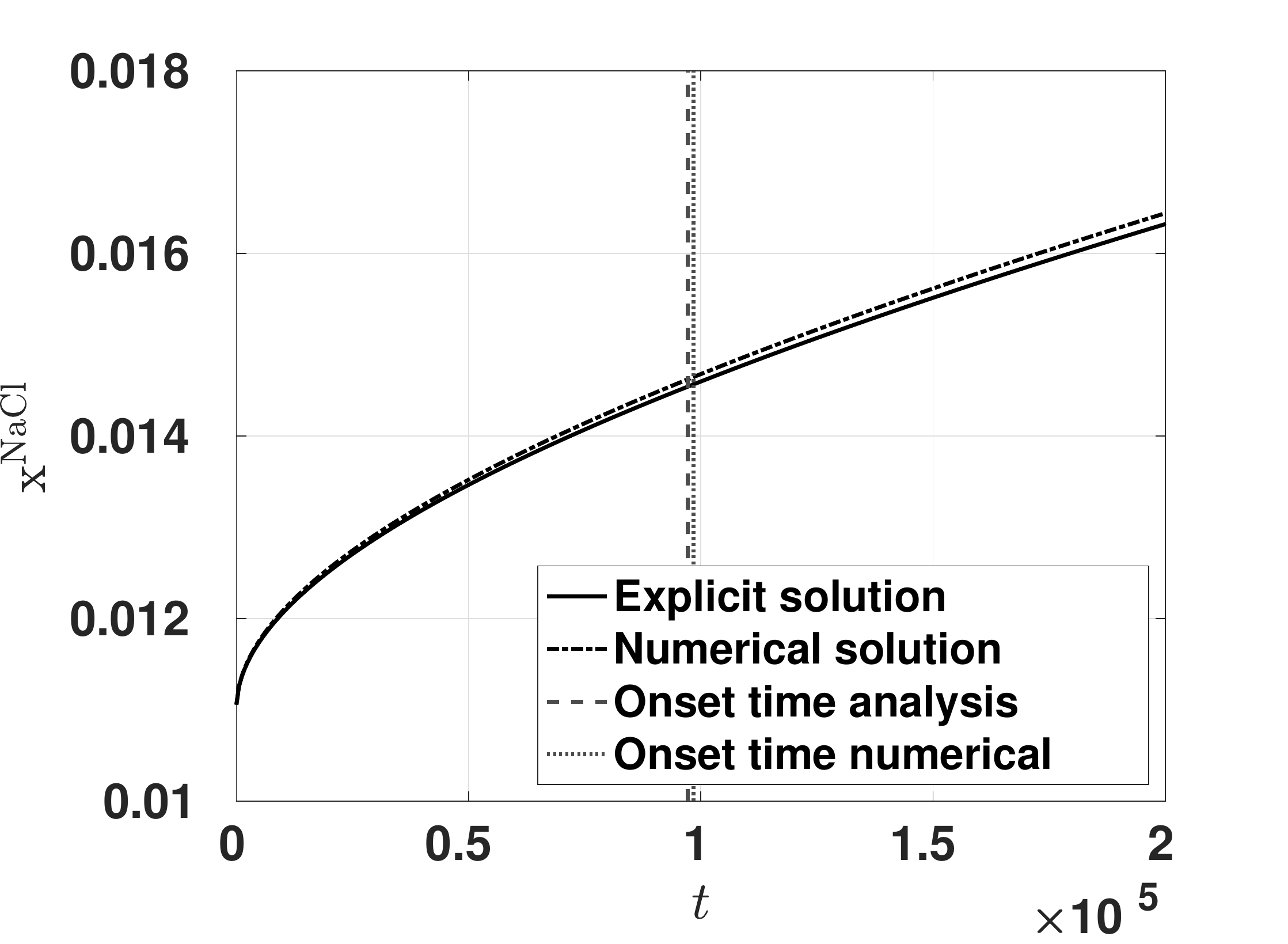}
        \subcaption{$K=10^{-11}$ m$^2$}
    \end{subfigure}
        \begin{subfigure}{0.49\textwidth}
        \includegraphics[width=\textwidth]{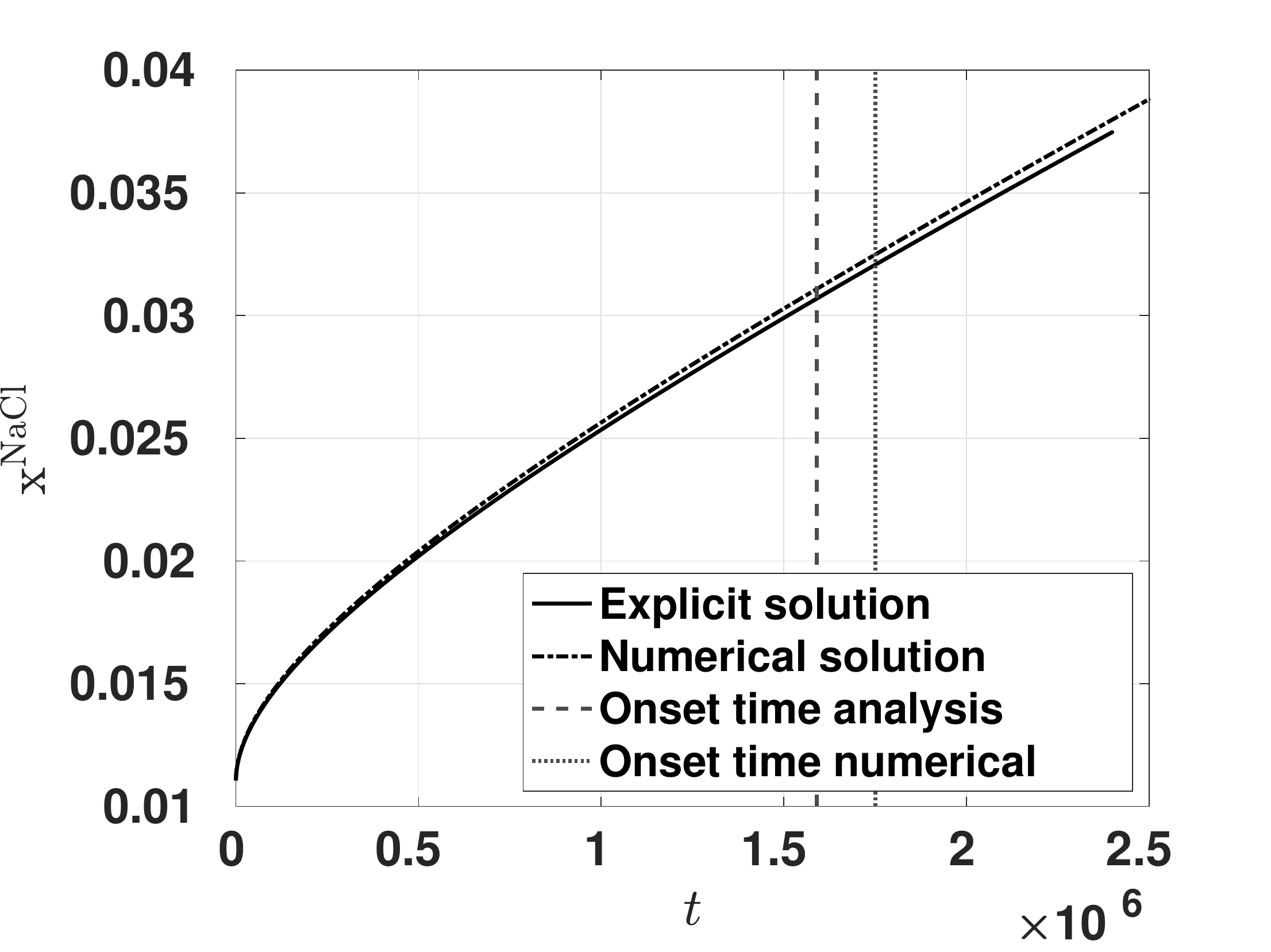}
        \subcaption{$K=10^{-12}$ m$^2$}
    \end{subfigure}
        \begin{subfigure}{0.49\textwidth}
        \includegraphics[width=\textwidth]{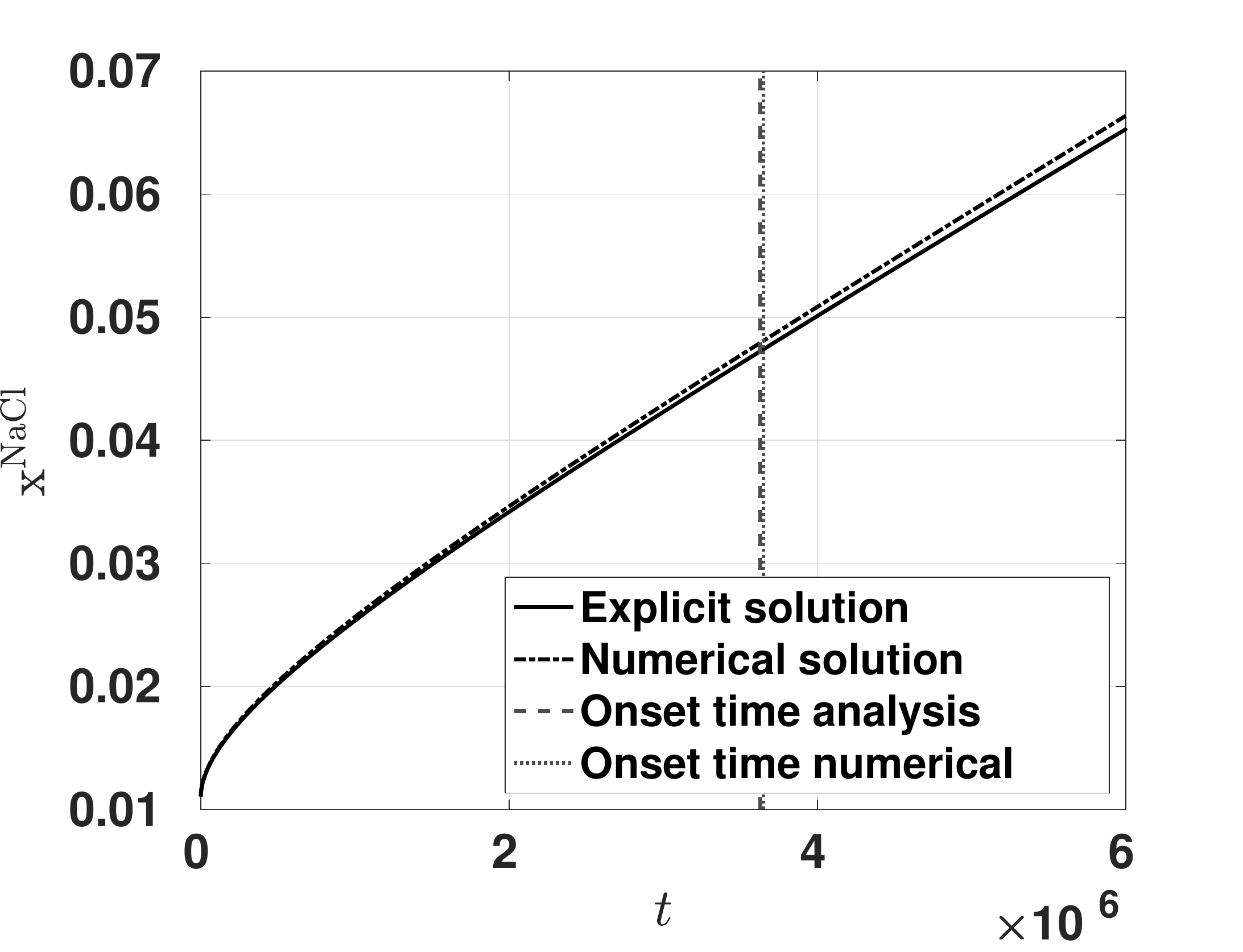}
        \subcaption{$K=10^{-13}$ m$^2$}
    \end{subfigure}
\caption{Salt mole fraction $\mathsf{x}^\mathrm{NaCl}$ at top of the domain as a function of time (in seconds), for $K=10^{-10}$ m$^2$ (top left), $K=10^{-11}$ m$^2$ (top right), $K=10^{-12}$ m$^2$ (bottom left), $K=10^{-13}$ m$^2$ (bottom right), when a specific wavelength is analyzed. The estimated onset times are marked as vertical lines. For $K=10^{-11}$ m$^2$ and $K=10^{-13}$ m$^2$ the vertical lines for onset are almost on top of each other.}
\label{fig:massfractioncompwavelength}       
\end{figure}

The time evolution of the salt mole fractions are found in Figure \ref{fig:massfractioncompwidth} for the bounded cases from Section \ref{sec:resultswidth}, and in Figure \ref{fig:massfractioncompwavelength} for the unbounded cases from Section \ref{sec:resultswavelength}. Since the development of the salt mole fraction is not varying with the applied perturbation, we only show the cases corresponding to wavelengths 0.01 m, 0.03 m, 0.06 m and 0.3 m for the permeabilities $K=10^{-10}$ m$^2$, $K=10^{-11}$ m$^2$, $K=10^{-12}$ m$^2$ and $K=10^{-13}$ m$^2$, respectively. The salt mole fractions coming from numerical simulations and explicit solutions are expected to coincide until onset of instabilities. The explicit solutions are plotted beyond the corresponding onset time for comparison, but develop as if instabilities do not occur. Hence, the numerical and explicit solutions should deviate after onset of instabilities. We see however, that especially for low permeabilities, the explicit and numerical salt mole fractions deviate also slightly before onset of instabilities. For the low permeabilities, the salt mole fraction increases much more before onset of instabilities, compared to the high-permeability cases. This also causes a larger change in the density at the top of the domain, which violates the Bousinessq approximation used to derive the explicit solution. We also see clearly that the simulated salt concentration at the top follows to a large extent the analytic solution also after onset of instabilities. This is due to the instabilities being so weak in the beginning, hence their capability to transport salt downwards is not developed.

\subsection{Development of the salt concentration after onset of instabilities}\label{sec:instabilityDevelopment}
In this section the development of instabilities before and after their onset is discussed, using the results of the numerical simulations. Here we also show the case of salt precipitation which occurs for some parameter sets. Different phases of the development are defined with help of the mean value $\mu_\mathrm{\mathsf{x}^\mathrm{NaCl}}^\mathrm{top}$ and the standard deviation $\sigma_\mathrm{\mathsf{x}^\mathrm{NaCl}}^\mathrm{top}$ of the salt mole fraction $\mathsf{x}^\mathrm{NaCl}$ of the grid cells in the top row. These phases can be explained and distinguished by the different dominant physical processes.

\begin{figure}
    \centering
    \begin{subfigure}{0.49\textwidth}
        \includegraphics[width=\textwidth]{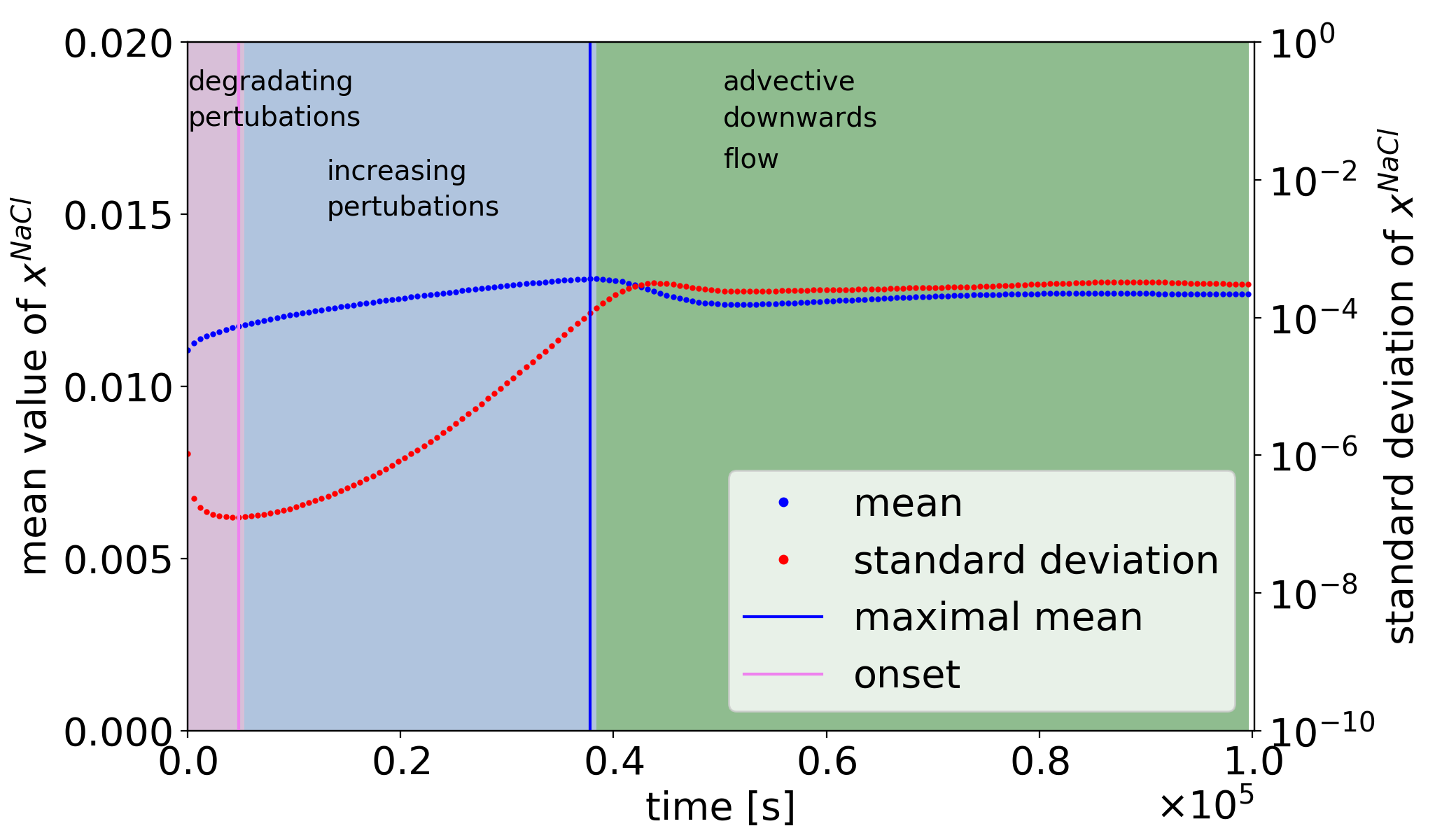}
        \subcaption{Fixed width for $K=10^{-10}$ m$^2$}
    \end{subfigure}
    \begin{subfigure}{0.49\textwidth}
        \includegraphics[width=\textwidth]{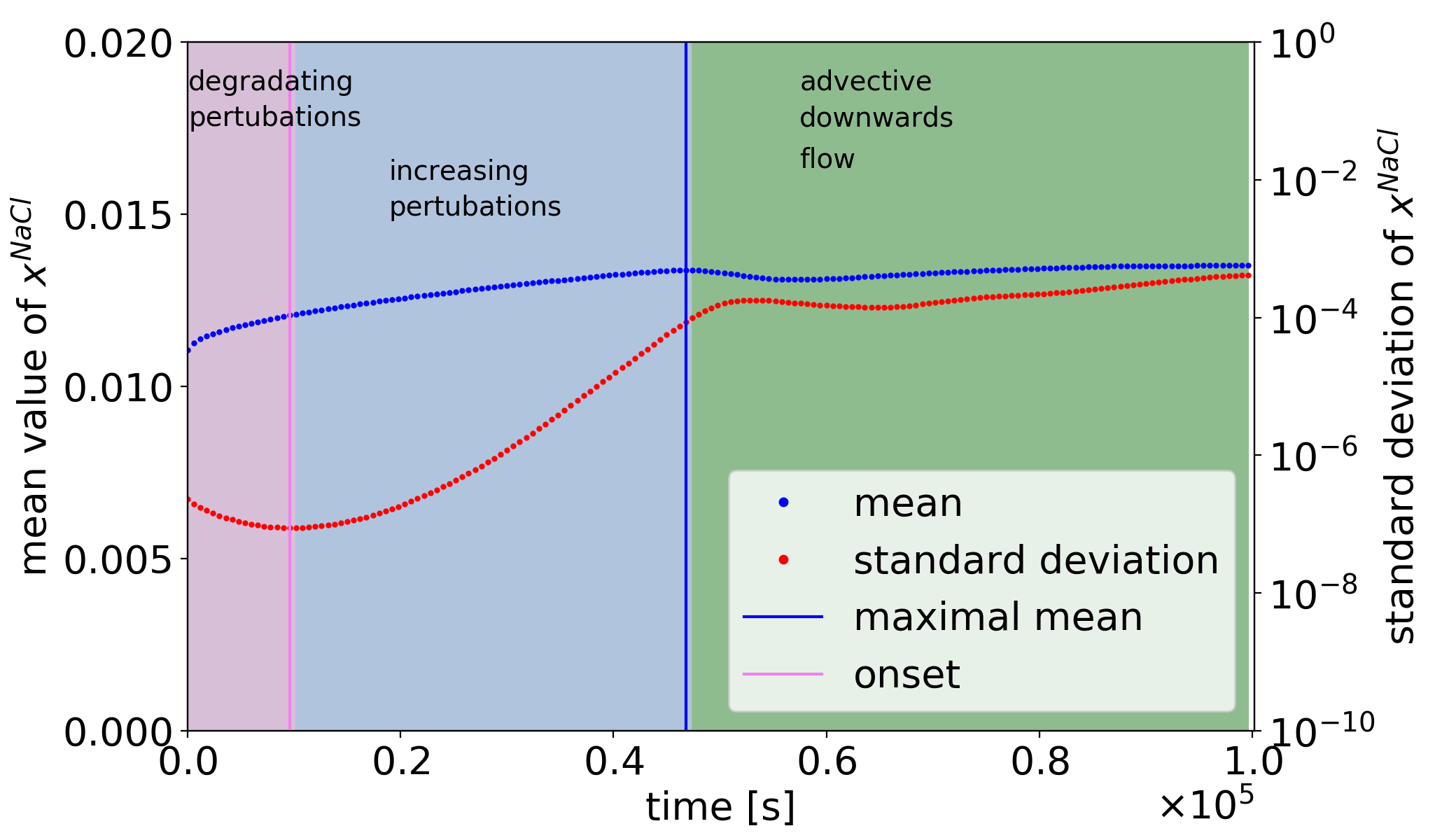}
        \subcaption{Fixed wavelength for $K=10^{-10}$ m$^2$}
    \end{subfigure}
    \begin{subfigure}{0.49\textwidth}
        \includegraphics[width=\textwidth]{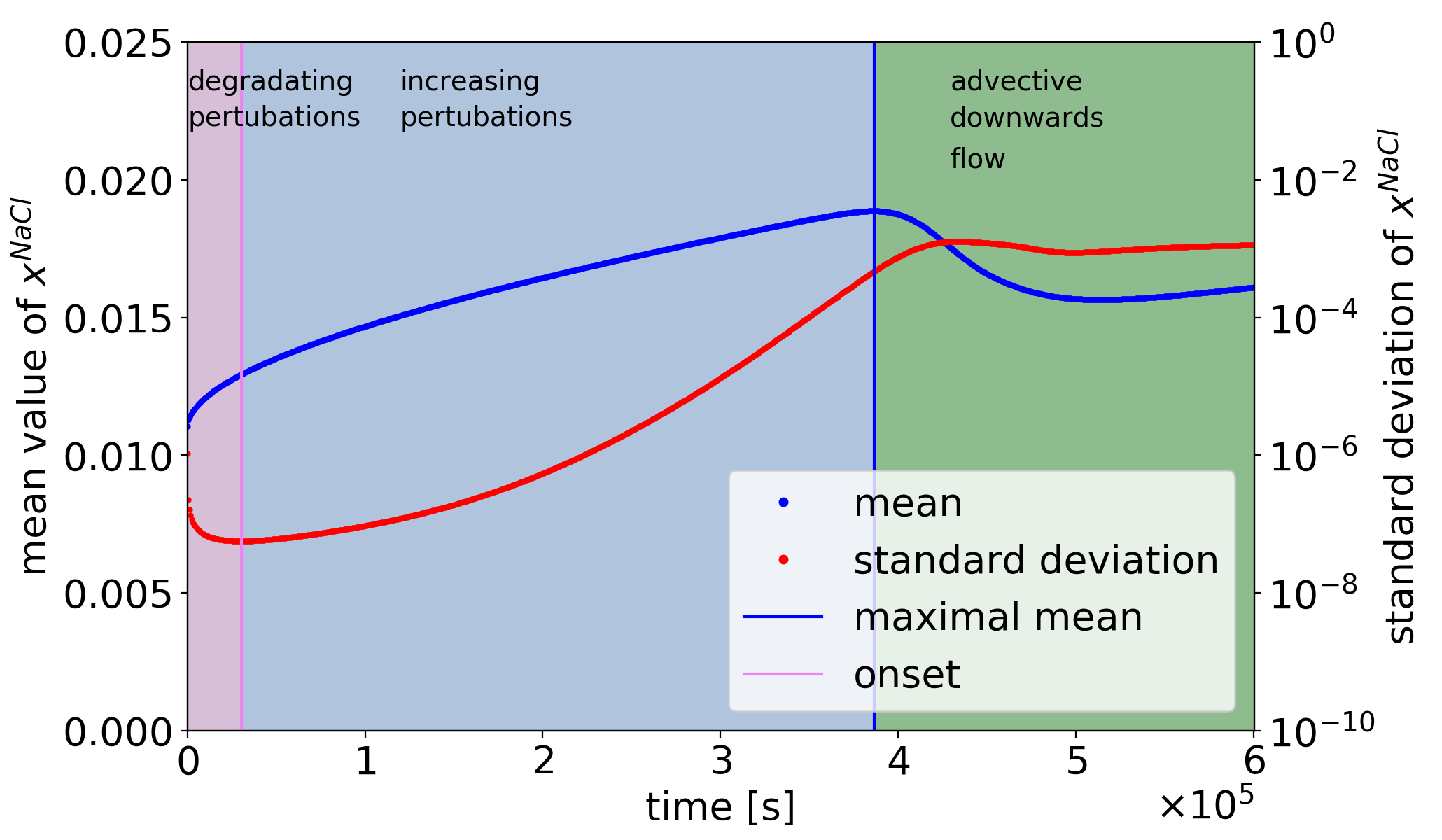}
        \subcaption{Fixed width for $K=10^{-11}$ m$^2$}
    \end{subfigure}
    \begin{subfigure}{0.49\textwidth}
        \includegraphics[width=\textwidth]{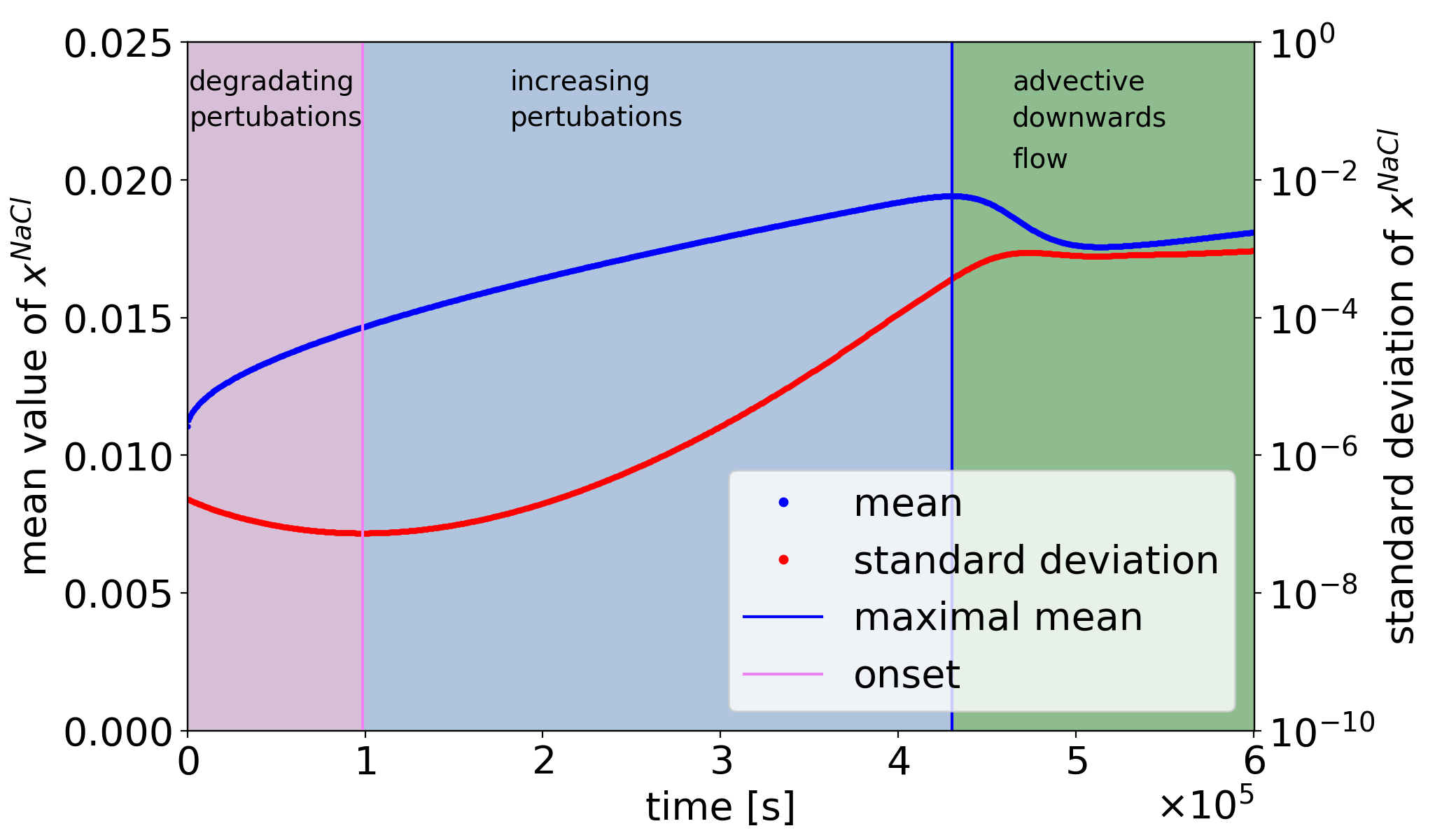}
        \subcaption{Fixed wavelength for $K=10^{-11}$ m$^2$}
    \end{subfigure}
    \begin{subfigure}{0.49\textwidth}
        \includegraphics[width=\textwidth]{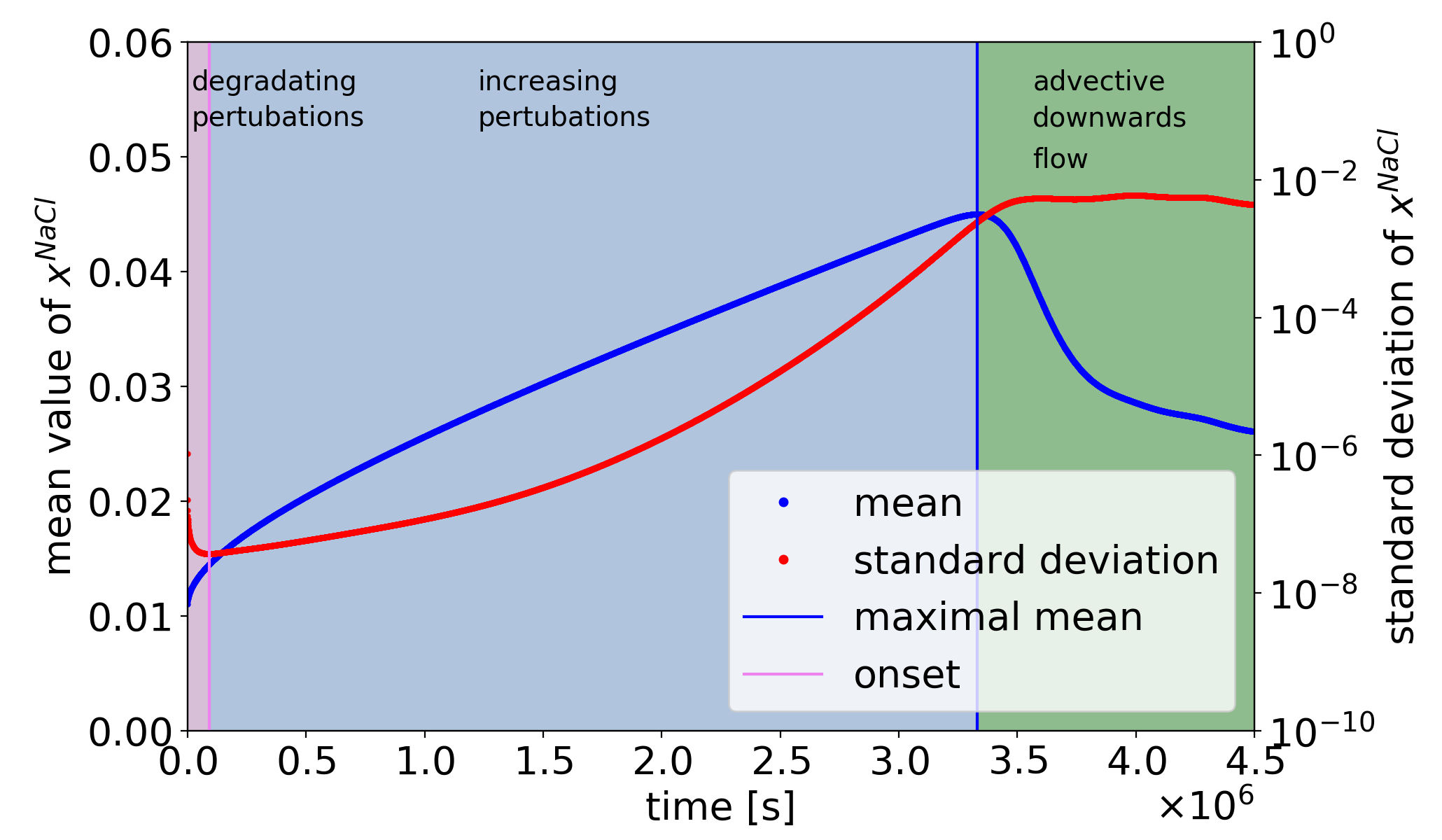}
        \subcaption{Fixed width for $K=10^{-12}$ m$^2$}
    \end{subfigure}
    \begin{subfigure}{0.49\textwidth}
        \includegraphics[width=\textwidth]{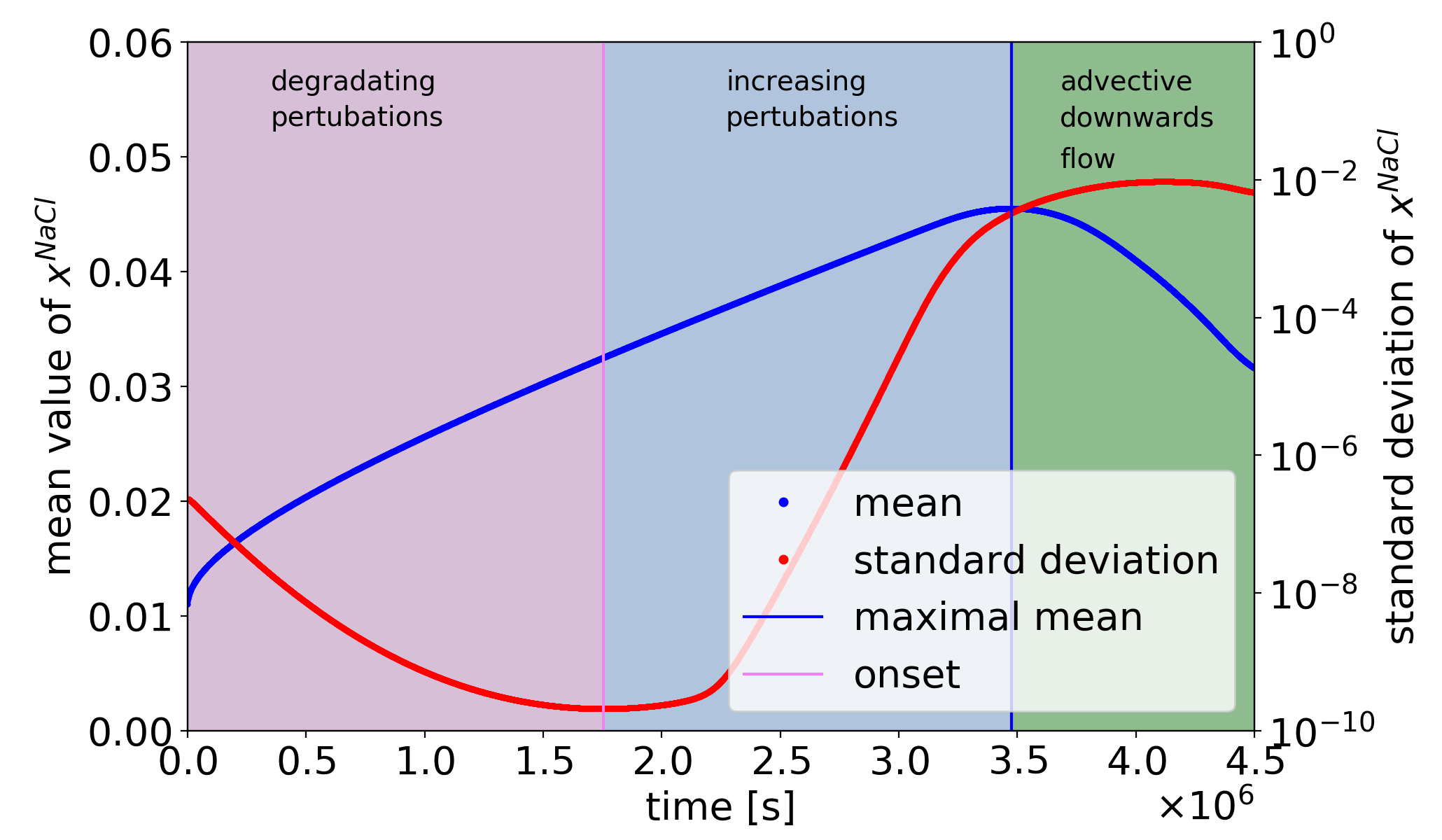}
        \subcaption{Fixed wavelength for $K=10^{-12}$ m$^2$}
    \end{subfigure}
    \begin{subfigure}{0.49\textwidth}
        \includegraphics[width=\textwidth]{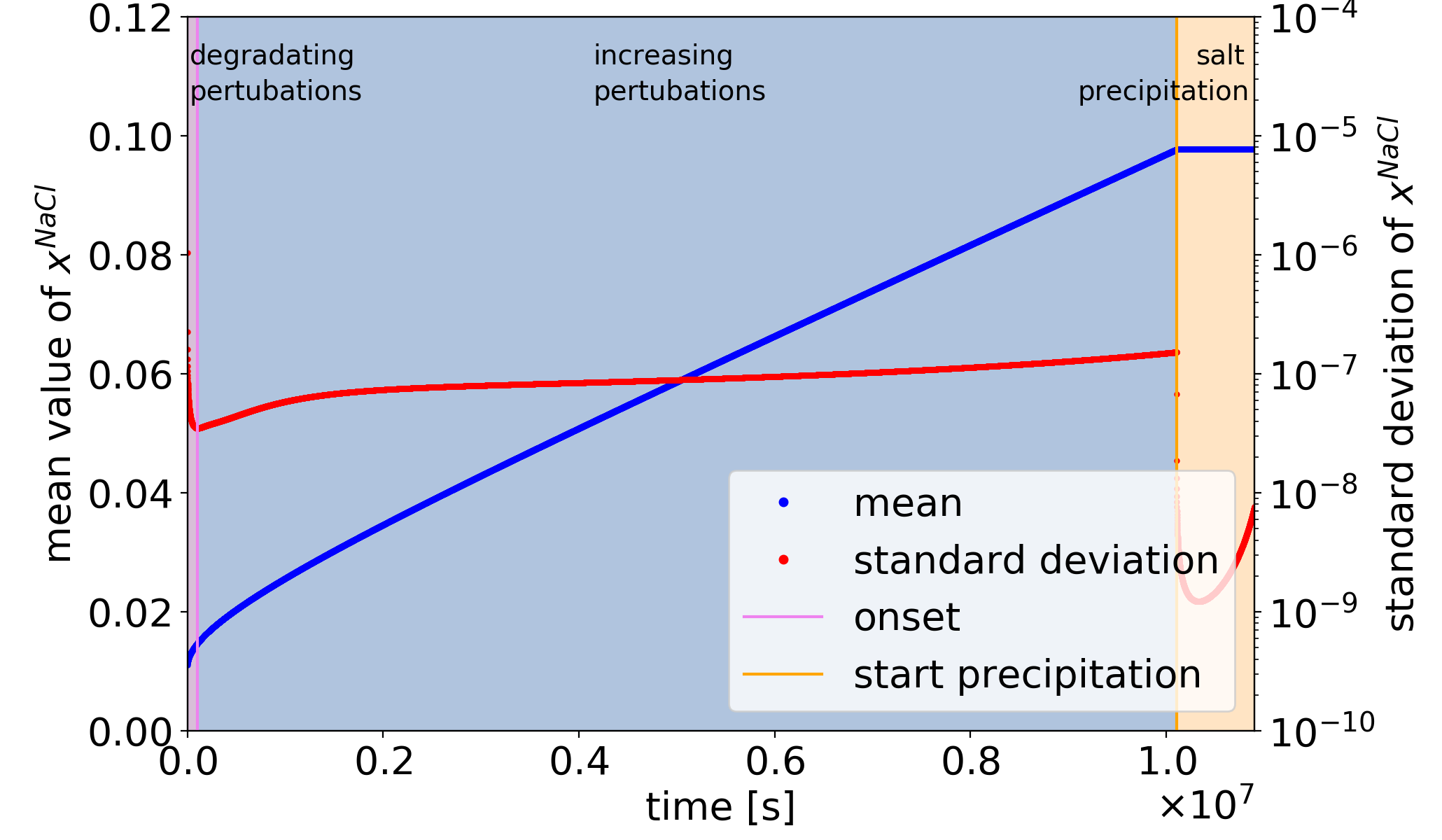}
        \subcaption{Fixed width for $K=10^{-13}$ m$^2$}
    \end{subfigure}
    \begin{subfigure}{0.49\textwidth}
        \includegraphics[width=\textwidth]{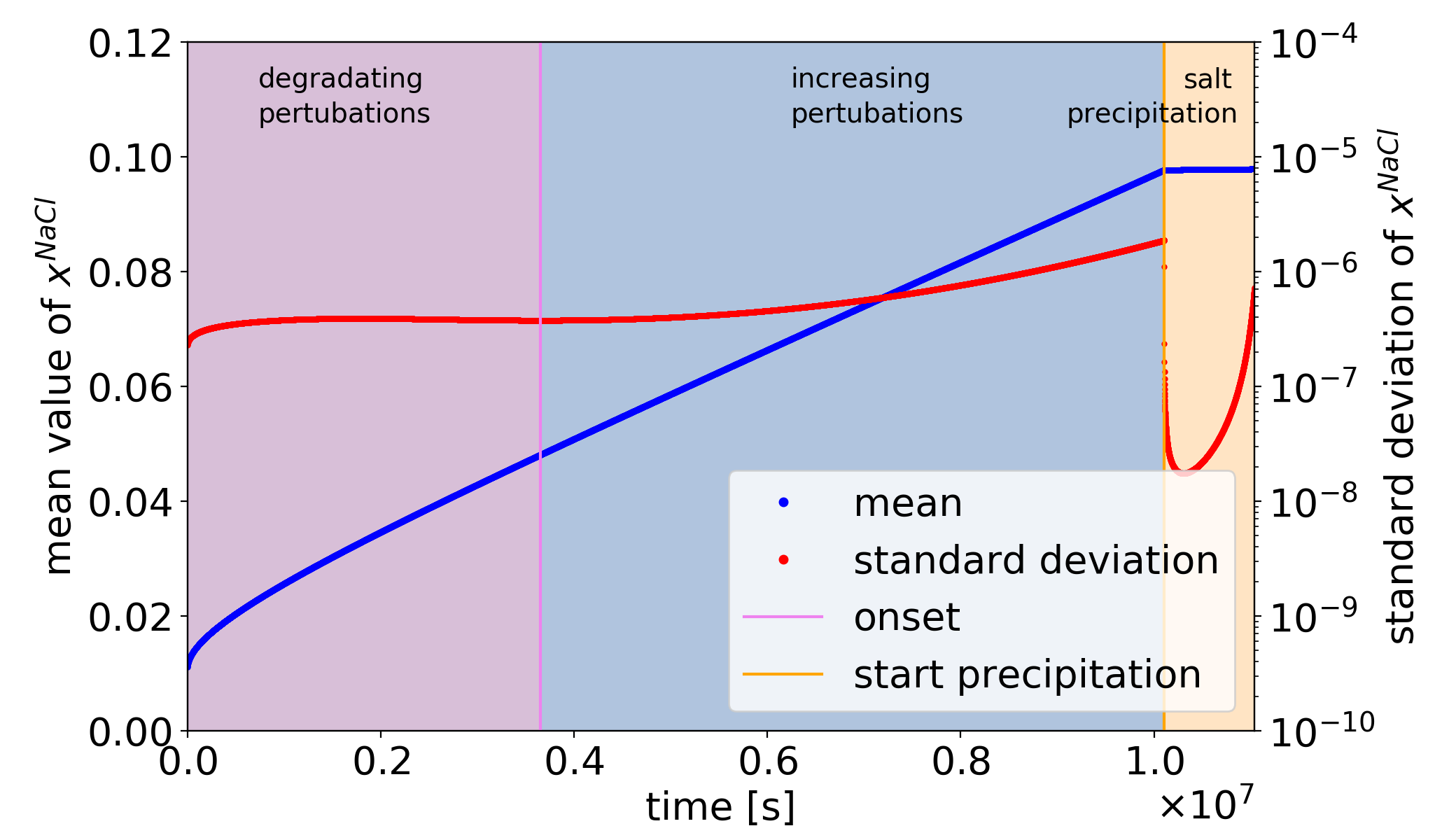}
        \subcaption{Fixed wavelength for $K=10^{-13}$ m$^2$}
    \end{subfigure}
    \caption{Result from the numerical simulations for a domain of fixed width (left) and for fixed wavelengths (right) and the different permeabilities. For each case the development of the mean value and standard deviation of the salt concentration is shown, as well as different phases of the development of instabilities.}
    \label{fig:NumericPhases}
\end{figure}

In Figure \ref{fig:NumericPhases} the evaluation of the numeric simulations is shown for the different permeabilities and initial perturbations. We here present only one wavelength (0.01 m, 0.03 m, 0.06 m or 0.3 m) per permeability for the periodic initial perturbations as they show a similar general behaviour.
The development of $\mu_\mathrm{\mathsf{x}^\mathrm{NaCl}}^\mathrm{top}$ and  $\sigma_\mathrm{\mathsf{x}^\mathrm{NaCl}}^\mathrm{top}$ over time indicates the different phases of the formation of instabilities.

\begin{figure}
\begin{subfigure}{\textwidth}
    \includegraphics[width=0.49\textwidth]{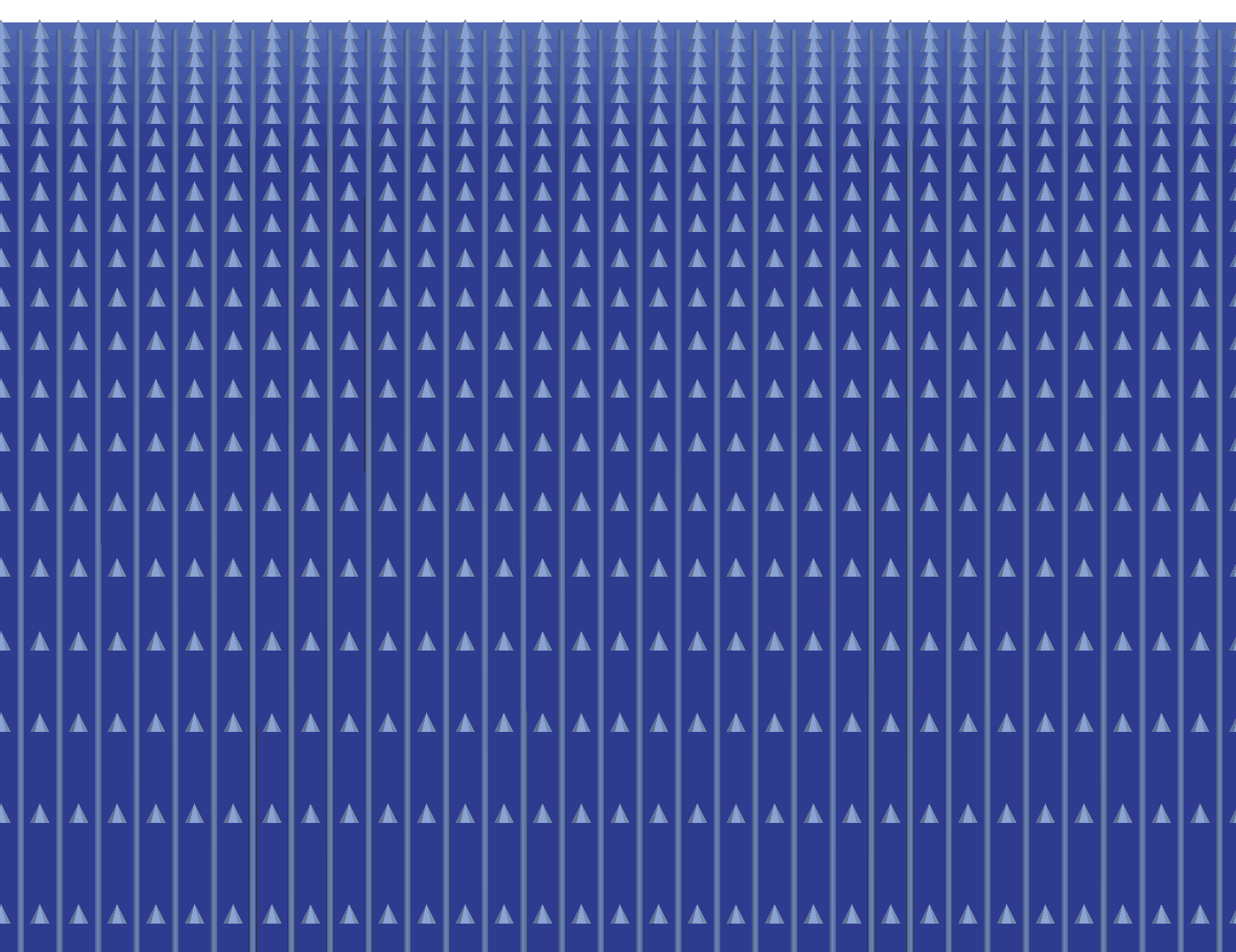}
    \includegraphics[width=0.49\textwidth]{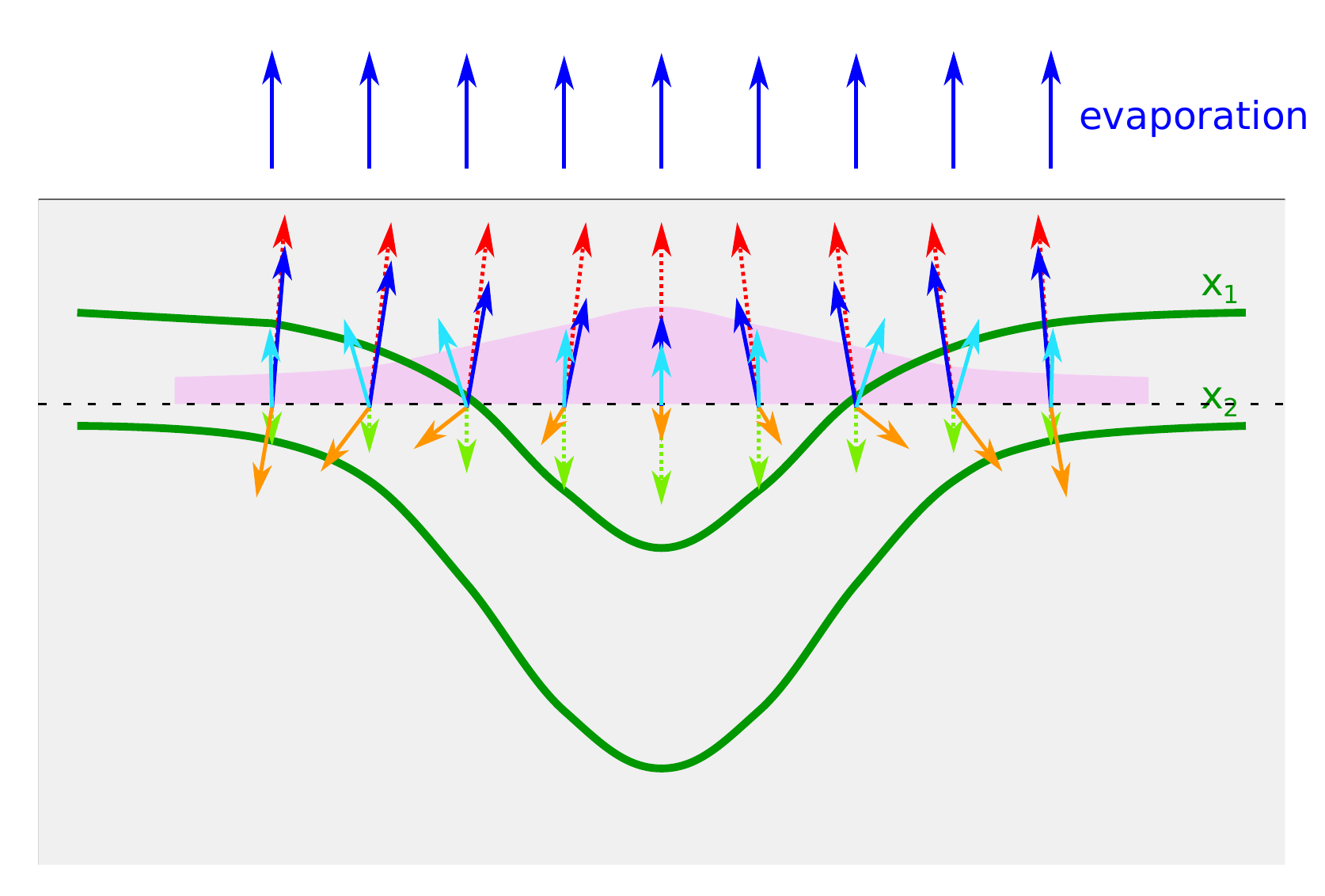}
    \subcaption{Phase of degradating pertubations}
    \label{fig:firstPhase}
\end{subfigure}
\begin{subfigure}{\textwidth}
    \includegraphics[width=0.49\textwidth]{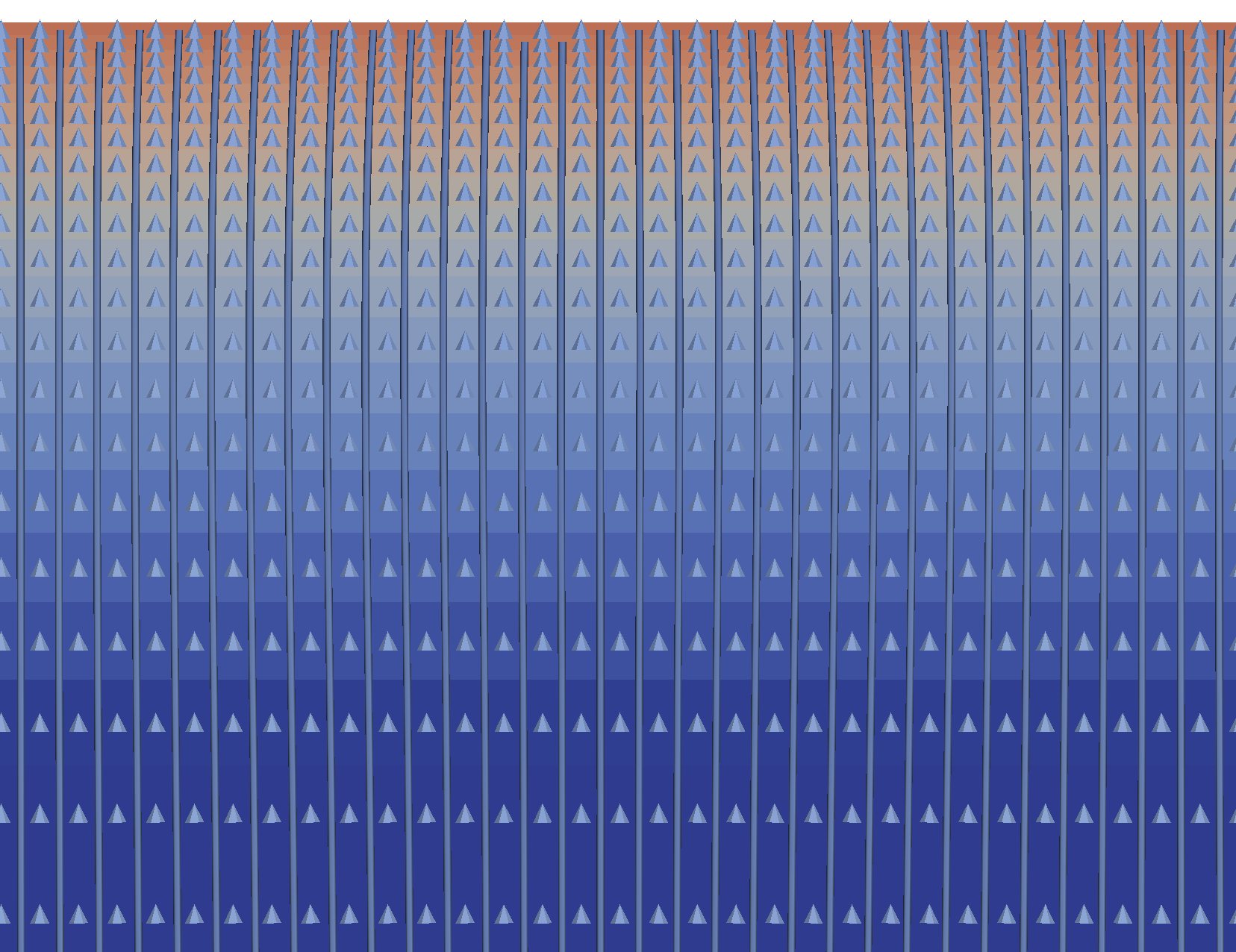}
    \includegraphics[width=0.49\textwidth]{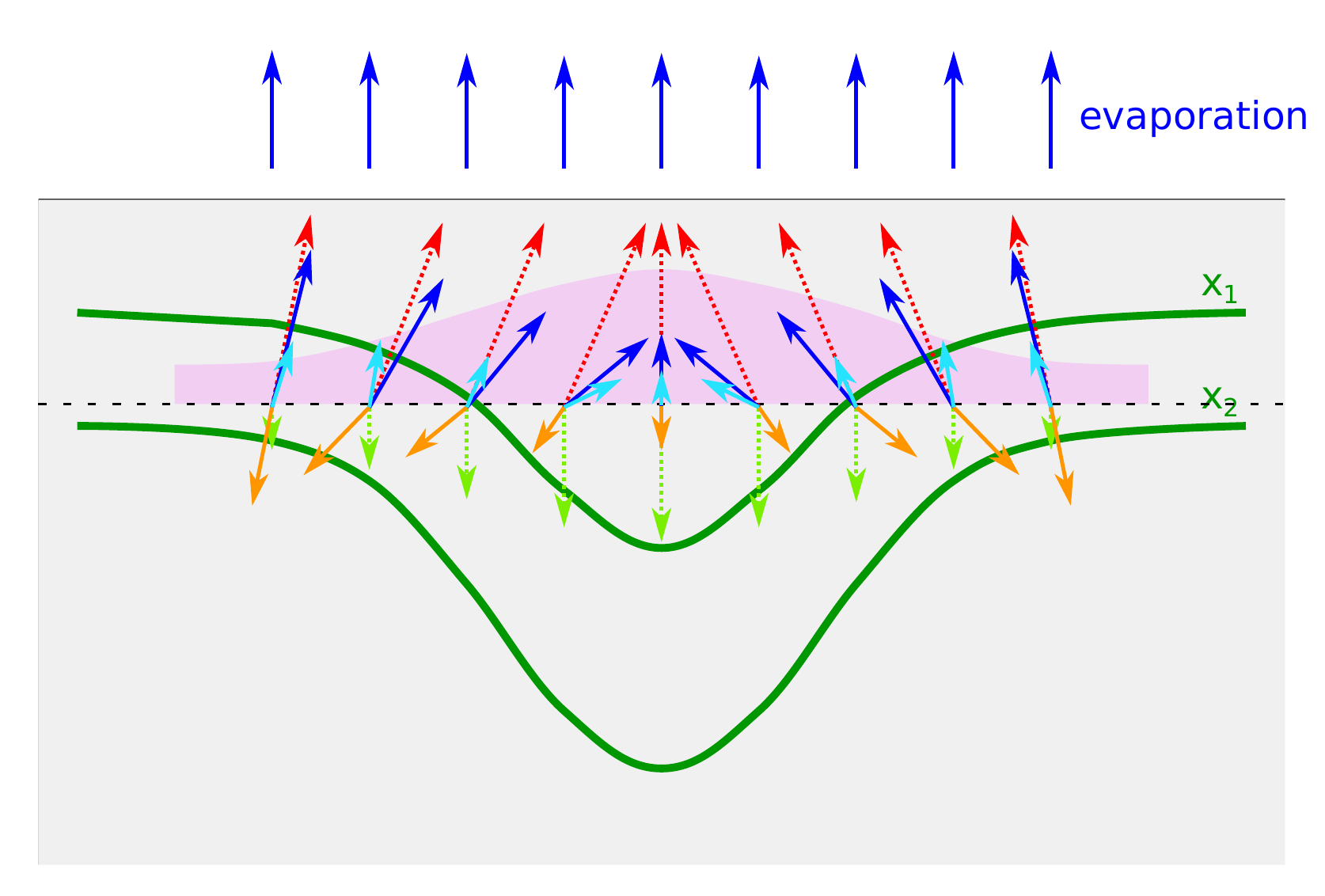}
    \subcaption{Phase of increasing pertubations}
    \label{fig:secondPhase}
\end{subfigure}
\begin{subfigure}{\textwidth}
    \includegraphics[width=0.49\textwidth]{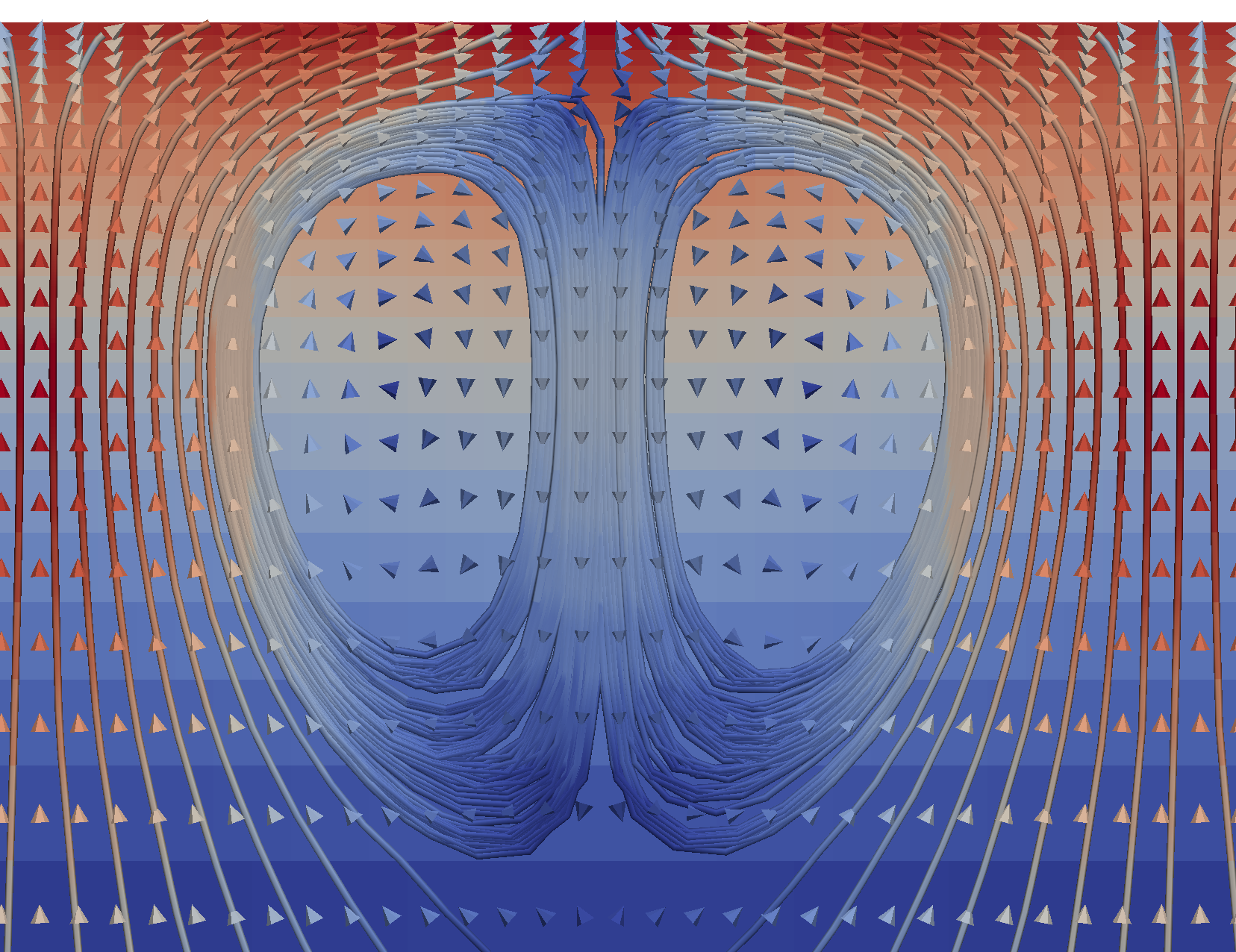}
    \includegraphics[width=0.49\textwidth]{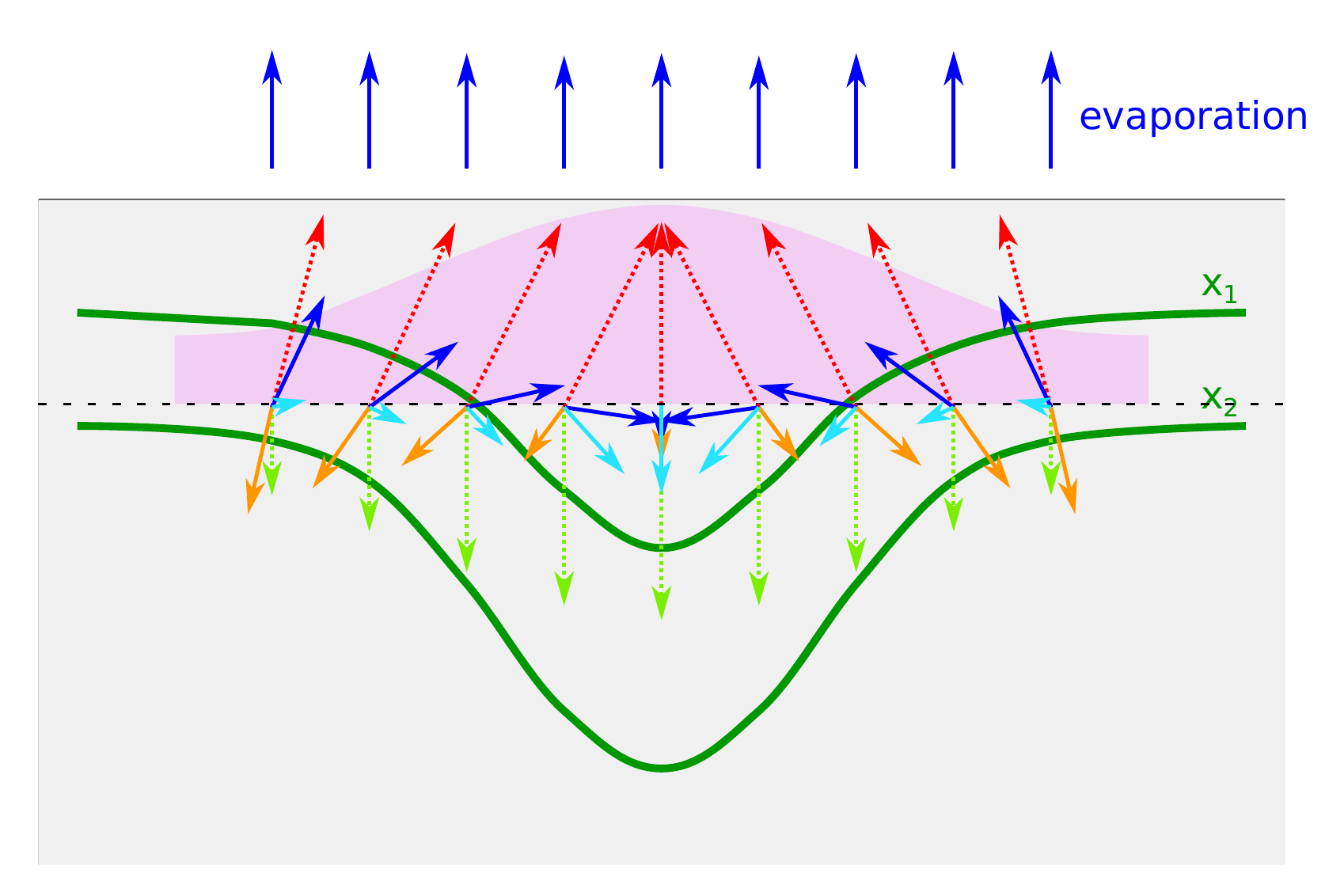}
    \subcaption{Phase of convective downwards flow}
    \label{fig:thirdPhase}
\end{subfigure}
    \includegraphics[width=0.49\textwidth]{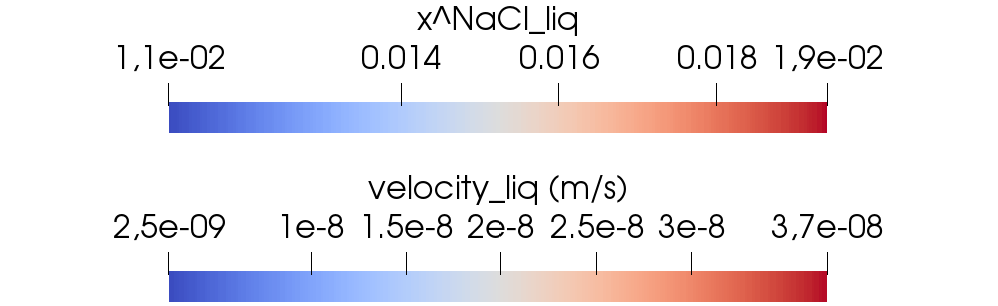}
    \includegraphics[width=0.49\textwidth]{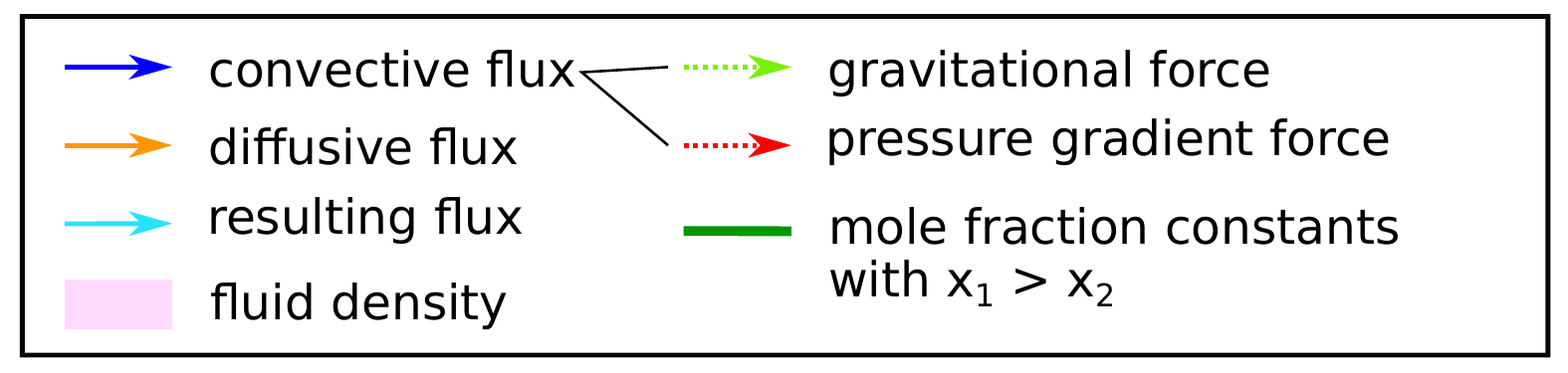}
    \caption{Important processes and forces in the different phases of instability development. Left the convective streamlines and velocity arrows as well as the salt mole fraction in the background are shown exemplary for one perturbation for $K=10^{-11}$ m$^2$ and periodic initial perturbations. Right a schematic overview of the fluxes and forces describes the formation of the resulting flux.}
    \label{fig:PhasesExplanation}
\end{figure}

In the first phase the initial standard deviation decreases due to the spreading of the initially applied perturbations by molecular diffusion. In Figure \ref{fig:firstPhase} the influence of the different fluxes on the resulting flux is shown for the first phase.
As already mentioned in Section \ref{sec:DiscussionProcesses} the higher density at the location of the perturbation leads to an increased gravitational force and a slowed down convective upwards flux. As we apply a constant evaporation rate a lower pressure develops at these locations.
This horizontal pressure gradient induces a horizontal component to the convective flux towards the perturbation. However, in the first phase the diffusive flux dominates the convective flux in the horizontal direction, which leads to a degradation of the perturbation. 
In the vertical direction the driving force of the pressure gradient outweighs the gravitational force. This results in an upwards convective flux which also dominates the diffusive flux in the vertical direction. This leads to an upwards transportation and accumulation of salt at the top during this phase.

The second phase starts at the time of onset, when the standard deviation reaches its minimum and the perturbations start to increase. The reason for that can be seen in Figure \ref{fig:secondPhase}. The increasing salt concentration and fluid density enhances the gravitational downward forces, which in the following enhances the horizontal component of the convective flux towards the perturbation. In this phase the convective flux dominates the diffusive flux in horizontal direction. This leads to an increased transport of salt towards the perturbation and consequently to its enhancement. The vertical direction of the resulting flux is still upwards, which increases the salt concentration at the top. This is also demonstrated by the continuous increase of $\mu_\mathrm{\mathsf{x}^\mathrm{NaCl}}^\mathrm{top}$ during this phase.

For the higher permeabilities $K=10^{-10} - 10^{-12}~\mathrm{m^2}$ the third phase is characterized by a resulting downwards flow and starts at the maximum value of $\mu_\mathrm{\mathsf{x}^\mathrm{NaCl}}^\mathrm{top}$. Until the start of the third phase the liquid density has increased so much at the top and especially at the location of perturbations, that the high gravitational force causes an convective and resulting flux downwards; so-called fingers. With a lower permeability a higher density difference is needed to overcome the resistance of the porous medium, resulting in higher maximum values for $\mu_\mathrm{\mathsf{x}^\mathrm{NaCl}}^\mathrm{top}$. The resulting flux transports the accumulated salt at the top downwards, which leads to a decrease of $\mu_\mathrm{\mathsf{x}^\mathrm{NaCl}}^\mathrm{top}$. Later in this phase, $\mu_\mathrm{\mathsf{x}^\mathrm{NaCl}}^\mathrm{top}$ stabilizes as the upwards transported salt equals the amount which is transported downwards with the fingers.
In these cases the solubility limit of the salt is never reached and thus no salt precipitation is observed in these systems.
For the lowest considered permeability $K = 10^{-13}~\mathrm{m^2}$, $\mu_\mathrm{\mathsf{x}^\mathrm{NaCl}}^\mathrm{top}$ reaches the solubility limit before a convective downwards flow develops. Here salt precipitates and $\mu_\mathrm{\mathsf{x}^\mathrm{NaCl}}^\mathrm{top}$ stays constant at the solubility limit as we use an equilibrium approach to simulate the precipitation reaction (see Equation \eqref{eq:reactionTerm}).

Note that in case of no precipitation, further phases for the instabilities can be defined. The fingers start to merge and form larger fingers with longer wavelengths. As we concentrate on the initial development of the instabilities we refer to Slim (2014) \cite{Slim2014} for a detailed description of these merging regimes. Slim describes similar phases, although there the instabilities are not evaporation-driven. 

\section{Final remarks}\label{sec:finalremarks}
As water evaporates from a porous medium saturated with saline water, the accumulated salt near the top boundary will either trigger density instabilities, or precipitate in form of a salt crust, or both. In this work we have addressed the onset of instabilities using two approaches: analytically by applying a linear stability analysis and numerically by performing simulations. The linear stability analysis simplifies the governing equations and formulates an eigenvalue problem giving conditions for whether and when instabilities can develop. The advantage of the linear stability analysis is that results for a large range of parameters can be obtained at very low costs. The numerical simulations can address the original governing equations, and can time-step these to address when instabilities develop. The computational costs are larger, but also information for the further development after the onset of instabilities can be obtained.

The onset times of instabilities depend not only on the physical parameters as the medium's permeability and the strength of the evaporation rate, but also on which type of instability is considered. The boundary conditions on the sidewalls and which wavelength develops, affect the development of the instabilities. We here considered two cases; either a bounded case with no-flux boundary conditions on the sidewalls, where we tried to trigger the most unstable wavelength, or an unbounded case using periodic boundary conditions, where we tried to trigger a specific wavelength. For the first case, the onset times predicted by the two approaches deviate as the applied perturbation is different and the two methods predict the onset of different instability modes. For the second case, the onset times largely coincide. In both cases, the development of the salt concentration up to onset of instabilities match up to the difference arising from using the Boussinesq approximation in the analytic case. This gives confidence that the two methods can correctly predict the onset of instabilities, when a specific wavelength is expected.

The numerical experiments show the development of the salt concentration also after onset of instabilities. In particular, we see how the salt concentration at the top of the domain continues to increase for some time after onset, as the instabilities are in the beginning too weak to cause a net downwards transport of salt. This means that salt can still precipitate even if instabilities have been triggered.

Our analysis open for comparison with column experiments considering evaporation from the top of a porous column saturated with saline water having different salts, e.g.~\cite{piortrowski2020crust}. However, the current analysis is performed under the assumption that the porous medium remains fully saturated. Hence, an extension to unsaturated porous media would give more accurate results.

\section*{Acknowledgements}
Funded by the Deutsche Forschungsgemeinschaft (DFG, German Research Foundation) – Project Number 327154368 – SFB 1313.

\appendix
\section{Explicit solution of the salt ground state}\label{app:A}
For the (non-dimensional) ground-state salt concentration $\hat X^0$, which is needed in Section \ref{sec:groundstate}, we here derive the explicit solution of 
\begin{align*}
    \partial_{\hat t} \hat X^0 = \partial_{\hat z}\hat X^0 + \partial^2_{\hat z}\hat X^0 \quad \hat z>0,\hat t>0, \\
    \hat X^0 +\partial_{\hat z}\hat X^0 = 0 \quad \hat z=0,\hat t>0,\\
    \hat X^0 = 1 \quad \hat z>0,\hat t=0.
\end{align*}
We rewrite this problem in terms of the flux; $f = \hat X^0 +\partial_{\hat z}\hat X^0$. Since $\partial_{\hat t} \hat X^0 = \partial_{\hat z} f$, we have
\begin{align*}
    \partial_{\hat t} f = \partial_{\hat t}\hat X^0 + \partial_{\hat t \hat z}\hat X^0 = \partial_{\hat z} f+\partial_{\hat z}^2 f \quad \hat z>0,\hat t>0, \\
    f = 0 \quad \hat z=0,\hat t>0,\\
    f = 1 \quad \hat z>0,\hat t=0.
\end{align*}
Writing in stead in terms of $g=1-f$, we have 
\begin{align*}
    \partial_{\hat t} g = \partial_{\hat z} g+\partial_{\hat z}^2 g \quad \hat z>0,\hat t>0, \\
    g = 1 \quad \hat z=0,\hat t>0,\\
    g = 0 \quad \hat z>0,\hat t=0,
\end{align*}
which has a known explicit solution \cite{bear2013dynamics}, namely
\begin{equation*}
    g(\hat t,\hat z) = \frac{1}{2}e^{-\hat z}\text{erfc}\Big(\frac{\hat z-\hat t}{2\sqrt{\hat t}}\Big)+\frac{1}{2}\text{erfc}\Big(\frac{\hat z+\hat t}{2\sqrt{\hat t}}\Big).
\end{equation*}
Hence,
\begin{equation*}
    f(\hat t,\hat z) = 1-\frac{1}{2}e^{-\hat z}\text{erfc}\Big(\frac{\hat z-\hat t}{2\sqrt{\hat t}}\Big)-\frac{1}{2}\text{erfc}\Big(\frac{\hat z+\hat t}{2\sqrt{\hat t}}\Big).
\end{equation*}
Using again that $\partial_{\hat t} \hat X^0 = \partial_{\hat z} f$, we obtain
\begin{equation*}
    \hat X^0(\hat t,\hat z) = 1+\int_0^{\hat t} \partial_{\hat z}f(\theta,\hat z)\ d\theta.
\end{equation*}

\section{Investigation of exponential growth rate}\label{app:B}
We here include the exponential growth rate $\sigma$ as a parameter in the eigenvalue problem, as considered in Section \ref{sec:eigenvalueproblem}. As introduced in \eqref{eq:sigma}, $\sigma>0$ corresponds to perturbations growing in time, while for $\sigma<0$ perturbations decay with time. When keeping $\sigma$ in the perturbed equations, we now obtain the following eigenvalue problem:\\
Given $\hat a>0$, $\hat t>0$, $\sigma\in\mathbb R$ and $\hat X^0=\hat X^0(\hat t,\hat z)$ by \eqref{eq:ustable}, find the smallest $R=R(\hat a,\hat t,\sigma)>0$ such that
\begin{equation}
    \label{eq:eigenvaluesigma}
    \tag{${\text{EP}}_\sigma$}
    \leqnomode
    \left\{\begin{array}{lr} 
    \hat w''+\hat a^2\hat \chi-\hat a^2\hat w = 0 & \hat z>0,\\
    \hat\chi' - R\hat w\partial_{\hat z}\hat X^0 + \hat\chi'' - (\hat a^2+\sigma)\hat\chi =0 & \hat z>0,\\
    \text{where } \hat w \text{ and } \hat \chi \text{ fulfill } & \\
    \hat w=0, \hat\chi+\hat\chi'=0 & \hat z=0,\\
    \hat w\to 0, \hat\chi\to 0 & \hat z\to\infty,
    \end{array}
    \right.
\end{equation}
has a non-trivial solution.

\begin{figure}
  \includegraphics[width=\textwidth]{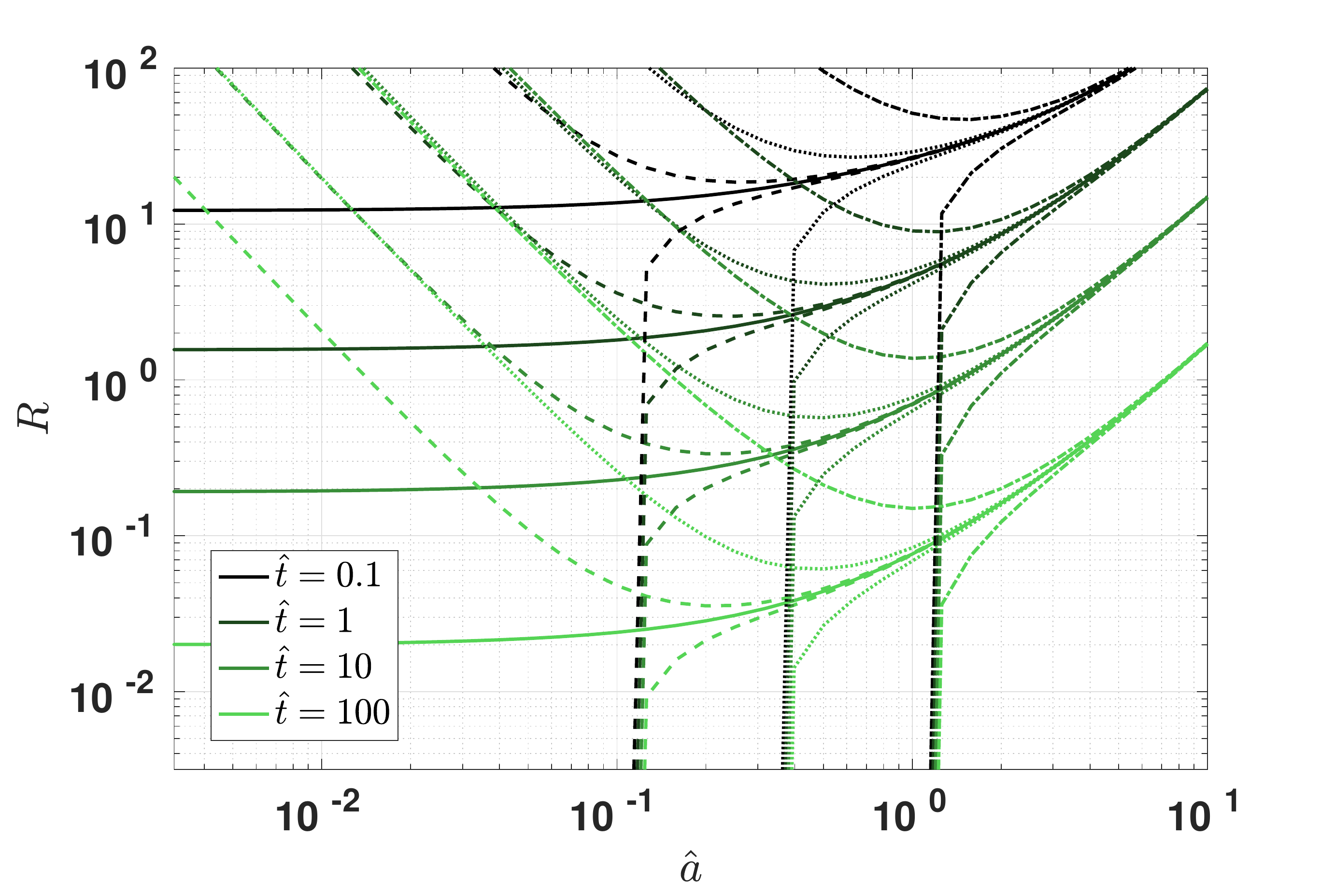}
\caption{Resulting eigenvalue $R$ (vertical axis) as a function of wavenumber $\hat a$ (horizontal axis) for various $\hat t$ and $\sigma$. Solid lines correspond to $\sigma=0$, while dashed lines correspond to $\sigma=\pm 0.01$, dotted lines to $\sigma=\pm0.1$ and dashed-dotted to $\sigma=\pm 1$. The curves lying below the corresponding solid curve are for negative $\sigma$ and the ones above the corresponding solid curve are for positive $\sigma$. Note that the eigenvalue problem degenerates when $(\hat a^2+\sigma)=0$, which is why the lines for negative $\sigma$ do not extend beyond $\hat a=\sqrt{-\sigma}$.}
\label{fig:Rvsa^0igma}     
\end{figure}

For given $\hat a,\hat t>0$ and $\sigma\in\mathbb R$, \eqref{eq:eigenvaluesigma} is an eigenvalue problem where $R$ is to be determined. We follow the same strategy as described in Section \ref{sec:linsolstrategy} to discretize and solve the eigenvalue problem. The corresponding version of Figure \ref{fig:RLvsa} when including $\sigma$ is shown in Figure \ref{fig:Rvsa^0igma}.

From Figure \ref{fig:Rvsa^0igma} we observe the following: For negative $\sigma$ (corresponding to perturbations decaying with time; i.e.~stability), we are always below the curves corresponding to $\sigma=0$ when looking at same $\hat t$ and $\hat a$. This means, when we are below a solid curve, which means that the Rayleigh number is lower than the one on the vertical axis, we have stability since a perturbation will decay with time. Similarly, for positive $\sigma$ (corresponding to perturbation growing with time), we are always above the curves corresponding to $\sigma=0$ when looking at same $\hat t$ and $\hat a$. This means, when we are above a solid curve, which means that the Rayleigh number is larger than the one on the vertical axis, a perturbation will grow exponentially with time. From this we can conclude that investigating $\sigma=0$ is the relevant case for the eigenvalue problem, as the eigenvalues found by using $\sigma=0$ correspond to the shift from perturbations growing or decaying. 

Note that the eigenvalue problem \eqref{eq:eigenvaluesigma} degenerates when $\hat a^2+\sigma=0$, which is why the curves for negative $\sigma$ are not extended beyond $\hat a=\sqrt{-\sigma}$. If one would like to investigate very small wavenumbers $\hat a$, one could overcome this by choosing a correspondingly small $\sigma$ to avoid (or, more correctly, shift) the degeneracy.

\section{The behavior for small wavenumbers $\hat a$}\label{app:c}
\begin{figure}
  \includegraphics[width=\textwidth]{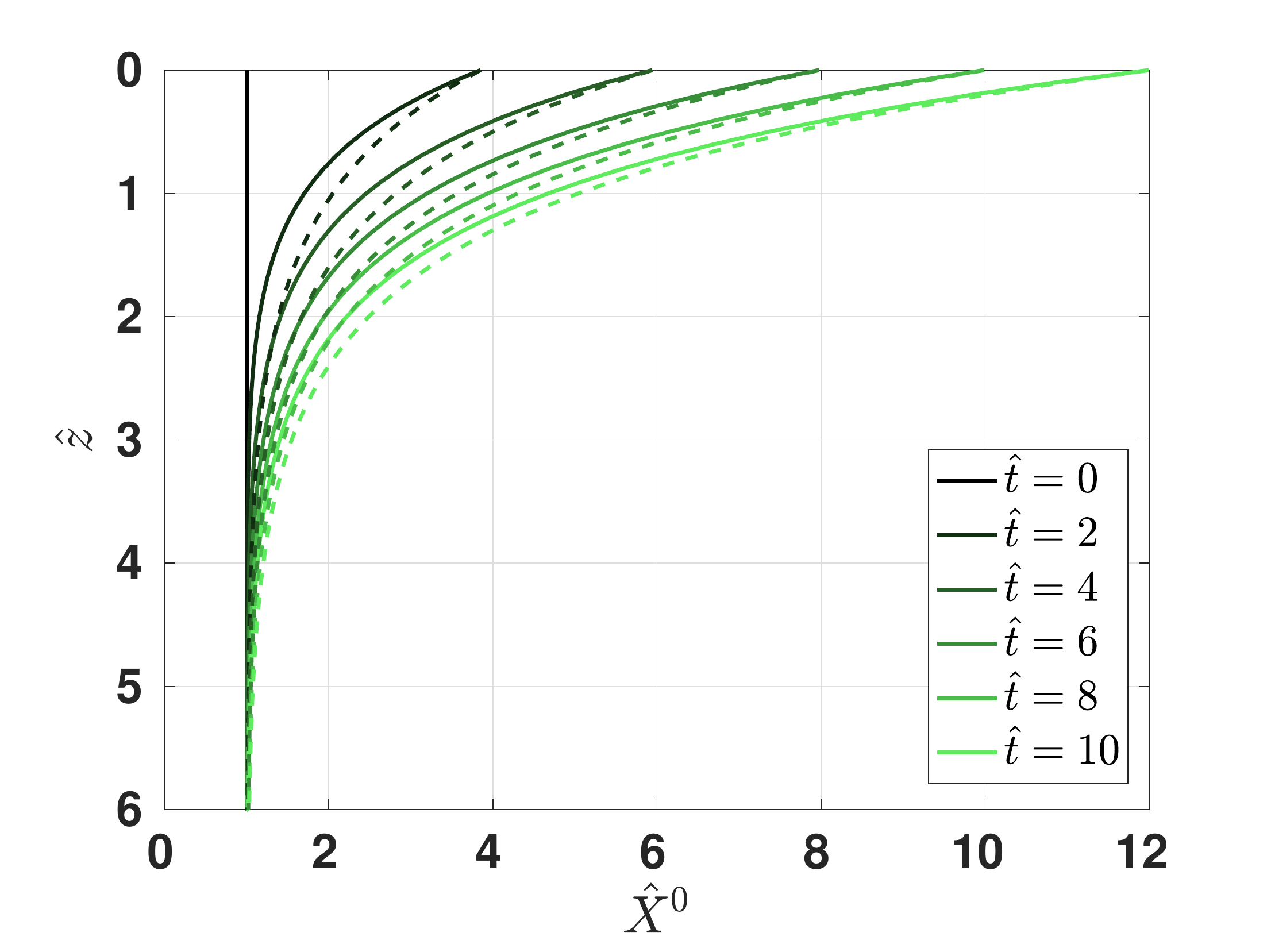}
\caption{Ground state solution for salt $\hat X^0$ using \eqref{eq:ustable} (solid lines) and approximate solution $\hat U^0$ using \eqref{eq:uapprox} (dashed lines).}
\label{fig:X0approx} 
\end{figure}

Figure \ref{fig:RLvsa} shows that for each $\hat t>0$, the Rayleigh number $R(\hat a,\hat t)$ has a finite value as $\hat a \searrow 0$. This behavior is different from other cases where $R(\hat a,\hat t)\to\infty$ as $\hat a\searrow 0$ and has a positive minimum at some critical wavenumber $\hat a_L>0$, e.g.~\cite{Riaz2006,vanduijn2002stability}. Here we provide an explanation for an approximate eigenvalue problem.
We first observe that for each $\hat t>0$, the ground state solution $\hat X^0(\hat t,\hat z)$ from \eqref{eq:ustable} has its maximum at $\hat z=0$ and decreases in a convex way towards $\hat X^0(\hat t,\infty)=1$. The idea is to replace $\hat X^0(\hat t,\hat z)$ by the expression
\begin{equation}\label{eq:uapprox}
    \hat U^0(\hat t,\hat z) = 1+e^{-\hat z}(\hat X^0(\hat t,0)-1).
\end{equation}
Figure \ref{fig:X0approx} shows the profiles of $\hat X^0$ and $\hat U^0$ for various $\hat t>0$. Note that they have the same qualitative behavior and that the relative error becomes smaller as $\hat t$ increases. We propose to use $\hat U^0(\hat t,\hat z)$ as ground state in the eigenvalue problem \eqref{eq:eigenvaluescaled}. Setting 
\begin{equation}\label{eq:Rstar}
    R^* = (\hat X^0(\hat t,0)-1)R,
\end{equation}
we obtain the neutral stability ($\sigma=0$) eigenvalue problem:\\
Given $\hat a>0$, find the smallest $R^*=R^*(\hat a)>0$ such that
\begin{equation}
    \label{eq:eigenvalueapprox}
    \tag{${\text{EP}}^*$}
    \leqnomode
    \left\{\begin{array}{lr} 
    \hat w''+\hat a^2\hat \chi-\hat a^2\hat w = 0 & \hat z>0,\\
    \hat\chi' + R^*e^{-\hat z}\hat w + \hat\chi'' - (\hat a^2+\sigma)\hat\chi =0 & \hat z>0,\\
    \text{where } \hat w \text{ and } \hat \chi \text{ fulfill } & \\
    \hat w=0, \hat\chi+\hat\chi'=0 & \hat z=0,\\
    \hat w\to 0, \hat\chi\to 0 & \hat z\to\infty,
    \end{array}
    \right.
\end{equation}
has a non-trivial solution.

The equations in \eqref{eq:eigenvalueapprox} can be combined into the fourth-order equation
\begin{equation}
    L[w] := (D_{\hat z}^2+D_{\hat z}-\hat a^2)(D_{\hat z}^2-\hat a^2) \hat w = \hat a^2R^*e^{-\hat z},\label{eq:Lw}
\end{equation}
where $D_{\hat z}$ denotes differentiation with respect to $\hat z$, resulting in the eigenvalue problem:\\
Given $\hat a>0$, find the smallest $R^*=R^*(\hat a)$ such that
\begin{equation}
    \label{eq:eigenvalueapprox4}
    \tag{$\widehat{{\text{EP}}}^*$}
    \leqnomode
    \left\{\begin{array}{lr} 
    L[\hat w] = \hat a^2R^*e^{-\hat z} \hat w & \hat z>0,\\
    \text{where } \hat w \text{ fulfills } & \\
    \hat w=0, (D_{\hat z}^2-\hat a^2)(D_{\hat z}+1)\hat w =0 & \hat z=0,\\
    \hat w\to 0 & \hat z\to\infty,
    \end{array}
    \right.
\end{equation}
has a non-trivial solution. 

Note that Equation \eqref{eq:Lw} also arises when studying the stability of the equilibrium state of the salt lake problem, see \cite{vanduijn2002stability} and references cited therein. In this study the eigenvalue problem differs from \eqref{eq:eigenvalueapprox4} only through the second boundary condition of $\hat w$ at $\hat z=0$. In the salt lake problem the second condition is $\hat\chi(0)=0$, implying that $D_{\hat z}^2\hat w(0)=0$.

\begin{figure}
    \begin{subfigure}{0.49\textwidth}
        \includegraphics[width=\textwidth]{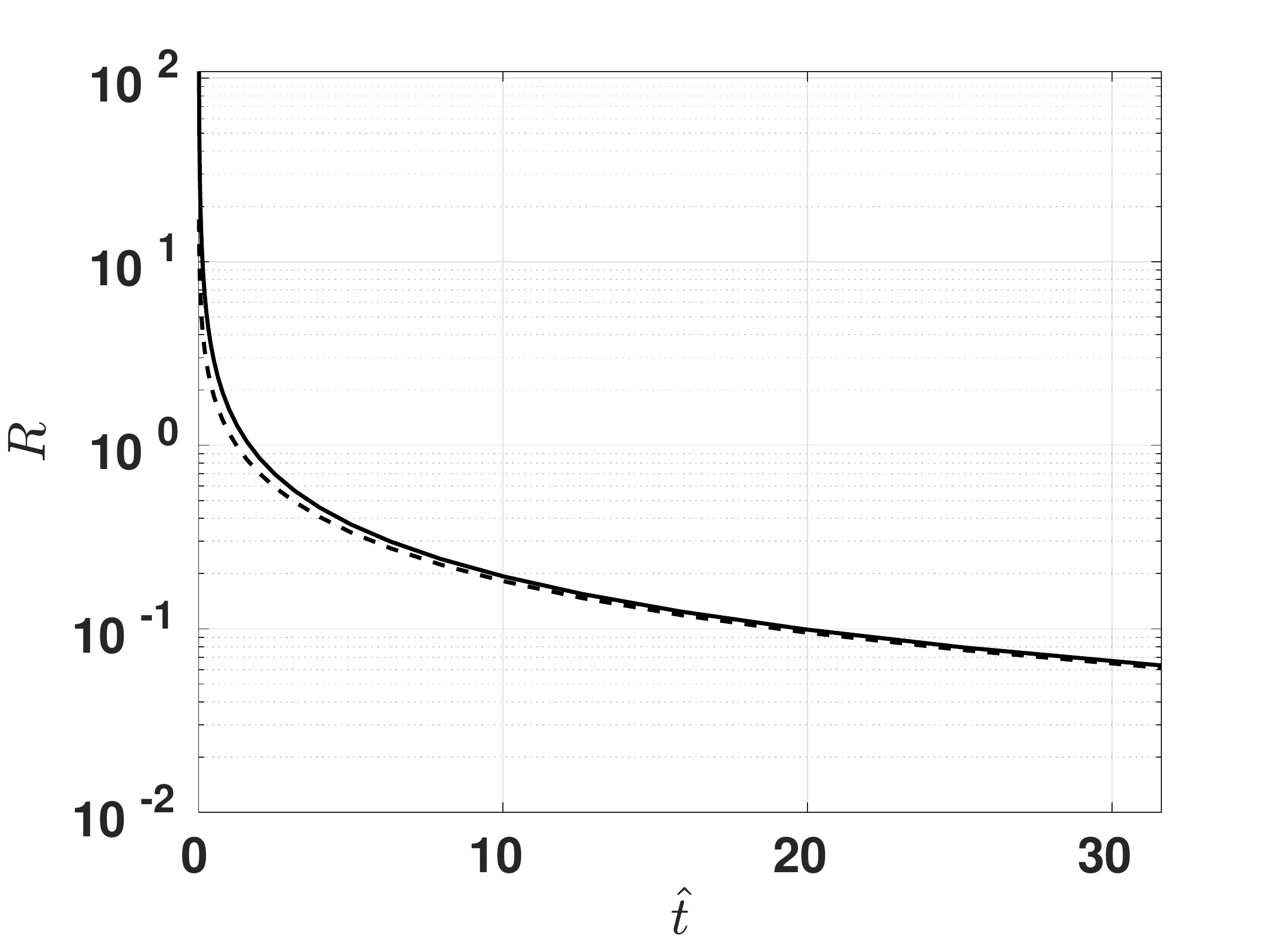}
        \subcaption{Critical (solid line) and approximate Rayleigh number (dashed line).}
    \end{subfigure}
        \begin{subfigure}{0.49\textwidth}
        \includegraphics[width=\textwidth]{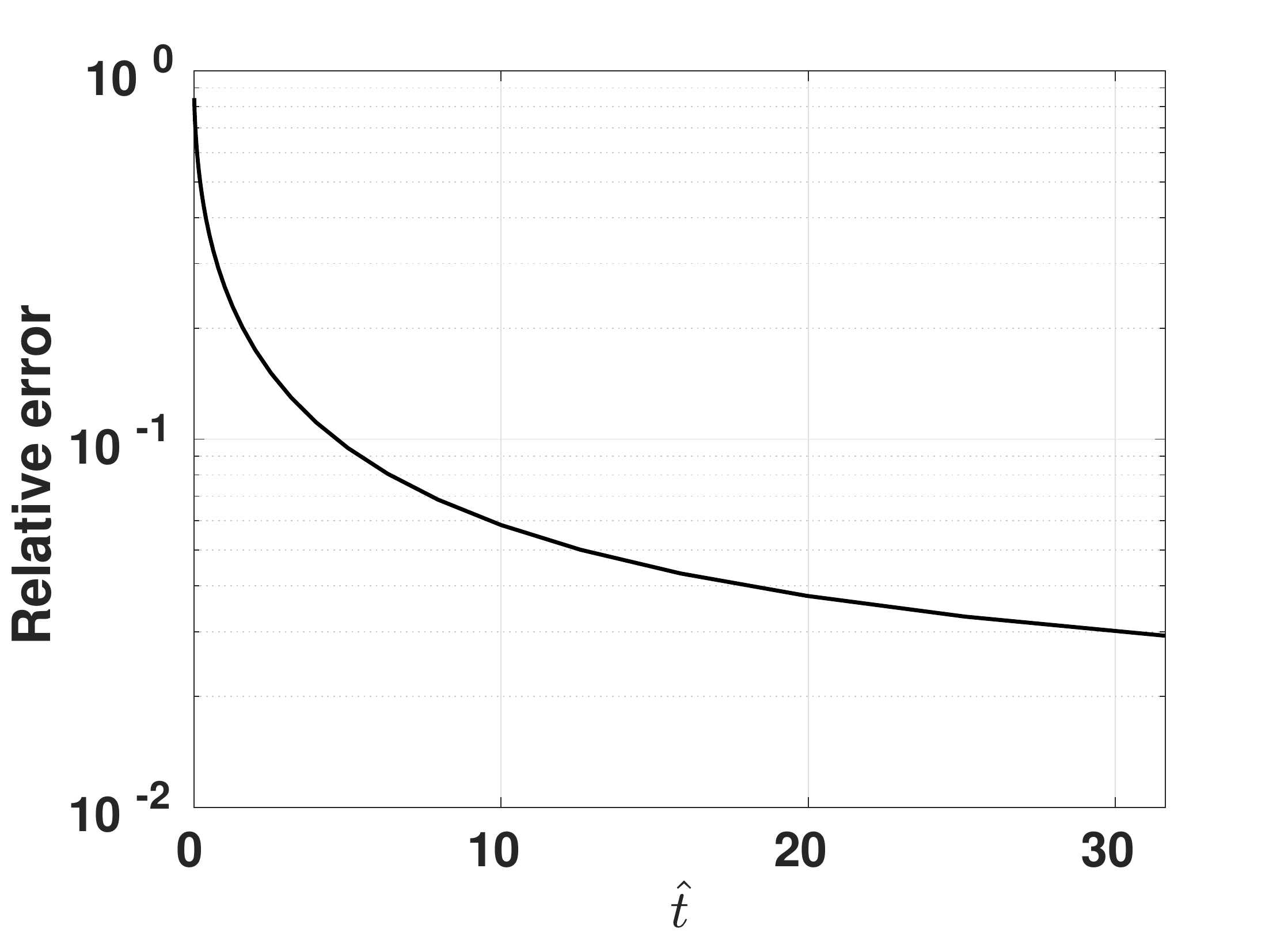}
        \subcaption{Relative error.}
    \end{subfigure}
\caption{Critical Rayleigh number from Section \ref{sec:linsolstrategy} using $\hat a=10^{-4}$ and approximate Rayleigh number from \eqref{eq:Rapprox}, and the relative error between them as a function of non-dimensional time $\hat t$.}
\label{fig:Rapprox} 
\end{figure}

Arguing as in \cite{vanduijn2002stability}, we treat \eqref{eq:eigenvalueapprox4} by a semi-analytical technique based on a Frobenius expansion in terms of descending exponential functions:
\begin{equation}\label{eq:Frobenius}
    \hat w(\hat z) = \sum_{n=0}^\infty A_n(R^*)^ne^{(c-n)\hat z},
\end{equation}
where $c<0$. Substituting \eqref{eq:Frobenius} into \eqref{eq:Lw} gives the indicial equation
\begin{equation*}
    (c^2+c-\hat a^2)(c^2-\hat a^2)=0,
\end{equation*}
yielding the negative roots 
\begin{equation*}
    c_1=-\hat a,\quad c_2 = -\frac{1}{2}-\sqrt{\frac{1}{4}+\hat a^2},
\end{equation*}
and the recurrence relation
\begin{equation*}
    \left\{ \begin{array}{lr}
       \frac{A_n^{(i)}}{A_{n-1}^{(i)}} = \frac{\hat a^2}{f_i(n;\hat a)}  & n\geq 1, \\
        A_0^{(i)}=1, &  
    \end{array} \right.
\end{equation*}
where
\begin{equation*}
    f_i(n;\hat a) = ((c_i-n)^2+(c_i-n)-\hat a^2)((c_i-n)^2-\hat a^2),\quad n\geq 1.
\end{equation*}
Hence, we obtain the power series solution
\begin{equation*}
    \hat w(\hat z) = A\hat w_1(\hat z) + B\hat w_2(\hat z),
\end{equation*}
where
\begin{equation*}
    \hat w_i (\hat z) = \sum_{n=0}^\infty A_n^{(i)}(R^*)^ne^{-(c_i-n)\hat z}, \quad i=1,2,\dots 
\end{equation*}
The boundary conditions at $\hat z=0$ are satisfied if
\begin{equation}\label{eq:BCFa}
    \big(\sum_{n=0}^\infty A_n^{(1)}(R^*)^n\big)A+\big(\sum_{n=0}^\infty A_n^{(2)}(R^*)^n\big)B=0
\end{equation}
and
\begin{equation}\label{eq:BCFb}
    \big(\sum_{n=0}^\infty A_n^{(1)}g_1(n;\hat a)(R^*)^n\big)A+\big(\sum_{n=0}^\infty A_n^{(2)}g_2(n,\hat a)(R^*)^n\big)B=0,
\end{equation}
where
\begin{equation*}
    g_i(n;\hat a) = (c_i-n)((c_i-n)^2+(c_i-n)-\hat a^2),\quad n\geq 0.
\end{equation*}
Writing \eqref{eq:BCFa} and \eqref{eq:BCFb} as
\begin{equation*}
    \begin{pmatrix}
    M
    \end{pmatrix}\begin{pmatrix}
    A\\ B
    \end{pmatrix}=\begin{pmatrix}
    0\\ 0
    \end{pmatrix},
\end{equation*}
we need to examine the characteristic equation $\det(M)=0$ to find the eigenvalues of \eqref{eq:eigenvalueapprox}. For $n\leq 2$ we have
\begin{align*}
    \det(M) &= \big(1+\frac{\hat a^2R^*}{f_1(1)}+\frac{(\hat a^2R^*)^2}{f_1(2)f_1(1)}\big)\big(\frac{g_2(1)}{f_2(1)}\hat a^2R^* + \frac{g_2(2)}{f_2(2)f_1(1)}(\hat aR^*)^2\big)\\ &- \big(1+\frac{\hat a^2R^*}{f_2(1)} + \frac{(\hat a^2R^*)^2}{f_2(2)f_2(1)}\big)\big(\hat a^2+\frac{g_1(1)}{f_1(1)}\hat a^2R^* + \frac{g_1(2)}{f_1(2)f_1(1)}(\hat a^2R^*)^2\big).
\end{align*}
For $\hat a\ll 1$, we take, to leading order, $n\leq 1$ and obtain
\begin{equation*}
     \frac{g_2(1)}{f_2(1)}\hat a^2R^* = \hat a^2 + \frac{g_1(1)}{f_1(1)}\hat a^2R^*,
\end{equation*}
or
\begin{equation*}
    R^*(\frac{g_2(1)}{f_2(1)}-\frac{g_1(1)}{f_1(1)}) = 1.
\end{equation*}
Evaluating the coefficients yields
\begin{equation*}
    \frac{g_2(1)}{f_2(1)}-\frac{g_1(1)}{f_1(1)} = \frac{-4+O(\hat a^2)}{8+O(\hat a^2)} - \frac{-\hat a(1+\hat a)}{\hat a(1+2\hat a)} = -\frac{1}{2} + O(\hat a^2) + \frac{1+\hat a}{1+2\hat a} = \frac{1}{2(1+2\hat a)}+O(\hat a^2). 
\end{equation*}
Hence
\begin{equation}
    R^*(\hat a) = 2(1+2\hat a) + O(\hat a^2) \text{ as } \hat a\searrow 0.\label{eq:Rabehavior}
\end{equation}
That means, in the limit as $\hat a$ approaches zero, we find
\begin{equation*}
    R^* = 2.
\end{equation*}
Using again \eqref{eq:Rstar}, we find that
\begin{equation}\label{eq:Rapprox}
    R = \frac{2}{\hat X^0(\hat t,0)-1}
\end{equation}
approximates the critical Rayleigh number for $\hat a$ approaching zero. The Rayleigh numbers using the strategy from Section \ref{sec:linsolstrategy} with $\hat a=10^{-4}$ together with the approximate ones from \eqref{eq:Rapprox} are shown in Figure \ref{fig:Rapprox}. The relative error between the approximate Rayleigh number and the Rayleigh number from Section \ref{sec:linsolstrategy} is larger for early times, as expected from Figure \ref{fig:X0approx} since the approximate ground state deviates more. However, the relative error is generally small, and decreases fast for later times. Hence, the approximate Rayleigh number \eqref{eq:Rapprox} represents the behavior for low wavenumbers $\hat a$ well. We also see from \eqref{eq:Rabehavior} that $R^*$ (and hence $R$) decreases monotonously for low $\hat a$ as $\hat a$ approaches zero. This shows that $\hat a=0$ corresponds to a (local) minimum for $R(\hat a,\hat t)$ for given $\hat t>0$.

\section{Grid and time step convergence study for the numerical simulations}\label{app:d}
\begin{figure}
    \centering
    \begin{subfigure}{0.49\textwidth}
        \includegraphics[width=\textwidth]{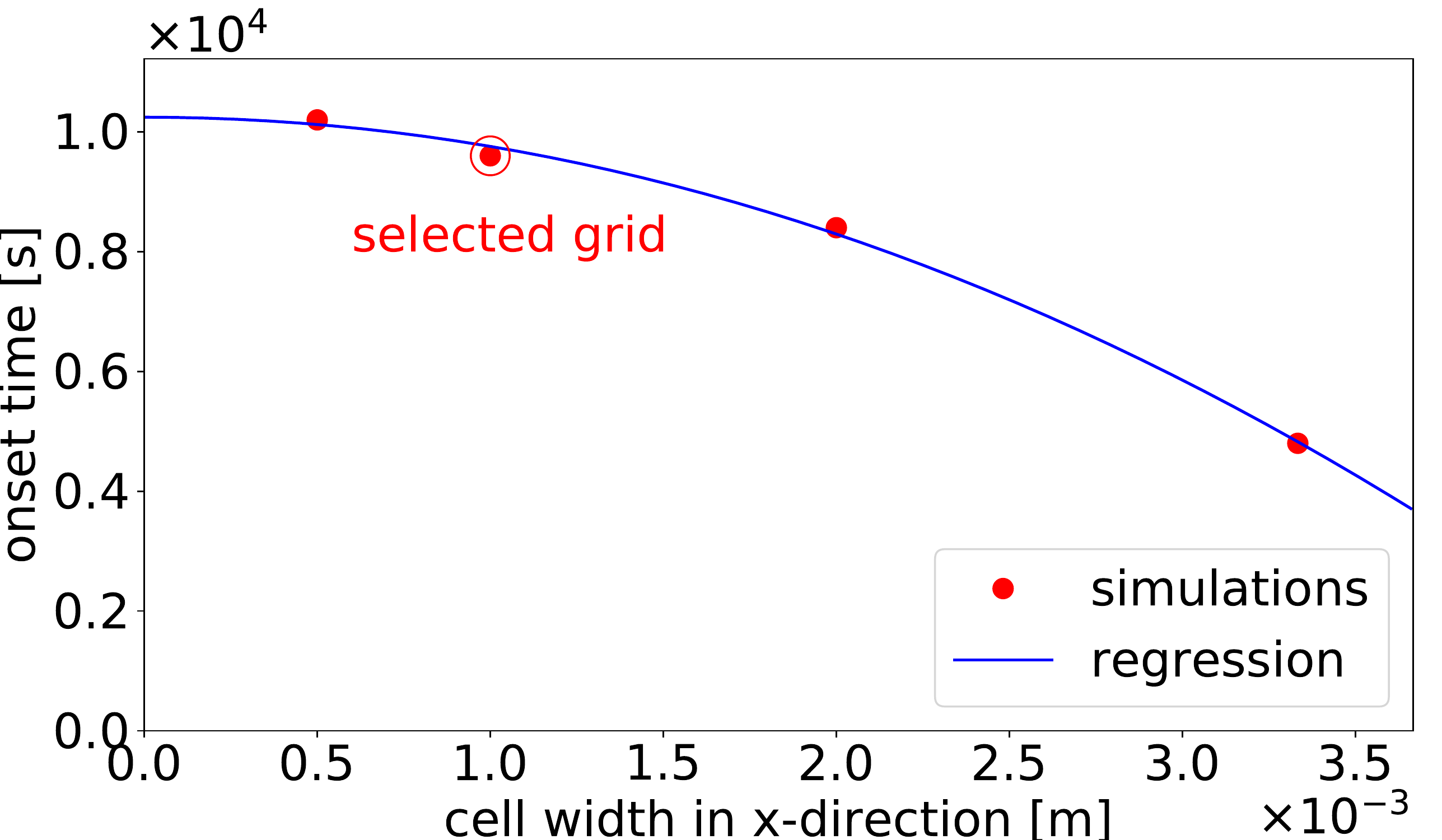}
        \subcaption{Onset time}
    \end{subfigure}
    \begin{subfigure}{0.49\textwidth}
        \includegraphics[width=\textwidth]{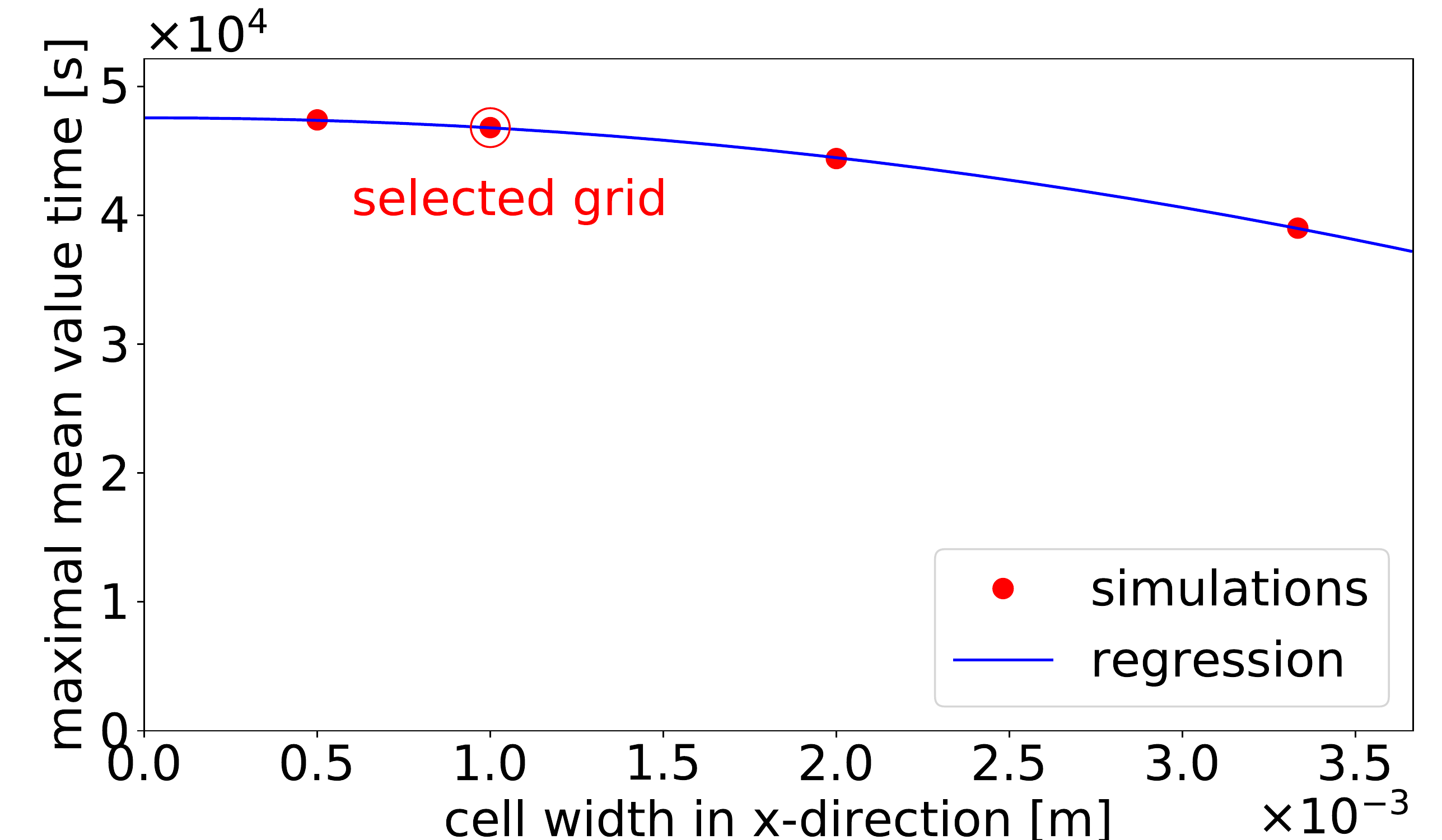}
        \subcaption{Time of maximal mean value $\mu_\mathrm{\mathsf{x}^\mathrm{NaCl}}^\mathrm{top}$}
    \end{subfigure}
    \caption{Influence of the grid cell size in $x$-direction on the onset time and the time of maximal mean value.}
    \label{fig:GridConvergence_x}
\end{figure}
\begin{figure}
    \centering
    \begin{subfigure}{0.49\textwidth}
        \includegraphics[width=\textwidth]{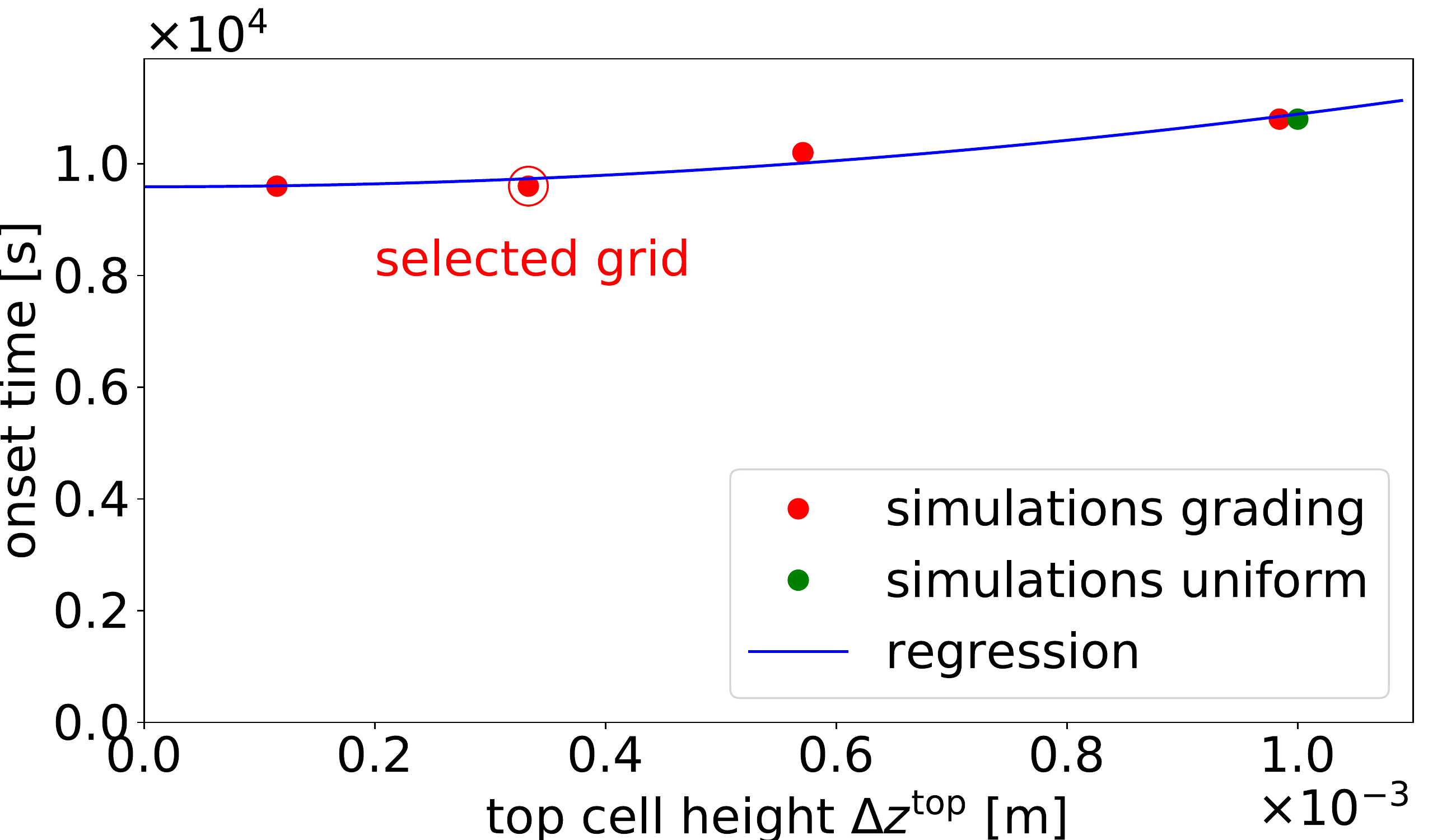}
        \subcaption{Onset time}
    \end{subfigure}
    \begin{subfigure}{0.49\textwidth}
        \includegraphics[width=\textwidth]{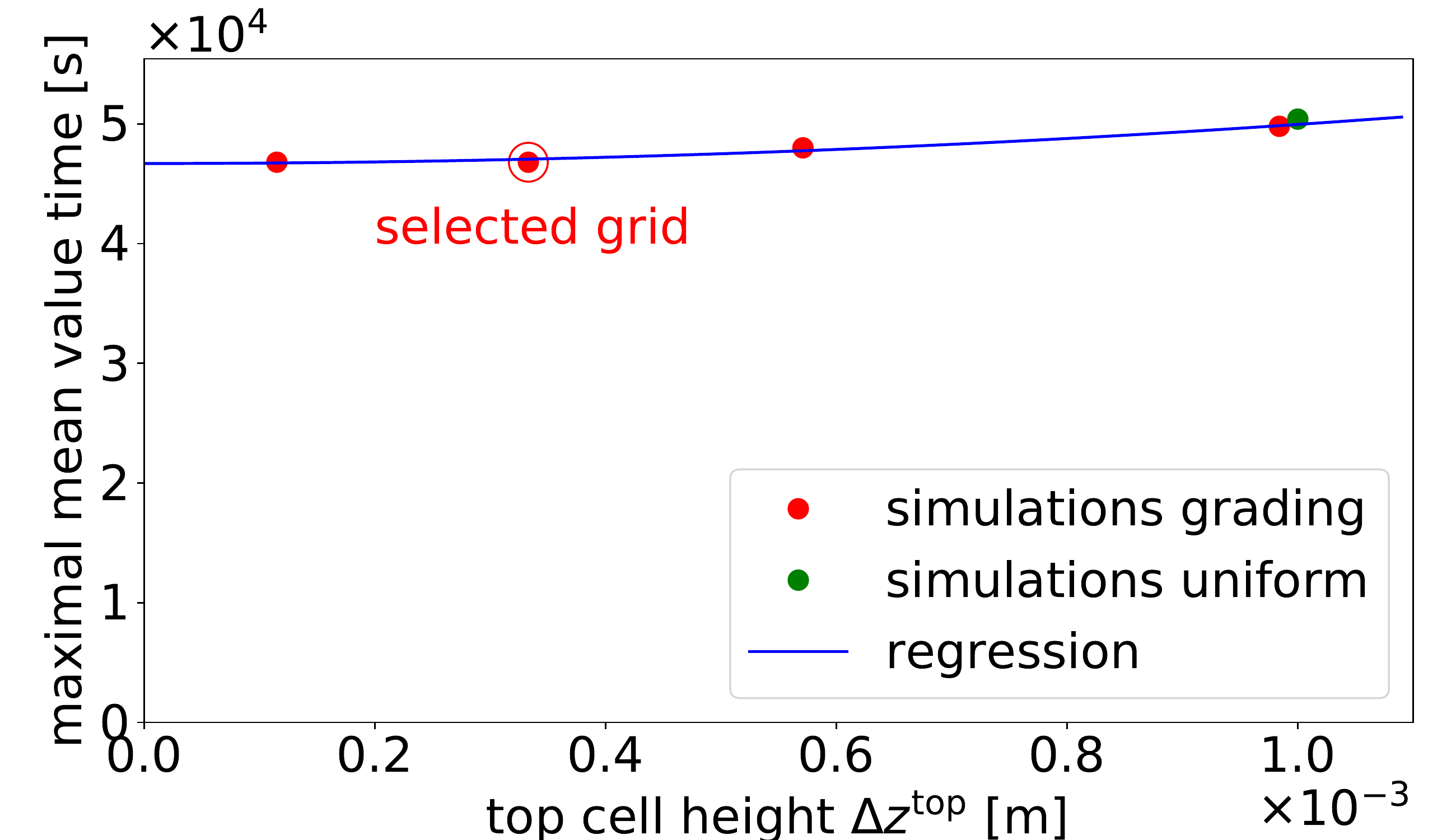}
        \subcaption{Time of maximal mean value $\mu_\mathrm{\mathsf{x}^\mathrm{NaCl}}^\mathrm{top}$}
    \end{subfigure}
    \caption{Influence of the height of the top cell $\Delta z^\mathrm{top}$ on the onset time and the time of maximal mean value.}
    \label{fig:GridConvergence_z}
\end{figure}
A convergence study was performed to ensure that the spatial and temporal discretization are able to capture the physical processes in sufficient detail and that the numerical diffusion has a negligible influence. The study was carried out using the same setup as described in Section~\ref{sec:results} exemplary for the case with a fixed wavelength of $\lambda=0.01$~m and a permeability of $K=10^{-10}~\mathrm{m^2}$. As this case has the lowest used wavelength and hence the lowest resolution per perturbation, the influence of the discretization should be even less for longer wavelengths. The influence of the cell width in $x$-direction, the discretization in $z$-direction and the time step size was investigated. The onset time and the time of the maximal mean value $\mu_\mathrm{\mathsf{x}^\mathrm{NaCl}}^\mathrm{top}$ are used to evaluate the influence. A quadratic regression is used to extrapolate the values for a theoretical infinitesimal small cell width, cell height and time step.

The influence of the cell widths in $x$-direction on the onset time and time of the maximal mean value is shown in Figure~\ref{fig:GridConvergence_x}. The discretization in $z$-direction equals the one of the selected grid as described below. 
The investigated cell widths $5 \cdot 10^{-4}$~m, $1 \cdot 10^{-3}$~m, $2 \cdot 10^{-3}$~m and $3.33 \cdot 10^{-3}$~m correlate with $20$, $10$, $5$ and $3$ cells per initially applied perturbation wavelength.
The selected grid uses 10 cells per perturbation wavelength. For longer wavelengths using lower permeabilities this results in more cells per perturbation wavelength. 
The deviation of the selected grid from the extrapolated value is $6.3~\%$ for the onset time and $1.6~\%$ for the time of maximal mean value.

Figure \ref{fig:GridConvergence_z} shows the investigation of the discretization in $z$-direction. In $x$-direction the discretization equals the one of the selected grid as described above. In $z$-direction a grading is used so that the top cell is the smallest and the cell height increases towards the bottom of the domain. The same grading factor of $0.9$ is used for the grids, but different numbers of cells. The different grids have $30$, $35$, $40$ and $50$ cells in $z$-direction, which correspond to a top cell height $\Delta z^\mathrm{top}$ of $9.84 \cdot 10^{-4}$~m, $5.71 \cdot 10^{-4}$~m, $3.33 \cdot 10^{-4}$~m and $1.15 \cdot 10^{-4}$~m.
Additionally a grid with uniform grid heights of $1.0 \cdot 10^{-3}$~m was simulated. It has in total $200$ cells in $z$-direction as it discretizes the lower parts of the domain finer, whereas $\Delta z^\mathrm{top}$ nearly matches the coarsest graded grid. Nearly no influence could be observed from the finer discretization in the lower domain, on the time of onset and in maximal mean value. It is found that the height of the top cell has the main influence. We therefor look at the height of the top cell $\Delta z^\mathrm{top}$ of the different grids. 
The selected grid uses grading and deviates $0.1~\%$ for the onset time and $0.2~\%$ for the time of maximal mean value from the respective theoretical extrapolated value. In comparison with the cell width, the cell height has less influence on the two evaluation times.

Further, an investigation of the time step size on the evaluation times was performed. It was found that for time steps from 25~s to 150~s, the evaluation is independent of the time discretization. For all simulations we use a time step of 50~s.

\bibliographystyle{plain} 
\bibliography{ms}  
\end{document}